%% file: ia0_prelaunch.tex
\documentclass[longauth]{aaEC}
\usepackage{graphicx}
\usepackage{euclid}
\usepackage{booktabs}
\usepackage{txfonts}
\usepackage[table]{xcolor}
\usepackage{natbib}
\usepackage[utf8]{inputenc}
\usepackage[switch, modulo]{lineno}
\usepackage[pdfencoding=auto,psdextra]{hyperref}
\usepackage{footmisc}
\hypersetup{
    colorlinks=true,
    linkcolor=blue,
    filecolor=magenta,      
    urlcolor=blue,
    citecolor=blue
}
\makeatletter
\renewcommand*\aa@pageof{, page \thepage{} of \pageref*{LastPage}}
\makeatother

\newcommand{\eq}[1]{\begin{equation}  #1 \end{equation}}
\newcommand{\eqa}[1]{\begin{eqnarray}   #1 \end{eqnarray}}
\newcommand{\sigMf}{$\sigma_{\rm MF}$\xspace}
\newcommand{\sigMfC}{$\sigma_{\rm MF}^{\rm cent}$\xspace}
\newcommand{\sigMfS}{$\sigma_{\rm MF}^{\rm sat}$\xspace}
\newcommand{\wgp}{$w_{\rm g+}$\xspace}
\newcommand{\wgg}{$w_{\rm gg}$\xspace}

\newcommand{\etaA}{$\eta_{\rm A}$\xspace}
\newcommand{\etaC}{$\eta_{\rm C}$\xspace}
\newcommand{\mpch}{$h^{-1}\mathrm{Mpc}$\xspace}
\newcommand{\chisqtot}{${\chi}^2_{\rm tot}$\xspace}
\newcommand{\cost}{$C$\xspace}
\newcommand{\flagship}{Flagship\xspace}

\begin{document} 
\title{\Euclid preparation}
\subtitle{Calibrated intrinsic galaxy alignments in the Euclid Flagship simulation}    

\newcommand{\orcid}[1]{} 
\author{Euclid Collaboration: K.~Hoffmann\orcid{0000-0002-7885-5274}\thanks{\email{kai.d.hoffmann@gmail.com}}\inst{\ref{aff1}}
\and R.~Paviot\orcid{0009-0002-8108-3460}\inst{\ref{aff2}}
\and B.~Joachimi\orcid{0000-0001-7494-1303}\inst{\ref{aff3}}
\and N.~Tessore\orcid{0000-0002-9696-7931}\inst{\ref{aff3}}
\and P.~Tallada-Cresp\'{i}\orcid{0000-0002-1336-8328}\inst{\ref{aff4},\ref{aff5}}
\and N.~E.~Chisari\orcid{0000-0003-4221-6718}\inst{\ref{aff6},\ref{aff7}}
\and E.~J.~Gonzalez\orcid{0000-0002-0226-9893}\inst{\ref{aff8},\ref{aff5},\ref{aff9}}
\and A.~Loureiro\orcid{0000-0002-4371-0876}\inst{\ref{aff10},\ref{aff11}}
\and P.~Fosalba\orcid{0000-0002-1510-5214}\inst{\ref{aff12},\ref{aff1}}
\and J.~Blazek\orcid{0000-0002-4687-4657}\inst{\ref{aff13}}
\and C.~Laigle\orcid{0009-0008-5926-818X}\inst{\ref{aff14}}
\and Y.~Dubois\orcid{0000-0003-0225-6387}\inst{\ref{aff14}}
\and C.~Pichon\orcid{0000-0003-0695-6735}\inst{\ref{aff14},\ref{aff15}}
\and B.~Altieri\orcid{0000-0003-3936-0284}\inst{\ref{aff16}}
\and S.~Andreon\orcid{0000-0002-2041-8784}\inst{\ref{aff17}}
\and N.~Auricchio\orcid{0000-0003-4444-8651}\inst{\ref{aff18}}
\and C.~Baccigalupi\orcid{0000-0002-8211-1630}\inst{\ref{aff19},\ref{aff20},\ref{aff21},\ref{aff22}}
\and M.~Baldi\orcid{0000-0003-4145-1943}\inst{\ref{aff23},\ref{aff18},\ref{aff24}}
\and S.~Bardelli\orcid{0000-0002-8900-0298}\inst{\ref{aff18}}
\and F.~Bernardeau\orcid{0009-0007-3015-2581}\inst{\ref{aff25},\ref{aff14}}
\and A.~Biviano\orcid{0000-0002-0857-0732}\inst{\ref{aff20},\ref{aff19}}
\and E.~Branchini\orcid{0000-0002-0808-6908}\inst{\ref{aff26},\ref{aff27},\ref{aff17}}
\and M.~Brescia\orcid{0000-0001-9506-5680}\inst{\ref{aff28},\ref{aff29}}
\and S.~Camera\orcid{0000-0003-3399-3574}\inst{\ref{aff30},\ref{aff31},\ref{aff32}}
\and G.~Ca\~nas-Herrera\orcid{0000-0003-2796-2149}\inst{\ref{aff33},\ref{aff34},\ref{aff7}}
\and V.~Capobianco\orcid{0000-0002-3309-7692}\inst{\ref{aff32}}
\and C.~Carbone\orcid{0000-0003-0125-3563}\inst{\ref{aff35}}
\and V.~F.~Cardone\inst{\ref{aff36},\ref{aff37}}
\and J.~Carretero\orcid{0000-0002-3130-0204}\inst{\ref{aff4},\ref{aff5}}
\and S.~Casas\orcid{0000-0002-4751-5138}\inst{\ref{aff38}}
\and F.~J.~Castander\orcid{0000-0001-7316-4573}\inst{\ref{aff1},\ref{aff12}}
\and M.~Castellano\orcid{0000-0001-9875-8263}\inst{\ref{aff36}}
\and G.~Castignani\orcid{0000-0001-6831-0687}\inst{\ref{aff18}}
\and S.~Cavuoti\orcid{0000-0002-3787-4196}\inst{\ref{aff29},\ref{aff39}}
\and K.~C.~Chambers\orcid{0000-0001-6965-7789}\inst{\ref{aff40}}
\and A.~Cimatti\inst{\ref{aff41}}
\and C.~Colodro-Conde\inst{\ref{aff42}}
\and G.~Congedo\orcid{0000-0003-2508-0046}\inst{\ref{aff43}}
\and L.~Conversi\orcid{0000-0002-6710-8476}\inst{\ref{aff44},\ref{aff16}}
\and Y.~Copin\orcid{0000-0002-5317-7518}\inst{\ref{aff45}}
\and F.~Courbin\orcid{0000-0003-0758-6510}\inst{\ref{aff46},\ref{aff47}}
\and H.~M.~Courtois\orcid{0000-0003-0509-1776}\inst{\ref{aff48}}
\and A.~Da~Silva\orcid{0000-0002-6385-1609}\inst{\ref{aff49},\ref{aff50}}
\and H.~Degaudenzi\orcid{0000-0002-5887-6799}\inst{\ref{aff51}}
\and G.~De~Lucia\orcid{0000-0002-6220-9104}\inst{\ref{aff20}}
\and H.~Dole\orcid{0000-0002-9767-3839}\inst{\ref{aff52}}
\and F.~Dubath\orcid{0000-0002-6533-2810}\inst{\ref{aff51}}
\and C.~A.~J.~Duncan\orcid{0009-0003-3573-0791}\inst{\ref{aff43},\ref{aff53}}
\and X.~Dupac\inst{\ref{aff16}}
\and S.~Dusini\orcid{0000-0002-1128-0664}\inst{\ref{aff54}}
\and S.~Escoffier\orcid{0000-0002-2847-7498}\inst{\ref{aff55}}
\and M.~Farina\orcid{0000-0002-3089-7846}\inst{\ref{aff56}}
\and R.~Farinelli\inst{\ref{aff18}}
\and S.~Farrens\orcid{0000-0002-9594-9387}\inst{\ref{aff2}}
\and S.~Ferriol\inst{\ref{aff45}}
\and F.~Finelli\orcid{0000-0002-6694-3269}\inst{\ref{aff18},\ref{aff57}}
\and N.~Fourmanoit\orcid{0009-0005-6816-6925}\inst{\ref{aff55}}
\and M.~Frailis\orcid{0000-0002-7400-2135}\inst{\ref{aff20}}
\and E.~Franceschi\orcid{0000-0002-0585-6591}\inst{\ref{aff18}}
\and M.~Fumana\orcid{0000-0001-6787-5950}\inst{\ref{aff35}}
\and S.~Galeotta\orcid{0000-0002-3748-5115}\inst{\ref{aff20}}
\and K.~George\orcid{0000-0002-1734-8455}\inst{\ref{aff58}}
\and B.~Gillis\orcid{0000-0002-4478-1270}\inst{\ref{aff43}}
\and C.~Giocoli\orcid{0000-0002-9590-7961}\inst{\ref{aff18},\ref{aff24}}
\and J.~Gracia-Carpio\inst{\ref{aff59}}
\and A.~Grazian\orcid{0000-0002-5688-0663}\inst{\ref{aff60}}
\and F.~Grupp\inst{\ref{aff59},\ref{aff61}}
\and S.~V.~H.~Haugan\orcid{0000-0001-9648-7260}\inst{\ref{aff62}}
\and H.~Hoekstra\orcid{0000-0002-0641-3231}\inst{\ref{aff7}}
\and W.~Holmes\inst{\ref{aff63}}
\and F.~Hormuth\inst{\ref{aff64}}
\and A.~Hornstrup\orcid{0000-0002-3363-0936}\inst{\ref{aff65},\ref{aff66}}
\and K.~Jahnke\orcid{0000-0003-3804-2137}\inst{\ref{aff67}}
\and M.~Jhabvala\inst{\ref{aff68}}
\and E.~Keih\"anen\orcid{0000-0003-1804-7715}\inst{\ref{aff69}}
\and S.~Kermiche\orcid{0000-0002-0302-5735}\inst{\ref{aff55}}
\and A.~Kiessling\orcid{0000-0002-2590-1273}\inst{\ref{aff63}}
\and M.~Kilbinger\orcid{0000-0001-9513-7138}\inst{\ref{aff2}}
\and B.~Kubik\orcid{0009-0006-5823-4880}\inst{\ref{aff45}}
\and M.~K\"ummel\orcid{0000-0003-2791-2117}\inst{\ref{aff61}}
\and M.~Kunz\orcid{0000-0002-3052-7394}\inst{\ref{aff70}}
\and H.~Kurki-Suonio\orcid{0000-0002-4618-3063}\inst{\ref{aff71},\ref{aff72}}
\and A.~M.~C.~Le~Brun\orcid{0000-0002-0936-4594}\inst{\ref{aff73}}
\and S.~Ligori\orcid{0000-0003-4172-4606}\inst{\ref{aff32}}
\and P.~B.~Lilje\orcid{0000-0003-4324-7794}\inst{\ref{aff62}}
\and V.~Lindholm\orcid{0000-0003-2317-5471}\inst{\ref{aff71},\ref{aff72}}
\and I.~Lloro\orcid{0000-0001-5966-1434}\inst{\ref{aff74}}
\and G.~Mainetti\orcid{0000-0003-2384-2377}\inst{\ref{aff75}}
\and D.~Maino\inst{\ref{aff76},\ref{aff35},\ref{aff77}}
\and E.~Maiorano\orcid{0000-0003-2593-4355}\inst{\ref{aff18}}
\and O.~Mansutti\orcid{0000-0001-5758-4658}\inst{\ref{aff20}}
\and S.~Marcin\inst{\ref{aff78}}
\and O.~Marggraf\orcid{0000-0001-7242-3852}\inst{\ref{aff79}}
\and M.~Martinelli\orcid{0000-0002-6943-7732}\inst{\ref{aff36},\ref{aff37}}
\and N.~Martinet\orcid{0000-0003-2786-7790}\inst{\ref{aff80}}
\and F.~Marulli\orcid{0000-0002-8850-0303}\inst{\ref{aff81},\ref{aff18},\ref{aff24}}
\and R.~J.~Massey\orcid{0000-0002-6085-3780}\inst{\ref{aff82}}
\and E.~Medinaceli\orcid{0000-0002-4040-7783}\inst{\ref{aff18}}
\and S.~Mei\orcid{0000-0002-2849-559X}\inst{\ref{aff83},\ref{aff84}}
\and Y.~Mellier\inst{\ref{aff85},\ref{aff14}}
\and M.~Meneghetti\orcid{0000-0003-1225-7084}\inst{\ref{aff18},\ref{aff24}}
\and E.~Merlin\orcid{0000-0001-6870-8900}\inst{\ref{aff36}}
\and G.~Meylan\inst{\ref{aff86}}
\and A.~Mora\orcid{0000-0002-1922-8529}\inst{\ref{aff87}}
\and M.~Moresco\orcid{0000-0002-7616-7136}\inst{\ref{aff81},\ref{aff18}}
\and L.~Moscardini\orcid{0000-0002-3473-6716}\inst{\ref{aff81},\ref{aff18},\ref{aff24}}
\and C.~Neissner\orcid{0000-0001-8524-4968}\inst{\ref{aff8},\ref{aff5}}
\and S.-M.~Niemi\orcid{0009-0005-0247-0086}\inst{\ref{aff33}}
\and C.~Padilla\orcid{0000-0001-7951-0166}\inst{\ref{aff8}}
\and S.~Paltani\orcid{0000-0002-8108-9179}\inst{\ref{aff51}}
\and F.~Pasian\orcid{0000-0002-4869-3227}\inst{\ref{aff20}}
\and K.~Pedersen\inst{\ref{aff88}}
\and V.~Pettorino\orcid{0000-0002-4203-9320}\inst{\ref{aff33}}
\and S.~Pires\orcid{0000-0002-0249-2104}\inst{\ref{aff2}}
\and G.~Polenta\orcid{0000-0003-4067-9196}\inst{\ref{aff89}}
\and M.~Poncet\inst{\ref{aff90}}
\and L.~A.~Popa\inst{\ref{aff91}}
\and L.~Pozzetti\orcid{0000-0001-7085-0412}\inst{\ref{aff18}}
\and F.~Raison\orcid{0000-0002-7819-6918}\inst{\ref{aff59}}
\and R.~Rebolo\orcid{0000-0003-3767-7085}\inst{\ref{aff42},\ref{aff92},\ref{aff93}}
\and A.~Renzi\orcid{0000-0001-9856-1970}\inst{\ref{aff94},\ref{aff54}}
\and J.~Rhodes\orcid{0000-0002-4485-8549}\inst{\ref{aff63}}
\and G.~Riccio\inst{\ref{aff29}}
\and E.~Romelli\orcid{0000-0003-3069-9222}\inst{\ref{aff20}}
\and M.~Roncarelli\orcid{0000-0001-9587-7822}\inst{\ref{aff18}}
\and R.~Saglia\orcid{0000-0003-0378-7032}\inst{\ref{aff61},\ref{aff59}}
\and Z.~Sakr\orcid{0000-0002-4823-3757}\inst{\ref{aff95},\ref{aff96},\ref{aff97}}
\and A.~G.~S\'anchez\orcid{0000-0003-1198-831X}\inst{\ref{aff59}}
\and D.~Sapone\orcid{0000-0001-7089-4503}\inst{\ref{aff98}}
\and B.~Sartoris\orcid{0000-0003-1337-5269}\inst{\ref{aff61},\ref{aff20}}
\and P.~Schneider\orcid{0000-0001-8561-2679}\inst{\ref{aff79}}
\and T.~Schrabback\orcid{0000-0002-6987-7834}\inst{\ref{aff99}}
\and A.~Secroun\orcid{0000-0003-0505-3710}\inst{\ref{aff55}}
\and E.~Sefusatti\orcid{0000-0003-0473-1567}\inst{\ref{aff20},\ref{aff19},\ref{aff21}}
\and G.~Seidel\orcid{0000-0003-2907-353X}\inst{\ref{aff67}}
\and S.~Serrano\orcid{0000-0002-0211-2861}\inst{\ref{aff12},\ref{aff100},\ref{aff1}}
\and P.~Simon\inst{\ref{aff79}}
\and C.~Sirignano\orcid{0000-0002-0995-7146}\inst{\ref{aff94},\ref{aff54}}
\and G.~Sirri\orcid{0000-0003-2626-2853}\inst{\ref{aff24}}
\and A.~Spurio~Mancini\orcid{0000-0001-5698-0990}\inst{\ref{aff101}}
\and L.~Stanco\orcid{0000-0002-9706-5104}\inst{\ref{aff54}}
\and J.~Steinwagner\orcid{0000-0001-7443-1047}\inst{\ref{aff59}}
\and A.~N.~Taylor\inst{\ref{aff43}}
\and I.~Tereno\orcid{0000-0002-4537-6218}\inst{\ref{aff49},\ref{aff102}}
\and S.~Toft\orcid{0000-0003-3631-7176}\inst{\ref{aff103},\ref{aff104}}
\and R.~Toledo-Moreo\orcid{0000-0002-2997-4859}\inst{\ref{aff105}}
\and F.~Torradeflot\orcid{0000-0003-1160-1517}\inst{\ref{aff5},\ref{aff4}}
\and I.~Tutusaus\orcid{0000-0002-3199-0399}\inst{\ref{aff1},\ref{aff12},\ref{aff96}}
\and L.~Valenziano\orcid{0000-0002-1170-0104}\inst{\ref{aff18},\ref{aff57}}
\and J.~Valiviita\orcid{0000-0001-6225-3693}\inst{\ref{aff71},\ref{aff72}}
\and T.~Vassallo\orcid{0000-0001-6512-6358}\inst{\ref{aff20}}
\and A.~Veropalumbo\orcid{0000-0003-2387-1194}\inst{\ref{aff17},\ref{aff27},\ref{aff26}}
\and Y.~Wang\orcid{0000-0002-4749-2984}\inst{\ref{aff106}}
\and J.~Weller\orcid{0000-0002-8282-2010}\inst{\ref{aff61},\ref{aff59}}
\and G.~Zamorani\orcid{0000-0002-2318-301X}\inst{\ref{aff18}}
\and F.~M.~Zerbi\inst{\ref{aff17}}
\and E.~Zucca\orcid{0000-0002-5845-8132}\inst{\ref{aff18}}
\and M.~Ballardini\orcid{0000-0003-4481-3559}\inst{\ref{aff107},\ref{aff108},\ref{aff18}}
\and E.~Bozzo\orcid{0000-0002-8201-1525}\inst{\ref{aff51}}
\and C.~Burigana\orcid{0000-0002-3005-5796}\inst{\ref{aff109},\ref{aff57}}
\and R.~Cabanac\orcid{0000-0001-6679-2600}\inst{\ref{aff96}}
\and M.~Calabrese\orcid{0000-0002-2637-2422}\inst{\ref{aff110},\ref{aff35}}
\and A.~Cappi\inst{\ref{aff18},\ref{aff111}}
\and D.~Di~Ferdinando\inst{\ref{aff24}}
\and J.~A.~Escartin~Vigo\inst{\ref{aff59}}
\and L.~Gabarra\orcid{0000-0002-8486-8856}\inst{\ref{aff112}}
\and W.~G.~Hartley\inst{\ref{aff51}}
\and S.~Matthew\orcid{0000-0001-8448-1697}\inst{\ref{aff43}}
\and M.~Maturi\orcid{0000-0002-3517-2422}\inst{\ref{aff95},\ref{aff113}}
\and N.~Mauri\orcid{0000-0001-8196-1548}\inst{\ref{aff41},\ref{aff24}}
\and R.~B.~Metcalf\orcid{0000-0003-3167-2574}\inst{\ref{aff81},\ref{aff18}}
\and A.~Pezzotta\orcid{0000-0003-0726-2268}\inst{\ref{aff17}}
\and M.~P\"ontinen\orcid{0000-0001-5442-2530}\inst{\ref{aff71}}
\and C.~Porciani\orcid{0000-0002-7797-2508}\inst{\ref{aff79}}
\and I.~Risso\orcid{0000-0003-2525-7761}\inst{\ref{aff17},\ref{aff27}}
\and V.~Scottez\orcid{0009-0008-3864-940X}\inst{\ref{aff85},\ref{aff114}}
\and M.~Sereno\orcid{0000-0003-0302-0325}\inst{\ref{aff18},\ref{aff24}}
\and M.~Tenti\orcid{0000-0002-4254-5901}\inst{\ref{aff24}}
\and M.~Viel\orcid{0000-0002-2642-5707}\inst{\ref{aff19},\ref{aff20},\ref{aff22},\ref{aff21},\ref{aff115}}
\and M.~Wiesmann\orcid{0009-0000-8199-5860}\inst{\ref{aff62}}
\and Y.~Akrami\orcid{0000-0002-2407-7956}\inst{\ref{aff116},\ref{aff117}}
\and S.~Alvi\orcid{0000-0001-5779-8568}\inst{\ref{aff107}}
\and I.~T.~Andika\orcid{0000-0001-6102-9526}\inst{\ref{aff118},\ref{aff119}}
\and S.~Anselmi\orcid{0000-0002-3579-9583}\inst{\ref{aff54},\ref{aff94},\ref{aff120}}
\and M.~Archidiacono\orcid{0000-0003-4952-9012}\inst{\ref{aff76},\ref{aff77}}
\and F.~Atrio-Barandela\orcid{0000-0002-2130-2513}\inst{\ref{aff121}}
\and D.~Bertacca\orcid{0000-0002-2490-7139}\inst{\ref{aff94},\ref{aff60},\ref{aff54}}
\and M.~Bethermin\orcid{0000-0002-3915-2015}\inst{\ref{aff122}}
\and A.~Blanchard\orcid{0000-0001-8555-9003}\inst{\ref{aff96}}
\and L.~Blot\orcid{0000-0002-9622-7167}\inst{\ref{aff123},\ref{aff73}}
\and M.~Bonici\orcid{0000-0002-8430-126X}\inst{\ref{aff124},\ref{aff35}}
\and S.~Borgani\orcid{0000-0001-6151-6439}\inst{\ref{aff125},\ref{aff19},\ref{aff20},\ref{aff21},\ref{aff115}}
\and M.~L.~Brown\orcid{0000-0002-0370-8077}\inst{\ref{aff53}}
\and S.~Bruton\orcid{0000-0002-6503-5218}\inst{\ref{aff126}}
\and A.~Calabro\orcid{0000-0003-2536-1614}\inst{\ref{aff36}}
\and B.~Camacho~Quevedo\orcid{0000-0002-8789-4232}\inst{\ref{aff19},\ref{aff22},\ref{aff20}}
\and F.~Caro\inst{\ref{aff36}}
\and C.~S.~Carvalho\inst{\ref{aff102}}
\and T.~Castro\orcid{0000-0002-6292-3228}\inst{\ref{aff20},\ref{aff21},\ref{aff19},\ref{aff115}}
\and F.~Cogato\orcid{0000-0003-4632-6113}\inst{\ref{aff81},\ref{aff18}}
\and S.~Conseil\orcid{0000-0002-3657-4191}\inst{\ref{aff45}}
\and A.~R.~Cooray\orcid{0000-0002-3892-0190}\inst{\ref{aff127}}
\and O.~Cucciati\orcid{0000-0002-9336-7551}\inst{\ref{aff18}}
\and S.~Davini\orcid{0000-0003-3269-1718}\inst{\ref{aff27}}
\and G.~Desprez\orcid{0000-0001-8325-1742}\inst{\ref{aff128}}
\and A.~D\'iaz-S\'anchez\orcid{0000-0003-0748-4768}\inst{\ref{aff129}}
\and J.~J.~Diaz\orcid{0000-0003-2101-1078}\inst{\ref{aff42}}
\and S.~Di~Domizio\orcid{0000-0003-2863-5895}\inst{\ref{aff26},\ref{aff27}}
\and J.~M.~Diego\orcid{0000-0001-9065-3926}\inst{\ref{aff130}}
\and M.~Y.~Elkhashab\orcid{0000-0001-9306-2603}\inst{\ref{aff20},\ref{aff21},\ref{aff125},\ref{aff19}}
\and A.~Enia\orcid{0000-0002-0200-2857}\inst{\ref{aff23},\ref{aff18}}
\and Y.~Fang\inst{\ref{aff61}}
\and A.~G.~Ferrari\orcid{0009-0005-5266-4110}\inst{\ref{aff24}}
\and A.~Finoguenov\orcid{0000-0002-4606-5403}\inst{\ref{aff71}}
\and A.~Fontana\orcid{0000-0003-3820-2823}\inst{\ref{aff36}}
\and A.~Franco\orcid{0000-0002-4761-366X}\inst{\ref{aff131},\ref{aff132},\ref{aff133}}
\and K.~Ganga\orcid{0000-0001-8159-8208}\inst{\ref{aff83}}
\and J.~Garc\'ia-Bellido\orcid{0000-0002-9370-8360}\inst{\ref{aff116}}
\and T.~Gasparetto\orcid{0000-0002-7913-4866}\inst{\ref{aff36}}
\and V.~Gautard\inst{\ref{aff134}}
\and R.~Gavazzi\orcid{0000-0002-5540-6935}\inst{\ref{aff80},\ref{aff14}}
\and E.~Gaztanaga\orcid{0000-0001-9632-0815}\inst{\ref{aff1},\ref{aff12},\ref{aff135}}
\and F.~Giacomini\orcid{0000-0002-3129-2814}\inst{\ref{aff24}}
\and F.~Gianotti\orcid{0000-0003-4666-119X}\inst{\ref{aff18}}
\and G.~Gozaliasl\orcid{0000-0002-0236-919X}\inst{\ref{aff136},\ref{aff71}}
\and M.~Guidi\orcid{0000-0001-9408-1101}\inst{\ref{aff23},\ref{aff18}}
\and C.~M.~Gutierrez\orcid{0000-0001-7854-783X}\inst{\ref{aff137}}
\and A.~Hall\orcid{0000-0002-3139-8651}\inst{\ref{aff43}}
\and S.~Hemmati\orcid{0000-0003-2226-5395}\inst{\ref{aff138}}
\and H.~Hildebrandt\orcid{0000-0002-9814-3338}\inst{\ref{aff139}}
\and J.~Hjorth\orcid{0000-0002-4571-2306}\inst{\ref{aff88}}
\and S.~Joudaki\orcid{0000-0001-8820-673X}\inst{\ref{aff4}}
\and J.~J.~E.~Kajava\orcid{0000-0002-3010-8333}\inst{\ref{aff140},\ref{aff141}}
\and Y.~Kang\orcid{0009-0000-8588-7250}\inst{\ref{aff51}}
\and V.~Kansal\orcid{0000-0002-4008-6078}\inst{\ref{aff142},\ref{aff143}}
\and D.~Karagiannis\orcid{0000-0002-4927-0816}\inst{\ref{aff107},\ref{aff144}}
\and K.~Kiiveri\inst{\ref{aff69}}
\and J.~Kim\orcid{0000-0003-2776-2761}\inst{\ref{aff112}}
\and C.~C.~Kirkpatrick\inst{\ref{aff69}}
\and S.~Kruk\orcid{0000-0001-8010-8879}\inst{\ref{aff16}}
\and J.~Le~Graet\orcid{0000-0001-6523-7971}\inst{\ref{aff55}}
\and L.~Legrand\orcid{0000-0003-0610-5252}\inst{\ref{aff145},\ref{aff146}}
\and M.~Lembo\orcid{0000-0002-5271-5070}\inst{\ref{aff14}}
\and F.~Lepori\orcid{0009-0000-5061-7138}\inst{\ref{aff147}}
\and G.~Leroy\orcid{0009-0004-2523-4425}\inst{\ref{aff148},\ref{aff82}}
\and G.~F.~Lesci\orcid{0000-0002-4607-2830}\inst{\ref{aff81},\ref{aff18}}
\and J.~Lesgourgues\orcid{0000-0001-7627-353X}\inst{\ref{aff38}}
\and L.~Leuzzi\orcid{0009-0006-4479-7017}\inst{\ref{aff18}}
\and T.~I.~Liaudat\orcid{0000-0002-9104-314X}\inst{\ref{aff149}}
\and J.~Macias-Perez\orcid{0000-0002-5385-2763}\inst{\ref{aff150}}
\and G.~Maggio\orcid{0000-0003-4020-4836}\inst{\ref{aff20}}
\and M.~Magliocchetti\orcid{0000-0001-9158-4838}\inst{\ref{aff56}}
\and F.~Mannucci\orcid{0000-0002-4803-2381}\inst{\ref{aff151}}
\and R.~Maoli\orcid{0000-0002-6065-3025}\inst{\ref{aff152},\ref{aff36}}
\and C.~J.~A.~P.~Martins\orcid{0000-0002-4886-9261}\inst{\ref{aff153},\ref{aff154}}
\and L.~Maurin\orcid{0000-0002-8406-0857}\inst{\ref{aff52}}
\and M.~Miluzio\inst{\ref{aff16},\ref{aff155}}
\and P.~Monaco\orcid{0000-0003-2083-7564}\inst{\ref{aff125},\ref{aff20},\ref{aff21},\ref{aff19},\ref{aff115}}
\and A.~Montoro\orcid{0000-0003-4730-8590}\inst{\ref{aff1},\ref{aff12}}
\and C.~Moretti\orcid{0000-0003-3314-8936}\inst{\ref{aff20},\ref{aff19},\ref{aff21},\ref{aff22}}
\and G.~Morgante\inst{\ref{aff18}}
\and S.~Nadathur\orcid{0000-0001-9070-3102}\inst{\ref{aff135}}
\and K.~Naidoo\orcid{0000-0002-9182-1802}\inst{\ref{aff135},\ref{aff3}}
\and A.~Navarro-Alsina\orcid{0000-0002-3173-2592}\inst{\ref{aff79}}
\and S.~Nesseris\orcid{0000-0002-0567-0324}\inst{\ref{aff116}}
\and D.~Paoletti\orcid{0000-0003-4761-6147}\inst{\ref{aff18},\ref{aff57}}
\and F.~Passalacqua\orcid{0000-0002-8606-4093}\inst{\ref{aff94},\ref{aff54}}
\and K.~Paterson\orcid{0000-0001-8340-3486}\inst{\ref{aff67}}
\and L.~Patrizii\inst{\ref{aff24}}
\and A.~Pisani\orcid{0000-0002-6146-4437}\inst{\ref{aff55}}
\and D.~Potter\orcid{0000-0002-0757-5195}\inst{\ref{aff147}}
\and S.~Quai\orcid{0000-0002-0449-8163}\inst{\ref{aff81},\ref{aff18}}
\and M.~Radovich\orcid{0000-0002-3585-866X}\inst{\ref{aff60}}
\and P.-F.~Rocci\inst{\ref{aff52}}
\and G.~Rodighiero\orcid{0000-0002-9415-2296}\inst{\ref{aff94},\ref{aff60}}
\and S.~Sacquegna\orcid{0000-0002-8433-6630}\inst{\ref{aff156},\ref{aff132},\ref{aff131}}
\and M.~Sahl\'en\orcid{0000-0003-0973-4804}\inst{\ref{aff157}}
\and D.~B.~Sanders\orcid{0000-0002-1233-9998}\inst{\ref{aff40}}
\and E.~Sarpa\orcid{0000-0002-1256-655X}\inst{\ref{aff22},\ref{aff115},\ref{aff21}}
\and J.~Schaye\orcid{0000-0002-0668-5560}\inst{\ref{aff7}}
\and A.~Schneider\orcid{0000-0001-7055-8104}\inst{\ref{aff147}}
\and M.~Schultheis\inst{\ref{aff111}}
\and D.~Sciotti\orcid{0009-0008-4519-2620}\inst{\ref{aff36},\ref{aff37}}
\and E.~Sellentin\inst{\ref{aff158},\ref{aff7}}
\and L.~C.~Smith\orcid{0000-0002-3259-2771}\inst{\ref{aff159}}
\and J.~G.~Sorce\orcid{0000-0002-2307-2432}\inst{\ref{aff160},\ref{aff52}}
\and K.~Tanidis\orcid{0000-0001-9843-5130}\inst{\ref{aff112}}
\and C.~Tao\orcid{0000-0001-7961-8177}\inst{\ref{aff55}}
\and G.~Testera\inst{\ref{aff27}}
\and R.~Teyssier\orcid{0000-0001-7689-0933}\inst{\ref{aff161}}
\and S.~Tosi\orcid{0000-0002-7275-9193}\inst{\ref{aff26},\ref{aff27},\ref{aff17}}
\and A.~Troja\orcid{0000-0003-0239-4595}\inst{\ref{aff94},\ref{aff54}}
\and M.~Tucci\inst{\ref{aff51}}
\and C.~Valieri\inst{\ref{aff24}}
\and A.~Venhola\orcid{0000-0001-6071-4564}\inst{\ref{aff162}}
\and D.~Vergani\orcid{0000-0003-0898-2216}\inst{\ref{aff18}}
\and F.~Vernizzi\orcid{0000-0003-3426-2802}\inst{\ref{aff25}}
\and G.~Verza\orcid{0000-0002-1886-8348}\inst{\ref{aff163}}
\and P.~Vielzeuf\orcid{0000-0003-2035-9339}\inst{\ref{aff55}}
\and N.~A.~Walton\orcid{0000-0003-3983-8778}\inst{\ref{aff159}}}
										   
\institute{Institute of Space Sciences (ICE, CSIC), Campus UAB, Carrer de Can Magrans, s/n, 08193 Barcelona, Spain\label{aff1}
\and
Universit\'e Paris-Saclay, Universit\'e Paris Cit\'e, CEA, CNRS, AIM, 91191, Gif-sur-Yvette, France\label{aff2}
\and
Department of Physics and Astronomy, University College London, Gower Street, London WC1E 6BT, UK\label{aff3}
\and
Centro de Investigaciones Energ\'eticas, Medioambientales y Tecnol\'ogicas (CIEMAT), Avenida Complutense 40, 28040 Madrid, Spain\label{aff4}
\and
Port d'Informaci\'{o} Cient\'{i}fica, Campus UAB, C. Albareda s/n, 08193 Bellaterra (Barcelona), Spain\label{aff5}
\and
Institute for Theoretical Physics, Utrecht University, Princetonplein 5, 3584 CE Utrecht, The Netherlands\label{aff6}
\and
Leiden Observatory, Leiden University, Einsteinweg 55, 2333 CC Leiden, The Netherlands\label{aff7}
\and
Institut de F\'{i}sica d'Altes Energies (IFAE), The Barcelona Institute of Science and Technology, Campus UAB, 08193 Bellaterra (Barcelona), Spain\label{aff8}
\and
Instituto de Astronomia Teorica y Experimental (IATE-CONICET), Laprida 854, X5000BGR, C\'ordoba, Argentina\label{aff9}
\and
Oskar Klein Centre for Cosmoparticle Physics, Department of Physics, Stockholm University, Stockholm, SE-106 91, Sweden\label{aff10}
\and
Astrophysics Group, Blackett Laboratory, Imperial College London, London SW7 2AZ, UK\label{aff11}
\and
Institut d'Estudis Espacials de Catalunya (IEEC),  Edifici RDIT, Campus UPC, 08860 Castelldefels, Barcelona, Spain\label{aff12}
\and
Department of Physics, Northeastern University, Boston, MA, 02115, USA\label{aff13}
\and
Institut d'Astrophysique de Paris, UMR 7095, CNRS, and Sorbonne Universit\'e, 98 bis boulevard Arago, 75014 Paris, France\label{aff14}
\and
Kyung Hee University, Dept. of Astronomy \& Space Science, Yongin-shi, Gyeonggi-do 17104, Republic of Korea\label{aff15}
\and
ESAC/ESA, Camino Bajo del Castillo, s/n., Urb. Villafranca del Castillo, 28692 Villanueva de la Ca\~nada, Madrid, Spain\label{aff16}
\and
INAF-Osservatorio Astronomico di Brera, Via Brera 28, 20122 Milano, Italy\label{aff17}
\and
INAF-Osservatorio di Astrofisica e Scienza dello Spazio di Bologna, Via Piero Gobetti 93/3, 40129 Bologna, Italy\label{aff18}
\and
IFPU, Institute for Fundamental Physics of the Universe, via Beirut 2, 34151 Trieste, Italy\label{aff19}
\and
INAF-Osservatorio Astronomico di Trieste, Via G. B. Tiepolo 11, 34143 Trieste, Italy\label{aff20}
\and
INFN, Sezione di Trieste, Via Valerio 2, 34127 Trieste TS, Italy\label{aff21}
\and
SISSA, International School for Advanced Studies, Via Bonomea 265, 34136 Trieste TS, Italy\label{aff22}
\and
Dipartimento di Fisica e Astronomia, Universit\`a di Bologna, Via Gobetti 93/2, 40129 Bologna, Italy\label{aff23}
\and
INFN-Sezione di Bologna, Viale Berti Pichat 6/2, 40127 Bologna, Italy\label{aff24}
\and
Institut de Physique Th\'eorique, CEA, CNRS, Universit\'e Paris-Saclay 91191 Gif-sur-Yvette Cedex, France\label{aff25}
\and
Dipartimento di Fisica, Universit\`a di Genova, Via Dodecaneso 33, 16146, Genova, Italy\label{aff26}
\and
INFN-Sezione di Genova, Via Dodecaneso 33, 16146, Genova, Italy\label{aff27}
\and
Department of Physics "E. Pancini", University Federico II, Via Cinthia 6, 80126, Napoli, Italy\label{aff28}
\and
INAF-Osservatorio Astronomico di Capodimonte, Via Moiariello 16, 80131 Napoli, Italy\label{aff29}
\and
Dipartimento di Fisica, Universit\`a degli Studi di Torino, Via P. Giuria 1, 10125 Torino, Italy\label{aff30}
\and
INFN-Sezione di Torino, Via P. Giuria 1, 10125 Torino, Italy\label{aff31}
\and
INAF-Osservatorio Astrofisico di Torino, Via Osservatorio 20, 10025 Pino Torinese (TO), Italy\label{aff32}
\and
European Space Agency/ESTEC, Keplerlaan 1, 2201 AZ Noordwijk, The Netherlands\label{aff33}
\and
Institute Lorentz, Leiden University, Niels Bohrweg 2, 2333 CA Leiden, The Netherlands\label{aff34}
\and
INAF-IASF Milano, Via Alfonso Corti 12, 20133 Milano, Italy\label{aff35}
\and
INAF-Osservatorio Astronomico di Roma, Via Frascati 33, 00078 Monteporzio Catone, Italy\label{aff36}
\and
INFN-Sezione di Roma, Piazzale Aldo Moro, 2 - c/o Dipartimento di Fisica, Edificio G. Marconi, 00185 Roma, Italy\label{aff37}
\and
Institute for Theoretical Particle Physics and Cosmology (TTK), RWTH Aachen University, 52056 Aachen, Germany\label{aff38}
\and
INFN section of Naples, Via Cinthia 6, 80126, Napoli, Italy\label{aff39}
\and
Institute for Astronomy, University of Hawaii, 2680 Woodlawn Drive, Honolulu, HI 96822, USA\label{aff40}
\and
Dipartimento di Fisica e Astronomia "Augusto Righi" - Alma Mater Studiorum Universit\`a di Bologna, Viale Berti Pichat 6/2, 40127 Bologna, Italy\label{aff41}
\and
Instituto de Astrof\'{\i}sica de Canarias, E-38205 La Laguna, Tenerife, Spain\label{aff42}
\and
Institute for Astronomy, University of Edinburgh, Royal Observatory, Blackford Hill, Edinburgh EH9 3HJ, UK\label{aff43}
\and
European Space Agency/ESRIN, Largo Galileo Galilei 1, 00044 Frascati, Roma, Italy\label{aff44}
\and
Universit\'e Claude Bernard Lyon 1, CNRS/IN2P3, IP2I Lyon, UMR 5822, Villeurbanne, F-69100, France\label{aff45}
\and
Institut de Ci\`{e}ncies del Cosmos (ICCUB), Universitat de Barcelona (IEEC-UB), Mart\'{i} i Franqu\`{e}s 1, 08028 Barcelona, Spain\label{aff46}
\and
Instituci\'o Catalana de Recerca i Estudis Avan\c{c}ats (ICREA), Passeig de Llu\'{\i}s Companys 23, 08010 Barcelona, Spain\label{aff47}
\and
UCB Lyon 1, CNRS/IN2P3, IUF, IP2I Lyon, 4 rue Enrico Fermi, 69622 Villeurbanne, France\label{aff48}
\and
Departamento de F\'isica, Faculdade de Ci\^encias, Universidade de Lisboa, Edif\'icio C8, Campo Grande, PT1749-016 Lisboa, Portugal\label{aff49}
\and
Instituto de Astrof\'isica e Ci\^encias do Espa\c{c}o, Faculdade de Ci\^encias, Universidade de Lisboa, Campo Grande, 1749-016 Lisboa, Portugal\label{aff50}
\and
Department of Astronomy, University of Geneva, ch. d'Ecogia 16, 1290 Versoix, Switzerland\label{aff51}
\and
Universit\'e Paris-Saclay, CNRS, Institut d'astrophysique spatiale, 91405, Orsay, France\label{aff52}
\and
Jodrell Bank Centre for Astrophysics, Department of Physics and Astronomy, University of Manchester, Oxford Road, Manchester M13 9PL, UK\label{aff53}
\and
INFN-Padova, Via Marzolo 8, 35131 Padova, Italy\label{aff54}
\and
Aix-Marseille Universit\'e, CNRS/IN2P3, CPPM, Marseille, France\label{aff55}
\and
INAF-Istituto di Astrofisica e Planetologia Spaziali, via del Fosso del Cavaliere, 100, 00100 Roma, Italy\label{aff56}
\and
INFN-Bologna, Via Irnerio 46, 40126 Bologna, Italy\label{aff57}
\and
University Observatory, LMU Faculty of Physics, Scheinerstr.~1, 81679 Munich, Germany\label{aff58}
\and
Max Planck Institute for Extraterrestrial Physics, Giessenbachstr. 1, 85748 Garching, Germany\label{aff59}
\and
INAF-Osservatorio Astronomico di Padova, Via dell'Osservatorio 5, 35122 Padova, Italy\label{aff60}
\and
Universit\"ats-Sternwarte M\"unchen, Fakult\"at f\"ur Physik, Ludwig-Maximilians-Universit\"at M\"unchen, Scheinerstr.~1, 81679 M\"unchen, Germany\label{aff61}
\and
Institute of Theoretical Astrophysics, University of Oslo, P.O. Box 1029 Blindern, 0315 Oslo, Norway\label{aff62}
\and
Jet Propulsion Laboratory, California Institute of Technology, 4800 Oak Grove Drive, Pasadena, CA, 91109, USA\label{aff63}
\and
Felix Hormuth Engineering, Goethestr. 17, 69181 Leimen, Germany\label{aff64}
\and
Technical University of Denmark, Elektrovej 327, 2800 Kgs. Lyngby, Denmark\label{aff65}
\and
Cosmic Dawn Center (DAWN), Denmark\label{aff66}
\and
Max-Planck-Institut f\"ur Astronomie, K\"onigstuhl 17, 69117 Heidelberg, Germany\label{aff67}
\and
NASA Goddard Space Flight Center, Greenbelt, MD 20771, USA\label{aff68}
\and
Department of Physics and Helsinki Institute of Physics, Gustaf H\"allstr\"omin katu 2, University of Helsinki, 00014 Helsinki, Finland\label{aff69}
\and
Universit\'e de Gen\`eve, D\'epartement de Physique Th\'eorique and Centre for Astroparticle Physics, 24 quai Ernest-Ansermet, CH-1211 Gen\`eve 4, Switzerland\label{aff70}
\and
Department of Physics, P.O. Box 64, University of Helsinki, 00014 Helsinki, Finland\label{aff71}
\and
Helsinki Institute of Physics, Gustaf H{\"a}llstr{\"o}min katu 2, University of Helsinki, 00014 Helsinki, Finland\label{aff72}
\and
Laboratoire d'etude de l'Univers et des phenomenes eXtremes, Observatoire de Paris, Universit\'e PSL, Sorbonne Universit\'e, CNRS, 92190 Meudon, France\label{aff73}
\and
SKAO, Jodrell Bank, Lower Withington, Macclesfield SK11 9FT, UK\label{aff74}
\and
Centre de Calcul de l'IN2P3/CNRS, 21 avenue Pierre de Coubertin 69627 Villeurbanne Cedex, France\label{aff75}
\and
Dipartimento di Fisica "Aldo Pontremoli", Universit\`a degli Studi di Milano, Via Celoria 16, 20133 Milano, Italy\label{aff76}
\and
INFN-Sezione di Milano, Via Celoria 16, 20133 Milano, Italy\label{aff77}
\and
University of Applied Sciences and Arts of Northwestern Switzerland, School of Computer Science, 5210 Windisch, Switzerland\label{aff78}
\and
Universit\"at Bonn, Argelander-Institut f\"ur Astronomie, Auf dem H\"ugel 71, 53121 Bonn, Germany\label{aff79}
\and
Aix-Marseille Universit\'e, CNRS, CNES, LAM, Marseille, France\label{aff80}
\and
Dipartimento di Fisica e Astronomia "Augusto Righi" - Alma Mater Studiorum Universit\`a di Bologna, via Piero Gobetti 93/2, 40129 Bologna, Italy\label{aff81}
\and
Department of Physics, Institute for Computational Cosmology, Durham University, South Road, Durham, DH1 3LE, UK\label{aff82}
\and
Universit\'e Paris Cit\'e, CNRS, Astroparticule et Cosmologie, 75013 Paris, France\label{aff83}
\and
CNRS-UCB International Research Laboratory, Centre Pierre Bin\'etruy, IRL2007, CPB-IN2P3, Berkeley, USA\label{aff84}
\and
Institut d'Astrophysique de Paris, 98bis Boulevard Arago, 75014, Paris, France\label{aff85}
\and
Institute of Physics, Laboratory of Astrophysics, Ecole Polytechnique F\'ed\'erale de Lausanne (EPFL), Observatoire de Sauverny, 1290 Versoix, Switzerland\label{aff86}
\and
Telespazio UK S.L. for European Space Agency (ESA), Camino bajo del Castillo, s/n, Urbanizacion Villafranca del Castillo, Villanueva de la Ca\~nada, 28692 Madrid, Spain\label{aff87}
\and
DARK, Niels Bohr Institute, University of Copenhagen, Jagtvej 155, 2200 Copenhagen, Denmark\label{aff88}
\and
Space Science Data Center, Italian Space Agency, via del Politecnico snc, 00133 Roma, Italy\label{aff89}
\and
Centre National d'Etudes Spatiales -- Centre spatial de Toulouse, 18 avenue Edouard Belin, 31401 Toulouse Cedex 9, France\label{aff90}
\and
Institute of Space Science, Str. Atomistilor, nr. 409 M\u{a}gurele, Ilfov, 077125, Romania\label{aff91}
\and
Consejo Superior de Investigaciones Cientificas, Calle Serrano 117, 28006 Madrid, Spain\label{aff92}
\and
Universidad de La Laguna, Dpto. Astrof\'\i sica, E-38206 La Laguna, Tenerife, Spain\label{aff93}
\and
Dipartimento di Fisica e Astronomia "G. Galilei", Universit\`a di Padova, Via Marzolo 8, 35131 Padova, Italy\label{aff94}
\and
Institut f\"ur Theoretische Physik, University of Heidelberg, Philosophenweg 16, 69120 Heidelberg, Germany\label{aff95}
\and
Institut de Recherche en Astrophysique et Plan\'etologie (IRAP), Universit\'e de Toulouse, CNRS, UPS, CNES, 14 Av. Edouard Belin, 31400 Toulouse, France\label{aff96}
\and
Universit\'e St Joseph; Faculty of Sciences, Beirut, Lebanon\label{aff97}
\and
Departamento de F\'isica, FCFM, Universidad de Chile, Blanco Encalada 2008, Santiago, Chile\label{aff98}
\and
Universit\"at Innsbruck, Institut f\"ur Astro- und Teilchenphysik, Technikerstr. 25/8, 6020 Innsbruck, Austria\label{aff99}
\and
Satlantis, University Science Park, Sede Bld 48940, Leioa-Bilbao, Spain\label{aff100}
\and
Department of Physics, Royal Holloway, University of London, Surrey TW20 0EX, UK\label{aff101}
\and
Instituto de Astrof\'isica e Ci\^encias do Espa\c{c}o, Faculdade de Ci\^encias, Universidade de Lisboa, Tapada da Ajuda, 1349-018 Lisboa, Portugal\label{aff102}
\and
Cosmic Dawn Center (DAWN)\label{aff103}
\and
Niels Bohr Institute, University of Copenhagen, Jagtvej 128, 2200 Copenhagen, Denmark\label{aff104}
\and
Universidad Polit\'ecnica de Cartagena, Departamento de Electr\'onica y Tecnolog\'ia de Computadoras,  Plaza del Hospital 1, 30202 Cartagena, Spain\label{aff105}
\and
Infrared Processing and Analysis Center, California Institute of Technology, Pasadena, CA 91125, USA\label{aff106}
\and
Dipartimento di Fisica e Scienze della Terra, Universit\`a degli Studi di Ferrara, Via Giuseppe Saragat 1, 44122 Ferrara, Italy\label{aff107}
\and
Istituto Nazionale di Fisica Nucleare, Sezione di Ferrara, Via Giuseppe Saragat 1, 44122 Ferrara, Italy\label{aff108}
\and
INAF, Istituto di Radioastronomia, Via Piero Gobetti 101, 40129 Bologna, Italy\label{aff109}
\and
Astronomical Observatory of the Autonomous Region of the Aosta Valley (OAVdA), Loc. Lignan 39, I-11020, Nus (Aosta Valley), Italy\label{aff110}
\and
Universit\'e C\^{o}te d'Azur, Observatoire de la C\^{o}te d'Azur, CNRS, Laboratoire Lagrange, Bd de l'Observatoire, CS 34229, 06304 Nice cedex 4, France\label{aff111}
\and
Department of Physics, Oxford University, Keble Road, Oxford OX1 3RH, UK\label{aff112}
\and
Zentrum f\"ur Astronomie, Universit\"at Heidelberg, Philosophenweg 12, 69120 Heidelberg, Germany\label{aff113}
\and
ICL, Junia, Universit\'e Catholique de Lille, LITL, 59000 Lille, France\label{aff114}
\and
ICSC - Centro Nazionale di Ricerca in High Performance Computing, Big Data e Quantum Computing, Via Magnanelli 2, Bologna, Italy\label{aff115}
\and
Instituto de F\'isica Te\'orica UAM-CSIC, Campus de Cantoblanco, 28049 Madrid, Spain\label{aff116}
\and
CERCA/ISO, Department of Physics, Case Western Reserve University, 10900 Euclid Avenue, Cleveland, OH 44106, USA\label{aff117}
\and
Technical University of Munich, TUM School of Natural Sciences, Physics Department, James-Franck-Str.~1, 85748 Garching, Germany\label{aff118}
\and
Max-Planck-Institut f\"ur Astrophysik, Karl-Schwarzschild-Str.~1, 85748 Garching, Germany\label{aff119}
\and
Laboratoire Univers et Th\'eorie, Observatoire de Paris, Universit\'e PSL, Universit\'e Paris Cit\'e, CNRS, 92190 Meudon, France\label{aff120}
\and
Departamento de F{\'\i}sica Fundamental. Universidad de Salamanca. Plaza de la Merced s/n. 37008 Salamanca, Spain\label{aff121}
\and
Universit\'e de Strasbourg, CNRS, Observatoire astronomique de Strasbourg, UMR 7550, 67000 Strasbourg, France\label{aff122}
\and
Center for Data-Driven Discovery, Kavli IPMU (WPI), UTIAS, The University of Tokyo, Kashiwa, Chiba 277-8583, Japan\label{aff123}
\and
Waterloo Centre for Astrophysics, University of Waterloo, Waterloo, Ontario N2L 3G1, Canada\label{aff124}
\and
Dipartimento di Fisica - Sezione di Astronomia, Universit\`a di Trieste, Via Tiepolo 11, 34131 Trieste, Italy\label{aff125}
\and
California Institute of Technology, 1200 E California Blvd, Pasadena, CA 91125, USA\label{aff126}
\and
Department of Physics \& Astronomy, University of California Irvine, Irvine CA 92697, USA\label{aff127}
\and
Kapteyn Astronomical Institute, University of Groningen, PO Box 800, 9700 AV Groningen, The Netherlands\label{aff128}
\and
Departamento F\'isica Aplicada, Universidad Polit\'ecnica de Cartagena, Campus Muralla del Mar, 30202 Cartagena, Murcia, Spain\label{aff129}
\and
Instituto de F\'isica de Cantabria, Edificio Juan Jord\'a, Avenida de los Castros, 39005 Santander, Spain\label{aff130}
\and
INFN, Sezione di Lecce, Via per Arnesano, CP-193, 73100, Lecce, Italy\label{aff131}
\and
Department of Mathematics and Physics E. De Giorgi, University of Salento, Via per Arnesano, CP-I93, 73100, Lecce, Italy\label{aff132}
\and
INAF-Sezione di Lecce, c/o Dipartimento Matematica e Fisica, Via per Arnesano, 73100, Lecce, Italy\label{aff133}
\and
CEA Saclay, DFR/IRFU, Service d'Astrophysique, Bat. 709, 91191 Gif-sur-Yvette, France\label{aff134}
\and
Institute of Cosmology and Gravitation, University of Portsmouth, Portsmouth PO1 3FX, UK\label{aff135}
\and
Department of Computer Science, Aalto University, PO Box 15400, Espoo, FI-00 076, Finland\label{aff136}
\and
 Instituto de Astrof\'{\i}sica de Canarias, E-38205 La Laguna; Universidad de La Laguna, Dpto. Astrof\'\i sica, E-38206 La Laguna, Tenerife, Spain\label{aff137}
\and
Caltech/IPAC, 1200 E. California Blvd., Pasadena, CA 91125, USA\label{aff138}
\and
Ruhr University Bochum, Faculty of Physics and Astronomy, Astronomical Institute (AIRUB), German Centre for Cosmological Lensing (GCCL), 44780 Bochum, Germany\label{aff139}
\and
Department of Physics and Astronomy, Vesilinnantie 5, University of Turku, 20014 Turku, Finland\label{aff140}
\and
Serco for European Space Agency (ESA), Camino bajo del Castillo, s/n, Urbanizacion Villafranca del Castillo, Villanueva de la Ca\~nada, 28692 Madrid, Spain\label{aff141}
\and
ARC Centre of Excellence for Dark Matter Particle Physics, Melbourne, Australia\label{aff142}
\and
Centre for Astrophysics \& Supercomputing, Swinburne University of Technology,  Hawthorn, Victoria 3122, Australia\label{aff143}
\and
Department of Physics and Astronomy, University of the Western Cape, Bellville, Cape Town, 7535, South Africa\label{aff144}
\and
DAMTP, Centre for Mathematical Sciences, Wilberforce Road, Cambridge CB3 0WA, UK\label{aff145}
\and
Kavli Institute for Cosmology Cambridge, Madingley Road, Cambridge, CB3 0HA, UK\label{aff146}
\and
Department of Astrophysics, University of Zurich, Winterthurerstrasse 190, 8057 Zurich, Switzerland\label{aff147}
\and
Department of Physics, Centre for Extragalactic Astronomy, Durham University, South Road, Durham, DH1 3LE, UK\label{aff148}
\and
IRFU, CEA, Universit\'e Paris-Saclay 91191 Gif-sur-Yvette Cedex, France\label{aff149}
\and
Univ. Grenoble Alpes, CNRS, Grenoble INP, LPSC-IN2P3, 53, Avenue des Martyrs, 38000, Grenoble, France\label{aff150}
\and
INAF-Osservatorio Astrofisico di Arcetri, Largo E. Fermi 5, 50125, Firenze, Italy\label{aff151}
\and
Dipartimento di Fisica, Sapienza Universit\`a di Roma, Piazzale Aldo Moro 2, 00185 Roma, Italy\label{aff152}
\and
Centro de Astrof\'{\i}sica da Universidade do Porto, Rua das Estrelas, 4150-762 Porto, Portugal\label{aff153}
\and
Instituto de Astrof\'isica e Ci\^encias do Espa\c{c}o, Universidade do Porto, CAUP, Rua das Estrelas, PT4150-762 Porto, Portugal\label{aff154}
\and
HE Space for European Space Agency (ESA), Camino bajo del Castillo, s/n, Urbanizacion Villafranca del Castillo, Villanueva de la Ca\~nada, 28692 Madrid, Spain\label{aff155}
\and
INAF - Osservatorio Astronomico d'Abruzzo, Via Maggini, 64100, Teramo, Italy\label{aff156}
\and
Theoretical astrophysics, Department of Physics and Astronomy, Uppsala University, Box 516, 751 37 Uppsala, Sweden\label{aff157}
\and
Mathematical Institute, University of Leiden, Einsteinweg 55, 2333 CA Leiden, The Netherlands\label{aff158}
\and
Institute of Astronomy, University of Cambridge, Madingley Road, Cambridge CB3 0HA, UK\label{aff159}
\and
Univ. Lille, CNRS, Centrale Lille, UMR 9189 CRIStAL, 59000 Lille, France\label{aff160}
\and
Department of Astrophysical Sciences, Peyton Hall, Princeton University, Princeton, NJ 08544, USA\label{aff161}
\and
Space physics and astronomy research unit, University of Oulu, Pentti Kaiteran katu 1, FI-90014 Oulu, Finland\label{aff162}
\and
Center for Computational Astrophysics, Flatiron Institute, 162 5th Avenue, 10010, New York, NY, USA\label{aff163}}

\date{Received ???; accepted ???}


\titlerunning{\Euclid preparation: Calibrated intrinsic alignments in Flagship}
\authorrunning{Euclid Collaboration: K. Hoffmann et al.}

\abstract
{
Intrinsic alignments of galaxies are potentially a major contaminant of cosmological analyses of weak gravitational lensing.
We construct a semi-analytic model of galaxy ellipticities and alignments in the \Euclid Flagship simulation to predict
this contamination in Euclid's weak lensing observations. Galaxy shapes and orientations are determined by the corresponding properties of the host haloes in the underlying $N$-body simulation, as well as the relative positions of galaxies within their halo. Alignment strengths are moderated via stochastic misalignments, separately for central and satellite galaxies and conditional on the galaxy's redshift, luminosity, and rest-frame colour. The resulting model is calibrated against galaxy ellipticity statistics from the COSMOS Survey, selected alignment measurements based on Sloan Digital Sky Survey samples, and galaxy orientations extracted from the Horizon-AGN hydrodynamic simulation at redshift $z=1$.
The best-fit model has a total of 12 alignment parameters and generally reproduces the calibration data sets well within the $1\sigma$ statistical uncertainties of the observations and the \flagship simulation, with notable exceptions for the most luminous sub-samples on small physical scales. The statistical power of the calibration data and the volume of the single \flagship realisation are still too small to provide informative prior ranges for intrinsic alignment amplitudes in relevant galaxy samples.
As a first application, we predict that \Euclid end-of-mission tomographic weak gravitational lensing two-point statistics are modified by up to order $10\,\%$ due to intrinsic alignments.
}


\maketitle
\nolinenumbers
\input{sections/introduction}

\input{sections/flagship_simulation}
\input{sections/mock_galaxy_catalogues}

\input{sections/ia_statistics}

\input{sections/modelling_galaxy_shapes}
\input{sections/modelling_galaxy_orientations}

\input{sections/euclid_predictions}

\input{sections/conclusions}

\begin{acknowledgements}
KH acknowledges support by the Swiss National Science Foundation (Grant No. 173716, 198674),
and from the Forschungskredit Grant of the University of Zurich (Projekt K-76106-01-01).
\AckEC
\AckCosmoHub
This publication is part of the project ``A rising tide: Galaxy intrinsic alignments as a new probe of cosmology and galaxy evolution'' (with project number VI.Vidi.203.011) of the Talent programme Vidi which is (partly) financed by the Dutch Research Council (NWO).
\end{acknowledgements}

\bibliography{references,references2,Euclid}

\label{LastPage}
\end{document}

%% file: sections/introduction.tex
\section{Introduction}\label{sec:intro}

The intrinsic alignments of galaxy ellipticities (IA) constitute a potentially limiting factor in the cosmological interpretation of weak gravitational lensing by the large-scale structure \citep{Kirk15,Krause16,Troxel15}. Alongside baryonic feedback on the dark matter distribution \citep{vanDaalen11,semboloni11}, IA are determined by astrophysical processes in and around galaxies that need to be understood to enable inference on cosmological parameters from weak gravitational lensing surveys. As opposed to baryonic feedback, however, IA cannot be readily isolated from the cosmological signal by selection on the dataset, e.g. through small-scale cuts.

Model-independent removal of intrinsic alignment (IA) is possible by marginalizing over IA model parameters due to its distinctive scaling with the redshifts of the source galaxy samples \citep[e.g.,][]{Wright25}. However, this comes at a significant cost to the cosmological constraining power, for example, reducing the figure of merit of dynamic dark energy constraints by an order of magnitude \citep{joachimi09}. Self-calibration algorithms, which rely on symmetries of the signal in redshift space, are being explored as an alternative path to IA removal, but are challenging to apply on photometric redshift samples as those used in Euclid \citep[e.g.,][]{Yao20, Chisari25}.

Therefore, current surveys rely on physically motivated models to account for IA contamination \citep{asgari21,secco22,dalal23}. The vast majority of models rely on the assumption that galaxy ellipticities and their orientations are determined by the local tidal field of the surrounding dark matter distribution (\citealp{Catelan01,Hirata04,Blazek19,vlah_eft_2019,Bakx23}; see \citealp{Joachimi15} for a review and discussion of alternative mechanisms and \citealp{IAGuide} for a practical introduction). Simple tidal field-based models have shown some success at explaining the scale and mass dependencies of IA signals for certain types of galaxies \citep{Joachimi11,Piras18,Johnston19,fortuna21a,Samuroff19}, but observational constraints remain scarce for the typical galaxy types, masses, and redshift ranges targeted by modern galaxy imaging surveys. Depending on the properties and statistical power of each survey, a careful balance has to be struck between conservative modelling of IA which minimises residual bias on cosmology and including robust prior knowledge on IA to preserve a maximum of cosmological information.

This work is the first of a series in which we explore this balance in the context of the ESA \Euclid mission, which features weak gravitational lensing (WL) as one of its primary cosmological probes \citep{EuclidSkyOverview}\footnote{\url{www.esa.int/Science\_Exploration/Space\_Science/Euclid}}. Here, we present a semi-analytic model of galaxy ellipticities and alignments for the \Euclid \flagship Simulation \citep[\flagship hereafter]{EuclidSkyFlagship}.
We introduce a semi-analytic model for galaxy ellipticities and alignments based on the \Euclid \flagship Simulation \citep[\flagship hereafter]{EuclidSkyFlagship}. Unlike simple tidal field-based models, this semi-analytic approach utilizes the high-resolution matter field as well as detailed galaxy photometry provided in \flagship to attribute IA characteristics to each galaxy in the simulation. It thereby matches multiple IA constraints simultaneously, enabling more physically grounded predictions for Euclid-like samples.
Using this model, we build mock datasets that will be used in companion papers to analyse the range of validity and the performance of analytic IA models to inform analysis choices particularly for \Euclid's first data release, which will be based on one year of observations.

Our semi-analytic model follows a similar approach to previous work on different simulations \citep[H22 hereafter]{Joachimi13a,joachimi13b,MICEIA}, but advances beyond these in three aspects: 1. we implement a flexible model for controlling the colour, luminosity, and redshift dependencies of the IA signal; 2. the model is calibrated against a selection of multiple constraints from observations, as well as from a hydro-dynamic simulation where observational data are not available; and 3. the resulting mock catalogue is unprecedented in its volume and depth, covering one octant of the sky up to redshift $z=3$, and containing
$3.4$ billion galaxies with magnitudes down to $H_{\mathrm{E}}=26$, thus allowing for the construction of realistic \Euclid-like galaxy samples.

This paper is organised as follows: in Sect.~\ref{sec:flagship_sim} we introduce the \flagship simulation and relevant data products for intrinsic alignment calibration, followed by details on the galaxy samples used for calibrating the IA model in Sect.~\ref{sec:gal_cats}. Section~\ref{sec:ia_statistics} provides an overview of the summary statistics employed in the calibration process.
In Sects.~\ref{sec:modelling_shapes} and~\ref{sec:modelling_orientations}, the modelling of the mock galaxy shapes and orientations are described, respectively.
The calibration of the orientations are described in Sect.~\ref{sec:param_calibration}.
In Sect.~\ref{sec:error_prop}, we investigate the statistical uncertainty of the semi-analytic model before computing the expected IA and WL signals for a \Euclid-like survey in Sect.~\ref{sec:ia_contamination_euclid}. We summarise our findings and conclude in Sect.~\ref{sec:Conclusions}.

%% file: sections/flagship_simulation.tex
\section{Flagship simulation}\label{sec:flagship_sim}

\subsection{Dark matter simulation}
The Flagship simulation is one of the largest cosmological $N$-body simulations of the Universe produced so far, with $(16\,000)^3$ particles over a simulation box of $3600\, h^{-1}\,{\rm Mpc}$, leading to a mass resolution of $10^9 \,h^{-1}\,M_{\odot}$. The simulation was run using \texttt{PKDGRAV3} \citep{PKDGRAV3-potter2016} with cosmological parameters similar to those of the $Planck$ 2015 cosmology~\citep{planck2015}. These are the density parameters of cold dark matter, $\Omega_{\rm CDM} = 0.270$, of baryons, $\Omega_{\rm b}= 0.049$, and of dark energy, $\Omega_{\Lambda} = 0.681 - \Omega_\mathrm{r} - \Omega_{\nu}$, where $\Omega_\mathrm{r}$ and $\Omega_{\nu}$ are the density parameters of radiation and massive neutrinos, respectively (where the neutrino normal mass hierarchy is assumed).
The other cosmological parameters are the equation of state parameter of dark energy, $w=-1$, the reduced Hubble constant, $h=0.67$, the (linear) matter density fluctuation amplitude on scales of $8\, h^{-1}\,{\rm Mpc}$, $\sigma_8 = 0.83$, and the slope of the primordial power spectrum of scalar density fluctuations, $n_{\rm s}=0.96$.
A light-cone up to
$z$ = 3 was produced on the fly during the simulation, covering one octant of the sky, around 5157 deg$^2$, and centred at approximately the North Galactic Pole.

\paragraph{Weak lensing maps}
All-sky lensing convergence maps were constructed by decomposing and rotating the dark matter light-cone
into a set of all-sky concentric spherical shells around the observer. Lensing observables
were then derived from the two-dimensional dark matter density maps of the different shells, using the Born approximation \citep{Fosalba08, lensing-fosalba2016}. The shells have a width
of $\Delta_r \approx 95.22$ Myr in lookback time, and an angular resolution of
$\Delta_{\theta} \approx \,$\ang{;0.43;}. Further details on the construction of the lensing maps are given in \citep{lensing-fosalba2016}.

By combining the dark-matter ``onion shells''  that make up the
light-cone, we can easily derive lensing observables, as explained in \cite{Fosalba08}.  This approach, based on approximating the
observables by a discrete sum of two-dimensional  dark-matter density maps
multiplied by the appropriate lensing weights, agrees with the much more complex and CPU time-consuming ray-tracing technique in the Born approximation limit, i.e. in the limit where lensing deflections are calculated using unperturbed light paths (see e.g., \citealt{hilbert_wlaccuracy_2020}). Following this technique, one can produce all-sky maps of the convergence field (which is simply related to the lensing potential in harmonic space), as well as maps for other lensing fields obtained from covariant derivatives of the lensing potential, such as the deflection angle, shear, flexion, etc. In particular, using the spherical transform of the lensing potential all-sky map one can obtain the corresponding maps for other lensing observables through simple relations (see~\citealt{Hu:00}). These weak lensing maps
allow us to predict the relative contribution of IA to the observed shear statistics.

\subsection{Halo detection and halo shape measurement}\label{sec:orientation_measurements}
Dark matter halos and sub-halos were identified in the particle distribution
in phase space using \texttt{ROCKSTAR} \citep{behroozi2012}, which provides positions, velocities, as well as
three-dimensional shapes and orientations for each object. The shapes and orientations are defined
via the eigenvectors and eigenvalues of the moment of inertia,
\eq{
    I_{i,j} = \sum_{n}^{N_{\rm p}} \frac{r_{n,i} r_{n,j}}{R_n^2}\,,
\label{eq:moment_of_inertia}
}
where $N_{\rm p}$ is the number of halo particles and $r_{n,i}$ are the components of the
three-dimensional position vector of the $n$th particle with respect to the halo centre
\citep[e.g.][]{Dubinski91, Allgood06}.
$R_n$ is a normalisation factor, which is set to $R_n=1$ in the standard definition of the moment of inertia and to
$R_n = ({r_{n,1}^2+r_{n,2}^2+r_{n,3}^2})^{1/2}$ for the so-called reduced moment of inertia.
The lengths of the halo's principal axes are given by the square root of the eigenvalues $\lambda$ of $I_{i,j}$,
$(A_{\rm 3D}, B_{\rm 3D}, C_{\rm 3D}) = (\sqrt{\lambda_A}, \sqrt{\lambda_B}, \sqrt{\lambda_C})$, and satisfy $A_{\rm 3D} \geq B_{\rm 3D} \geq C_{\rm 3D}$.
The halo shapes are characterized by two of the three axis ratios
\begin{equation}
q_{\rm 3D} \equiv \frac{B_{\rm 3D}}{A_{\rm 3D}}, ~~\mbox{ }~~
r_{\rm 3D} \equiv \frac{C_{\rm 3D}}{B_{\rm 3D}}, ~~\mbox{ }~~
s_{\rm 3D} \equiv \frac{C_{\rm 3D}}{A_{\rm 3D}},
\label{eq:3D_axes_ratios}
\end{equation}
while the halo orientations are given by the normalized eigenvectors
$\hat{\bf A}_{\rm 3D}$, $\hat{\bf B}_{\rm 3D}$, and $\hat{\bf C}_{\rm 3D}$.
In this study, we use halo orientations measured via the reduced moment of
inertia, as it assigns more weight to the mass distribution in the halo's centre,
where the central galaxies - whose orientations we aim to model - are located.
Measurements of galaxy orientations from the stellar particle distribution in
the hydrodynamical simulation (described in Sect.~\ref{sec:hagn}), on the other hand,
are based on the standard moment of inertia, as it provides a higher signal-to-noise ratio
\citep{Chisari15}.

Orientations measured from discrete particle distributions are subject to sampling noise, which increases
for decreasing numbers of particles. In the galaxy and halo samples used in this work, the average number of particles per object ranges from about $1000$ to $18000$, for which we expect only moderate randomization of below 5 degree \citep{Herle2025}. However, the minimum number of particles per object can be as low as 10, in which case a strong randomization of the orientation is expected. H22 demonstrated that randomly down-sampling halos to $10$ particles reduces the alignment signal by roughly $30$ percent, while increasing the particle limit to $100$ lowers the impact of noise on the alignment signal to the percent level. For this reason, we do not expect noise in the measured orientations to have a dominant effect on the predicted IA statistics.

\subsection{Galaxy modelling}

Based on the dark matter halos, a galaxy catalogue was built
using the \texttt{SciPIC} Scientific pipeline at Port d'Informació Científica \citep{scipic-carretero2017}. \texttt{SciPIC} combines the halo occupation distribution technique with halo abundance matching (\citealt{micemock-carretero2015}, \citealt{EuclidSkyFlagship}) to distribute
central and satellite galaxy positions in the halos and then assign
velocities as well as initial magnitudes and colours. The parameters
of the model were calibrated such that the simulation reproduces Sloan Digital Sky Survey \citep[SDSS]{York00}
observations of the luminosity function, the colour-magnitude distribution
and the clustering in redshift space as a function of colour and luminosity \citep{Blanton03, Zehavi11}. In a subsequent step,
spectral energy distributions (SEDs) are derived for each galaxy based
on its initial magnitude and colour. From these SEDs, additional magnitudes
can then be derived for any filter, which enables the construction of
mock galaxy catalogues of any desired galaxy survey.

%% file: sections/mock_galaxy_catalogues.tex
\section{Mock Galaxy Catalogs}\label{sec:gal_cats}

\begin{figure}
        \begin{center}
        \includegraphics[width=0.45\textwidth]{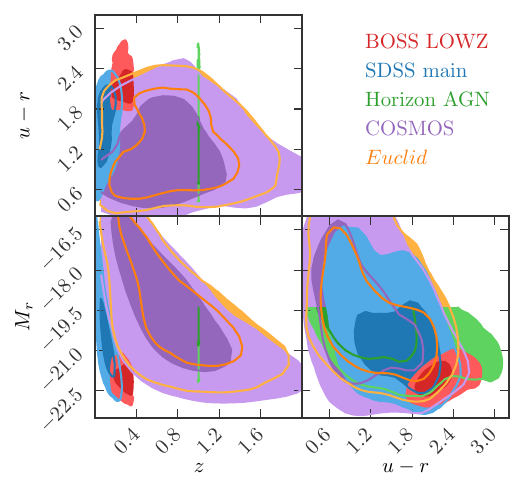}
        \caption{Redshift, magnitude, colour distributions of mock samples, used for calibrating the IA
        model parameters for galaxy misalignment. Contours enclose $68\%$ and $95\%$ of the distributions.}
        \label{fig:zmc_distribution}
        \end{center}
    \end{figure}

We construct mock galaxy catalogues to calibrate the IA simulation against measurements
from reference samples. Specifically, we use a mock catalogue of the COSMOS \citep{scoville07}
survey to adjust the definition of red and blue galaxies, used in the shape component of our IA
model and to validate the shape distribution predicted by that model.
In addition, we use mock catalogues of the SDSS main sample and the BOSS LOWZ sample
for calibrating the galaxy orientation model component at low redshifts.
These low-redshift samples are complemented by a mock catalogue of the Horizon-AGN simulation \citep{Dubois14,Kaviraj17},
which we use for the calibration of the orientations at $z=1$.
In Fig.~\ref{fig:zmc_distribution} we compare the colour-magnitude-redshift distribution 
of these different mock catalogues, while the main characteristics of each mock sample used in this work
are summarised in Table~\ref{tab:mocks_summary}.
In the following, we give an overview of how the reference samples are selected and detail the
construction of the corresponding mocks.

\begin{table}[]
\caption{Mean redshift $z$, mean absolute $r$-band magnitude $M_r$, mean rest-frame colour index $u-r$ and satellite fraction $f_{\rm sat}$ of the mock samples used to calibrate the \flagship simulation.}
\begin{tabular}{lllllll}
sample & & $N_{\rm gal}$ & $\langle z \rangle$ & $\langle M_r \rangle$ &$\langle u - r \rangle$ & $f_{\rm sat}$ \\\hline\hline 
\noalign{\vskip 1pt}
SDSS & main & $495502$  & $0.10$ & $-20.50$  & $1.59$ & $0.35$ \\
     & blue & $198674$ & $0.09$ & $-19.96$ & $1.16$ & $0.25$ \\
     & red & $296828$ & $0.11$ & $-20.87$ & $1.87$ & $0.43$ \\\hline
     \noalign{\vskip 1pt}
BOSS & main & $155290$ & $0.28$ & $-22.06$ & $2.17$ & $0.25$ \\
LOWZ & L1 & $31058$ & $0.29$ & $-22.60$ & $2.08$ & $0.22$ \\
     & L2 & $31058$ & $0.28$ & $-22.25$ & $2.11$ & $0.23$ \\
     & L3 & $31058$ & $0.28$ & $-22.05$ & $2.15$ & $0.24$ \\
     & L4 & $62116$ & $0.26$ & $-21.69$ & $2.26$ & $0.28$ \\\hline
     \noalign{\vskip 1pt}
HAGN & main & $757638$ & $1.00$ & $-20.41$ & $1.76$ & $0.35$ \\
     & H1   & $33311$  & $1.00$ & $-22.49$ & $2.41$ & $0.20$ \\
     & H2   & $183915$ & $1.00$ & $-21.35$ & $2.08$ & $0.31$ \\
     & H3   & $540412$ & $1.00$ & $-19.96$ &$1.62$ & $0.38$
\end{tabular}
\label{tab:mocks_summary}
\end{table}

\subsection{COSMOS}\label{sec:gal_cats:cosmos}

The COSMOS survey provides space-based imaging from the Hubble Space Telescope (HST) together with ground-based photometry in various bands
which allows for highly accurate measurements of galaxy shapes and redshifts. Primarily due to its relatively
small area of roughly one square degree, two-point IA statistics have not been measured in COSMOS to date.
However, it is large enough to provide constraints on galaxy shape
distributions over a wide range of galaxy magnitudes, colours, and redshifts \citep[e.g.,][]{Joachimi13a}.
The reference sample of the COSMOS survey used in this work has been composed from the COSMOS2015 catalogue\footnote{\url{https://www.eso.org/qi}} \citep{Laigle16} and the Advanced Camera for Surveys General Catalog
\citep[ACS-GC\footnote{vizier.u-strasbg.fr/viz-bin/VizieR-3?-source=J/ApJS/200/9/acs-gc},][]{Griffith12}.
The COSMOS15 catalogue provides photometry in $30$ bands together with photometric
redshift estimates. The ACS-GC catalogue provides measurements of galaxy
shapes from imaging in the HST ACS F814W $i$-band which are corrected for effects
of the HST point spread function. 
For our analysis, we select objects with apparent magnitudes brighter than $i=24$ to ensure that
the redshifts and shape measurements are reliable
\citep{Griffith12, Laigle16}.
The corresponding mock sample is constructed from \flagship by applying the same cut.
Based on this mock, we expect this matched COSMOS catalogue to overlap almost entirely
with the \Euclid sample in the colour-magnitude-redshift space, as we show in Fig. \ref{fig:zmc_distribution}.
A more detailed summary of the COSMOS15 and the ACS-GC catalogues, as well as the procedure for matching these catalogues, can be found in \citet{Hoffmann22_cosmos_discs}.

\subsubsection{Colour-selected sub-samples}
\label{sec:color_cuts_cosmos}

\begin{figure}
    \begin{center}
    \includegraphics[width=0.45\textwidth]{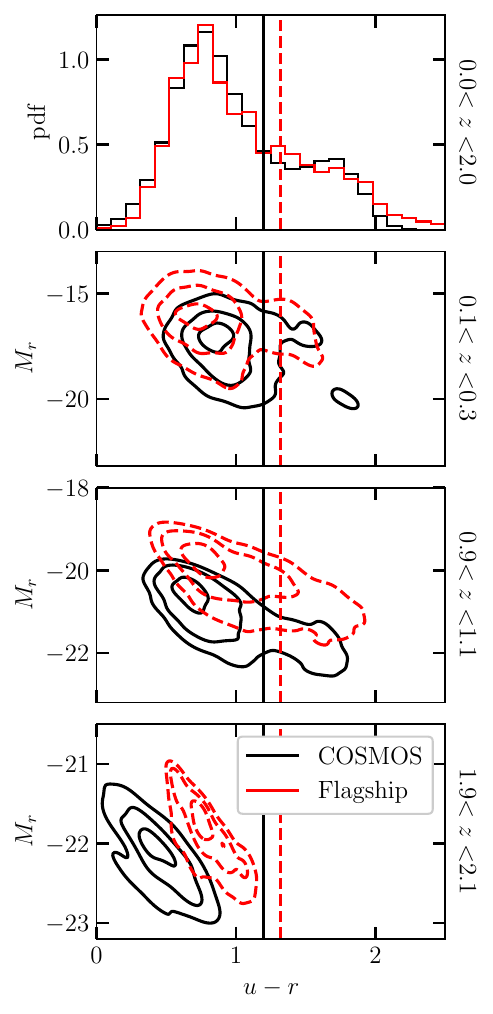}
    \caption{
        {\it Top:} Distribution of the rest-frame colour $u-r \equiv M_u-M_r$ for galaxies in a \Euclid-like sample,
        selected by $i < 24$ from COSMOS (black) and \flagship (red) across the full redshift range covered by \Euclid.
        {\it Center to Bottom:} Joint distribution of the rest-frame colour and the absolute rest-frame magnitude
        in the Subaru $r$-band in three redshift bins. Contours enclose the central $10$, $40$, and $70$ percent of
        the distributions, while black solid and red dashed lines represent results for COSMOS and \flagship respectively. 
        Vertical lines at $u-r=1.2$ and $u-r=1.32$ indicate the colour cuts used to define red and blue sub-samples
        in COSMOS and \flagship, respectively. The cut for \flagship is shifted such that
        the global fraction of blue galaxies matches that of COSMOS.
        }
    \label{fig:cmz_distribution_cosmos_vs_fs2}
    \end{center}
\end{figure}

To introduce a colour dependence of galaxy shapes in our model, we assign a colour type
to each object, classifying it as either red or blue. Following H22, we define this
colour type by a cut in the $u-r := M_u-M_r$ colour, where $M_u$ and
$M_r$ are absolute rest-frame magnitudes in the CFHT $u$-band and the Subaru $r$-band,
respectively. We compare the colour distribution in \flagship and COSMOS catalogues in the
top panel of Fig. \ref{fig:cmz_distribution_cosmos_vs_fs2}, for \Euclid-like samples of galaxies
with $i<24$ and $z<2.0$ and find a good agreement between both datasets.

For a more detailed validation of \flagship, we display the joint distribution of
the $u-r$ colour and the $r$-band magnitude in three redshift bins in the central and bottom panels
of Fig. \ref{fig:cmz_distribution_cosmos_vs_fs2}.
We find an overall good agreement between both datasets, with both showing similar
trends of increasing brightness and blueness with redshift as well as comparable shapes of the distributions.
However, our comparison also reveals a shift of the distribution in \flagship towards
fainter magnitudes at all redshifts compared to COSMOS, while the colours in \flagship
are slightly bluer in the lowest redshift bin ($0.1<z<0.3$) and significantly redder in the
highest redshift bin ($1.9<z<2.1$). 

\begin{figure}
    \begin{center}
    \includegraphics[width=0.45\textwidth]{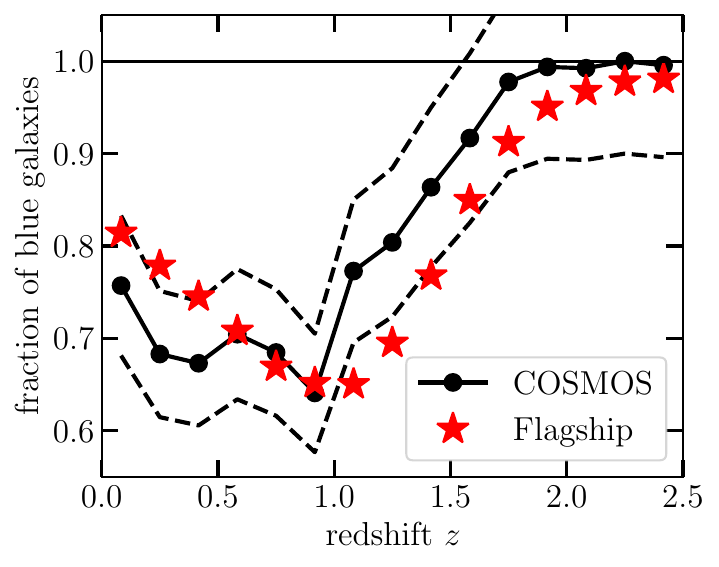}
    \caption{Fraction of blue galaxies in \flagship and COSMOS, selected by the colour cuts shown in Fig.~\ref{fig:cmz_distribution_cosmos_vs_fs2}.
        The dashed lines mark $\pm 10\%$ deviations from the COSMOS data.}
    \label{fig:frac_blue_cosmos_vs_fs2}
    \end{center}
\end{figure}

For the colour type definition in COSMOS, we adopt the cut from H22 at $u-r=1.2$, which leads to a
global fraction of blue galaxies of $f_{\rm blue} = 0.72$ for galaxies with $i<24$. We adjust this
cut in \flagship to $u-r=1.32$, such that $f_{\rm blue}$ has the same value as in COSMOS.
In Fig. \ref{fig:cmz_distribution_cosmos_vs_fs2} we show that these simple cuts provide a
reasonable separation between the red and blue sequences in colour-magnitude space in COSMOS
as well as in \flagship within the redshift range probed by \Euclid.
A more quantitative validation of the colour cut is shown in Fig. \ref{fig:frac_blue_cosmos_vs_fs2},
where we compare $f_{\rm blue}$ in \flagship and COSMOS in different redshift
bins, showing that \flagship replicates $f_{\rm blue}$ from COSMOS typically to within $10$ percent or less.

\subsection{SDSS Main sample}
The SDSS Main sample\footnote{\url{https://classic.sdss.org}}
consists of galaxies with an apparent SDSS $r$-band magnitude brighter than
$r_{\rm lim} = 17.77$. Spectroscopic redshifts from this sample
as well as estimates of the absolute rest-frame magnitudes are publicly
available in the New York University Value-Added Galaxy Catalog of the SDSS 
Data Release 7 \citep[NYU-VAGC]{Blanton05}\footnote{\url{http://sdss.physics.nyu.edu/vagc}},
which we use in this work as an observational reference when constructing the mock sample.
This catalogue consists of $741 \, 874$ galaxies and covers $7966$ square degree, which results
in a galaxy number density of $93\,{\rm deg}^{-2}$.

When constructing the mock SDSS main sample from \flagship, we adjust the limiting magnitude
by adding the constant $\Delta r = -0.15$ (i.e. $r_{\rm lim} + \Delta r$),
which is chosen such that the mock catalogue has the same number density
as the reference sample for objects with redshifts above $z=0.15$. Lower redshifts have been excluded for this
adjustment to reduce the impact of sampling variance when comparing the number densities.

\begin{figure*}
        \begin{center}
        \includegraphics[width=0.9\textwidth]{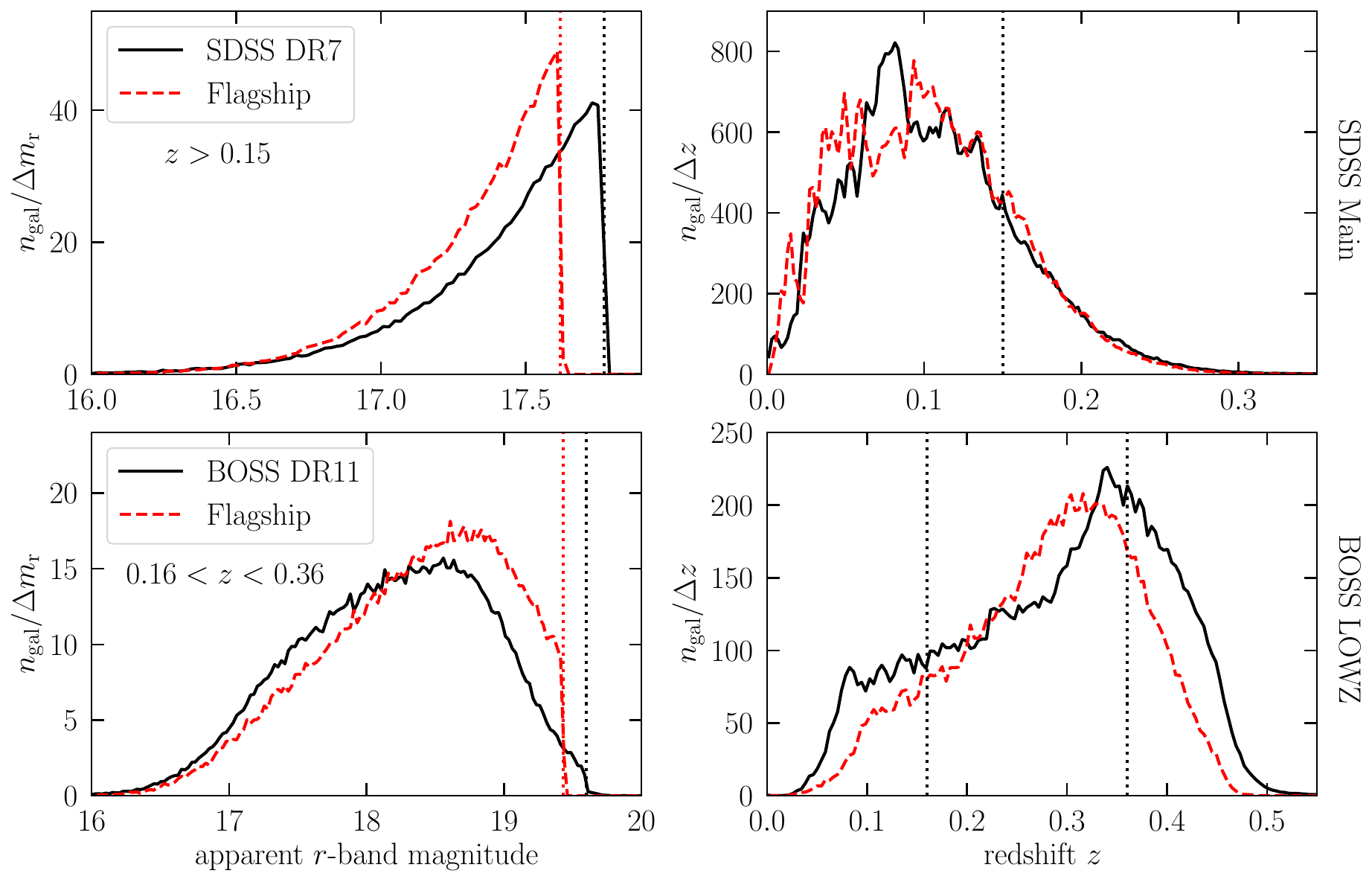}
        \caption{
            Distribution of galaxy counts per square degree ($n_{\mathrm{gal}}$) as a function of apparent $r$-band magnitude ($m_{\rm r}$, left) and redshift ($z$, right).
            Results are normalized by the bin widths $\Delta m_{\rm r}$ and $\Delta z$ to allow comparisons independent of bin size.
            Top and bottom panels show results for the SDSS Main and BOSS LOWZ samples, respectively.
            Black solid and red dashed lines represent observational data and the \flagship\ mock samples, respectively.
            Vertical dotted lines indicate the selection cuts in magnitude (left) and redshift (right) applied in our sample selection. The redshift cuts in Flagship coincide with those used for the observed samples.
        }
        \label{fig:mr-z_sdss_lowz_vs_fs2}
        \end{center}
    \end{figure*}

In the top right panel of Fig.~\ref{fig:mr-z_sdss_lowz_vs_fs2} we compare the differential redshift distributions
of the galaxy number density in NYU-VAGC to the \flagship mock sample. We find a reasonable agreement between both datasets for $z>0.15$ with absolute deviations around $15\%$. At lower redshifts both distributions appear to be similar but are strongly affected by fluctuations,
which we attribute to sampling variance. The top left panel of Fig. \ref{fig:mr-z_sdss_lowz_vs_fs2} 
shows a comparison between the differential galaxy number densities in the NYU-VAGC and \flagship as a function of the
apparent $r$-band magnitude for objects above $z=0.15$. We find the differential 
number density to be higher in the \flagship than in NYU-VAGC, which explains the 
necessity of reducing the limiting magnitude in order to obtain the observed cumulative number density.

\subsubsection{Colour-selected sub-samples}
\label{sec:color_cuts_sdss}

\begin{figure}
    \begin{center}
    \includegraphics[width=0.45\textwidth]{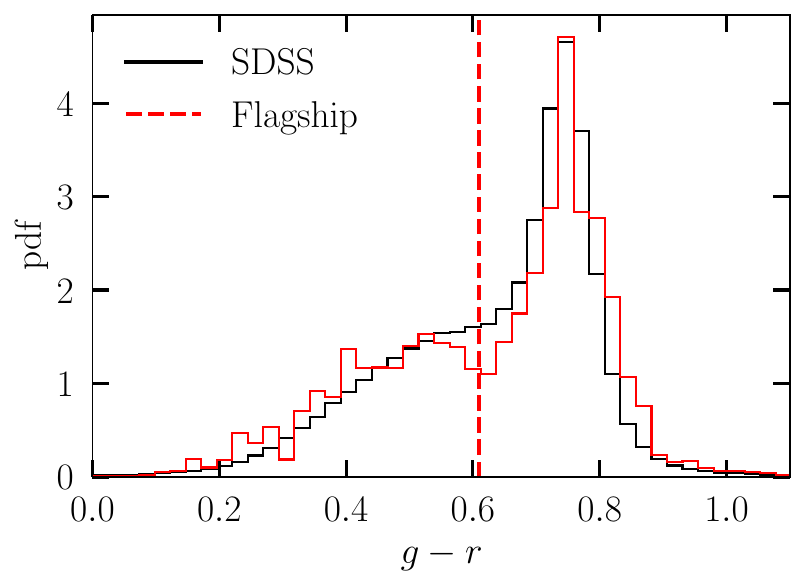}
    \caption{Distribution of the rest-frame colour $g-r := M_g-M_r$ rest-frame colour in the SDSS Main sample and the corresponding \flagship mock
    catalogue. The vertical red line shows the cut at $u-r=0.61$ where we split the mock SDSS catalogue
    into a red and blue sub-sample. This cut is chosen such that we reproduce the fraction of blue galaxies
    from J19 of $f_{\rm blue}=0.4$.}
    \label{fig:ur_sdss_fs2}
    \end{center}
\end{figure}

In order to calibrate our IA model against the IA statistics measured in the
red and blue sub-samples of the SDSS Main sample used by \citet[][hereafter referred to as J19]{Johnston19}, we adopt the authors’ methodology by
splitting our mock SDSS main sample with a cut on the $g-r := M_{g}-M_{r}$
rest-frame colour. While J19 applied a cut at $g-r=0.66$, we again adjust our cut
in \flagship to $g-r=0.61$ in order to obtain the same fraction of blue objects of $f_{\rm blue}=0.4$ as J19.
In Fig.~\ref{fig:ur_sdss_fs2} we compare the distribution of the colours in
the NYU-VAGC with the \flagship mock, showing that both datasets agree well with each other.

As an additional validation of the \flagship mock selection, we compare the two-point galaxy clustering statistics \wgg (described in Sect.~\ref{sec:ia_statistics}) with the observational measurements of the red and blue sub-samples from J19, as shown in Fig.~\ref{fig:wgg_lowz_sdss}. We find good agreement between the two datasets, indicating that our mock selection is suitable for calibrating our IA model; see Sect.~\ref{sec:wgg_fs_vs_obs} for a detailed discussion of the clustering comparison.

\subsection{BOSS LOWZ sample}
While the SDSS Main sample provides IA constraints over a wide range of absolute magnitudes and colours,
it is limited to relatively low redshifts. We therefore use direct IA
measurements based on the LOWZ sample of the BOSS survey from \citet[][hereafter referred to as SM16]{Singh16}, which
provide constraints at higher redshifts when calibrating the IA model
(see Fig.~\ref{fig:zmc_distribution} and \ref{fig:mr-z_sdss_lowz_vs_fs2}). The LOWZ sample consists of
luminous red galaxies that were selected by multiple cuts in the apparent
colour-magnitude space,
\begin{align}
    \label{eq:lowzcuts1}
    r &< 13.5 + \frac{c_\parallel}{0.3} + \Delta r \;;\\  16.0  &< r < 19.6 + \Delta r \notag \;;\\
    |c_\perp| &< 0.2 \notag \;,
\end{align}
with
\begin{align}
    \label{eq:lowzcuts2}
    c_\parallel &= 0.7(g-r) + 1.2[(r-i) - 0.18] \;; \\
    c_\perp &= (r-i) - (g-r) / 4.0 -0.18 \notag \,.
\end{align}
In addition to these cuts, we select galaxies within the redshift range $0.16 < z < 0.36$
in which SM16 performed their analysis. As for the SDSS selection, we introduce a parameter
$\Delta r$, which is zero in the selection of observed data and is adjusted in the
simulation to match the observed galaxy number density. For that adjustment, we use as a reference the 
LOWZ sample from the BOSS Data Release $12$, which covers an area of 
$8337$ deg$^2$ (here we use the effective area from \citealp{Reid16}, table~2).
It contains $249 \,938$ galaxies within the SM16 redshift range, which leads to a galaxy number density
of $29.98\,{\rm deg}^{-2}$. In order to achieve the same density in \flagship we set 
$\Delta r=-0.17$, which leads to $155\,179$ galaxies in the mock catalogue in the $5157$ deg$^2$
octant of the \flagship simulation.
In the bottom right panel of Fig. \ref{fig:mr-z_sdss_lowz_vs_fs2} we compare the differential
redshift distributions of the galaxy number density in the LOWZ sample from BOSS and the corresponding
\flagship mock sample. We find both data sets to agree reasonably well
with absolute deviations around $15\%$ within the SM16 redshift range,
given the complexity of the selection cuts and the fact that the
galaxy colours and magnitudes in \flagship have not been calibrated against the LOWZ sample.

The bottom left panel of Fig.~\ref{fig:mr-z_sdss_lowz_vs_fs2} 
shows a comparison between the differential galaxy number densities in BOSS and \flagship as a function of the
apparent $r$-band magnitude for objects within the redshift range analysed by SM16. For apparent magnitudes brighter
than $m_{r} < 18$ we find the differential galaxy number density in \flagship to be slightly below the observational data,
while it is significantly higher at dimmer magnitudes. This figure illustrates how reducing the limiting
magnitude by $\Delta r$ allows us to match the observed cumulative number density.

\subsubsection{Magnitude-selected sub-samples}
\begin{figure}
    \begin{center}
    \includegraphics[width=0.45\textwidth]{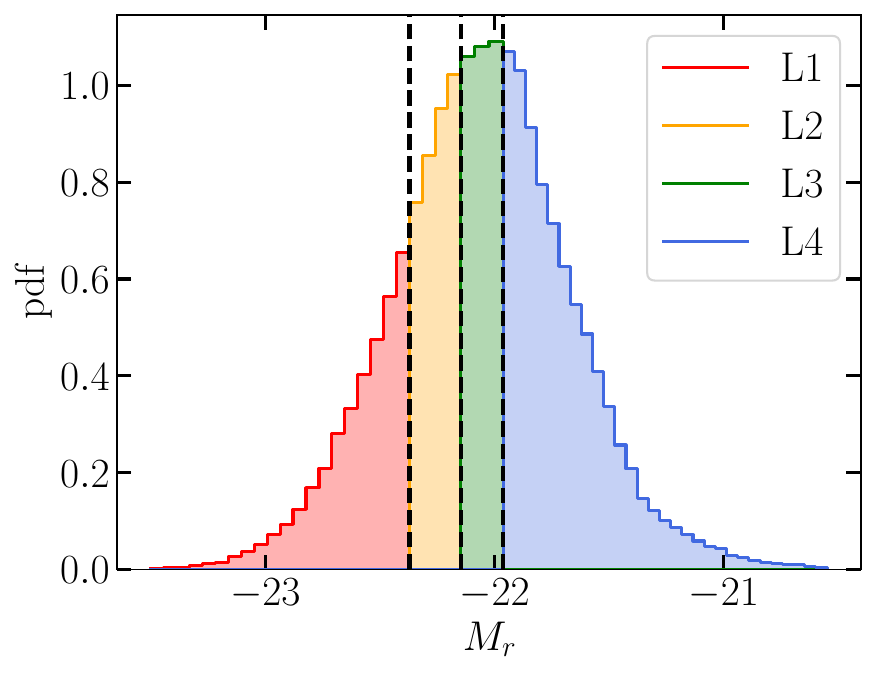}
    \caption{Distribution of absolute SDSS $r$-band magnitude of the mock
    BOSS LOWZ sample constructed from \flagship. Vertical lines indicate the limits
    of the four luminosity sub-samples L1 to L4.}
    \label{fig:lowz_subsample_selection}
    \end{center}
\end{figure}

Following SM16, we split the LOWZ samples into four luminosity sub-samples that are 
labelled from bright to faint L1 to L4. The samples are selected by cuts in the
absolute SDSS $r$-band magnitude, which are chosen so that each of the L1 to L3 samples contains $20\%$ of all galaxies,
while L4 contains the faintest $40\%$. The distribution of absolute magnitudes is shown for the four sub-samples from the \flagship mock in Fig.~\ref{fig:lowz_subsample_selection} while the main characteristics of
these sub-samples are given in Table~\ref{tab:mocks_summary}.

\begin{figure*}
	\includegraphics[width=\textwidth]{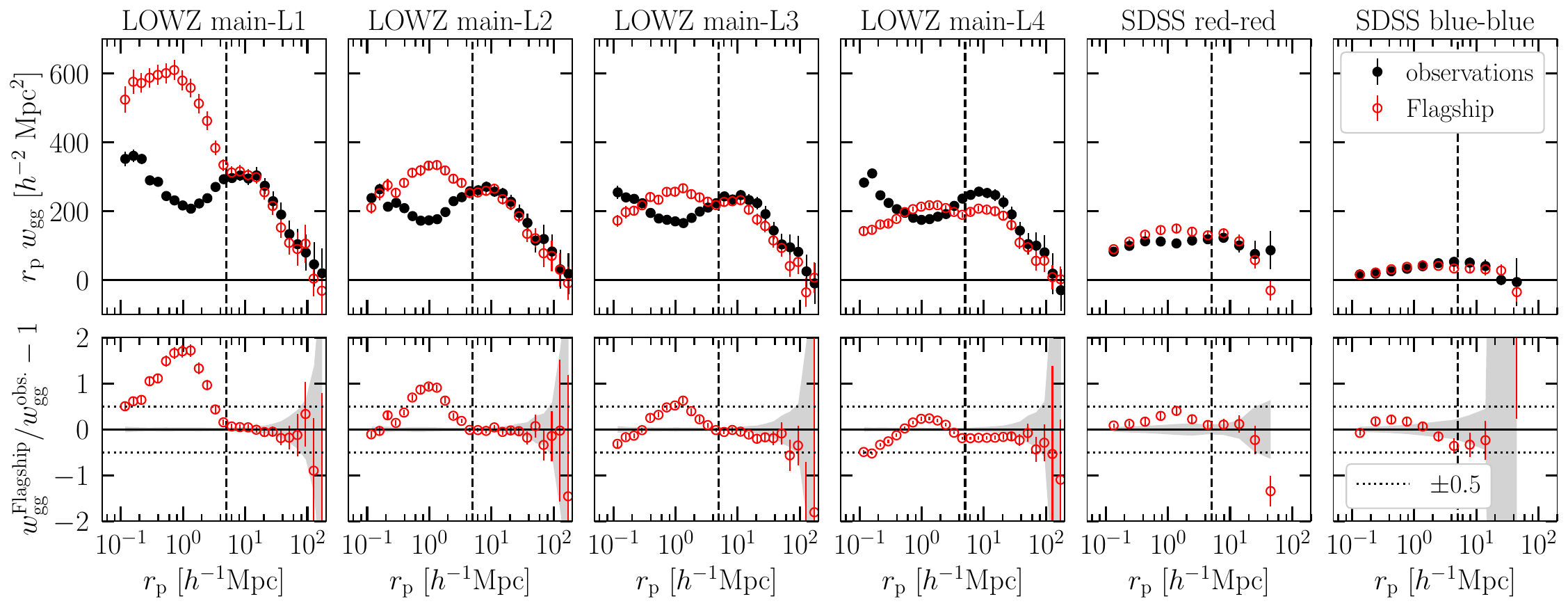}
   \caption{
   {\it Top}: Projected galaxy-galaxy cross-correlations between
   different density and shape samples against the projected scale
   for different observations and corresponding mock samples from the
   calibrated \flagship simulation (black and red symbols, respectively).
   The four left panels show results for the BOSS LOWZ sample, where
   the density sample is the LOWZ main sample, while the shape samples
   are the magnitude-selected sub-samples L1 to L4 (from bright to faint).
   The two right panels show results for SDSS, where both,
   density and shape samples are red and blue sub-samples.
   Error bars indicate jackknife estimates of the standard deviation.
   {\it Bottom}: Relative deviations of the \flagship results from the
   observations. Grey areas show the standard deviation of the
   observational data. Dotted lines mark $50$ \% deviations.
   The vertical dashed lines in the top and bottom panels 
   at $5$ \mpch indicate the approximate scale below (above) which we expect
   the alignment signal to be dominated by the alignment of satellites (centrals).
   }
    \label{fig:wgg_lowz_sdss}
\end{figure*}

To further validate the \flagship mock selection, we compare the two-point galaxy clustering statistics \wgg (see Sect.~\ref{sec:ia_statistics}) with observational measurements of the luminosity sub-samples from SM16, as shown in Fig.~\ref{fig:wgg_lowz_sdss}. We find overall good agreement, although significant deviations appear at small scales for the brightest samples (L1 and L2), which are taken into account when calibrating our IA model. Further details on this clustering comparison are provided in Sect.~\ref{sec:wgg_fs_vs_obs}.

\subsection{Horizon AGN}\label{sec:hagn}
The observational samples described in the previous sub-sections provide
IA constraints up to redshifts of $z=0.36$, which is well below the
bulk of redshifts probed by \Euclid (see Fig. \ref{fig:zmc_distribution}).
Since observations currently do not provide tight constraints at higher redshifts, we use data from the Horizon-AGN (HAGN, \citealt{Dubois14, Kaviraj17}) simulation at $z=1.0$ instead.

HAGN is a hydrodynamic simulation of cosmic structure formation within
a $100$ \mpch cube. It takes into account key processes in the
formation of galaxies: cooling, heating, and the chemical enrichment of gas,
the formation and evolution of stars and black holes, as well as feedback
from stellar winds, supernovae, and active galactic nuclei as described in \cite{Dubois14}.
Galaxies were identified in the simulation as groups of gravitationally bound stellar particles using the AdaptaHOP finder \citep{Aubert04}.

The HAGN simulation has been used for extensive studies on IA \citep{Chisari15, Codis15,chisari_redshift_2016, Chisari17, Codis18, Soussana20, Bate2020}. Galaxy shapes and
orientations were derived from the standard moment of inertia of the galaxies' stellar particle distribution, analogously to the measurements of halo shapes and orientations in \flagship from the reduced
moment of inertia, as detailed in Sect. \ref{sec:flagship_sim}. 

For the calibration of the \flagship IA model against the predictions from HAGN, we select a main reference sample of galaxies from HAGN that consists of objects at $z=1.0$ with absolute SDSS $r$-band magnitudes brighter than $M_r=-20$.
We further require stellar particle groups to be reliably detected at the AdaptaHOP tree level equal to $1$. This selection leads to a minimum of $621$ stellar particles per galaxy.
Note that the HAGN simulation output is available at multiple redshifts, which could have been incorporated into the calibration for a more comprehensive coverage of the \Euclid light-cone. However, due to the limited resources available for this work, we focus on the output at $z=1.0$, as it is the redshift closest to the mean redshift of our mock \Euclid sample when selecting objects at redshifts higher than those probed by the observational reference samples ($z>0.36$).

We construct a mock HAGN sample from \flagship by first selecting galaxies in a thin shell of the light-cone around $z=1.0$ with the width of $\Delta z = 0.02$. The comoving volume of this shell
is $33$ times larger than the volume covered by HAGN, leading to significantly smaller errors on the IA statistics measured in \flagship compared to HAGN. Galaxies in the \flagship redshift shell are required to be
brighter than the absolute SDSS $r$-band magnitude $M_r=-20 + \Delta M_r$. 
Here, we have introduced the constant $\Delta M_r=0.7$
which we adjust to match the comoving galaxy number density of the HAGN main reference sample of $0.023 \,h^{3}\textrm{Mpc}^{-3}$, analogously to the adjustment of the sample selections of the mock observational samples.

\begin{figure}
        \begin{center}
        \includegraphics[width=0.45\textwidth]{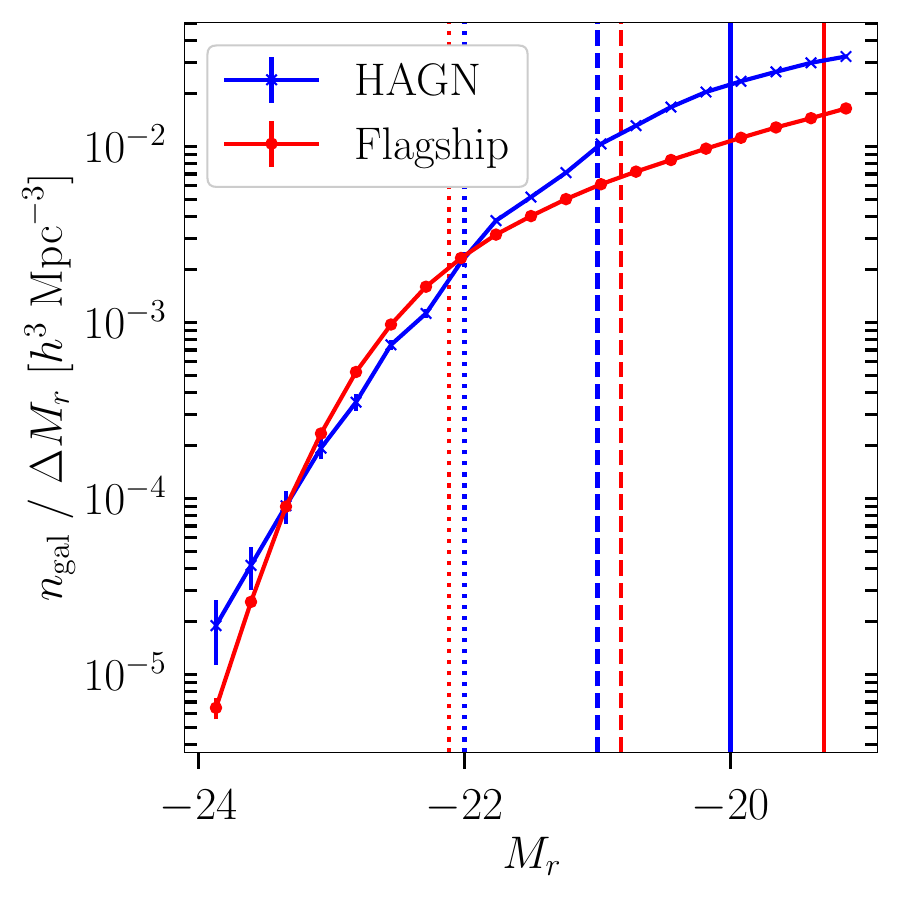}
        \caption{Comoving galaxy number density, $n_{\mathrm{gal}}$, as a function of absolute
            $r$-band magnitude ($M_r$). Values are normalised by the bin width $\Delta M_r$ to
            remove bin-size dependence. Results are shown at redshift $z = 1.0$
            for \flagship and HAGN (red and blue lines, respectively).  Error bars enclose the central $68\%$ of a Poisson distribution. Blue vertical lines at $M_r = [-20, -21, -22]$ denote the lower limits of HAGN sub-samples H1 to H3, used as high-redshift references for IA model calibration. The corresponding limits in \flagship (red vertical lines) are adjusted to $M_r \in [-19.3, -20.8, -22.1]$ to match the number densities in HAGN.
            }
        \label{fig:pdf_mr_hagn_vs_fs2}
        \end{center}
    \end{figure}

In Fig. \ref{fig:pdf_mr_hagn_vs_fs2} we compare the luminosity functions of HAGN and \flagship at $z=1$ to each other as functions of $M_r$,. 
At magnitudes brighter than $M_r=-22$, we find a similar 
decrease of the luminosity functions with magnitude for both simulations. At fainter magnitudes the luminosity function in HAGN is significantly higher than in \flagship, which explains why the limiting magnitude had to be set fainter in \flagship in order to reproduce
the cumulative number density of HAGN. Please note here that \flagship as well as HAGN have been calibrated against luminosity functions
at low redshifts, while the high-redshift luminosity functions are predictions which may differ due to the
fundamental differences in the methods used in these simulations.

In order to probe the luminosity dependence of the IA signal in HAGN we
split the main sample into three luminosity sub-samples at $M_r=-21$ and $-22$, as illustrated in Fig. \ref{fig:pdf_mr_hagn_vs_fs2}. The resulting samples, labelled H1 (bright) through H3 (faint) contain $4.4\%$, $24.3\%$, and $71.3\%$ of the galaxies, respectively. The corresponding selection cuts in \flagship are adjusted to $M_r=-20.8$ and $-22.1$ in order to obtain the same fractions of galaxies per sub-sample. The main characteristics of the mock HAGN samples are summarised in Table~\ref{tab:mocks_summary}.

%% file: sections/ia_statistics.tex
\section{Clustering and IA statistics}\label{sec:ia_statistics}

For the calibration of the \flagship IA model, we quantify the IA signal using two types of two-point statistics. First, we use the projected two-point correlation function \wgp to compare the IA signal predicted by the \flagship mock with direct measurements from observational data. For the calibration based on the HAGN simulation, we compute the two-point correlation of three-dimensional galaxy orientations, $\eta$, and compare these directly to the corresponding measurements in HAGN.

\subsection{Projected correlation functions \texorpdfstring{\wgg}{wgg} and \texorpdfstring{\wgp}{wgp}}

We measure the projected galaxy-ellipticity correlation to calibrate the IA model against observational constraints derived from the BOSS LOWZ sample by SM16 and the SDSS Main sample by J19. In addition, we measure projected galaxy-galaxy correlations to validate the galaxy clustering in the mock samples constructed from \flagship that are used for the calibration. When measuring these correlations we follow SM16 and J19 by studying the cross-correlation between a shape sample $\rm S$, consisting of the galaxies whose IA signal we want to measure and a density sample $\rm D$, which is used as a tracer for the underlying matter distribution.

The galaxy-galaxy cross-correlation function between two samples is defined as
$\xi_{\rm gg}(r_{\rm p},\Pi) := \langle \delta_{\rm g}^{\rm S} \delta_{\rm g}^{\rm D} \rangle (r_{\rm p},\Pi)$, where $\delta_{\rm g}^{\rm S}$ and $\delta_{\rm g}^{\rm D}$ are the galaxy density contrasts
of the shape and density samples, respectively, and $\langle \ldots \rangle$
indicates an ensemble average.
Measurements are done in bins centred on a transverse comoving separation $r_{\rm p}$ and a line-of-sight comoving separation $\Pi$.
We measure this correlation from the data using the estimator from \citet{Landy93},
\begin{equation}
{\hat \xi_{\rm gg}}(r_{\rm p},\Pi) =  \frac{\rm SD - D R_{\rm S} - R_{\rm D} S + R_{\rm D} R_{\rm S} }{R_{\rm S} R_{\rm D}}\,.
\label{eq:xi_gg_LS}
\end{equation}
Each term in the numerator and denominator on the right-hand side of
this equation stands for the counts of galaxy pairs in bins separated
by $r_{\rm p}$ and $\Pi$. The symbols $\rm R_S$ and $\rm R_D$ denote samples of randomly distributed points
that are constructed to follow the redshift distributions and sky footprints of the $\rm S$ and $\rm D$ samples, respectively.

Analogously to the galaxy-galaxy correlation one can define the galaxy-ellipticity correlation function as
$\xi_{\rm g+,\times}(r_{\mathrm p},\Pi) = \langle \delta_{\rm g}^{\rm D} \epsilon_{+,\times} \rangle (r_{\rm p},\Pi)$. The
ellipticity is defined for each position-ellipticity galaxy pair in the average
$\langle \ldots \rangle$ such that the orientation is the angle between the galaxies' major axis
and the separation vector on the sky. An estimator for this correlation function can be defined as \citep{Mandelbaum06}
\begin{equation}
\hat{\xi}_{\rm g+,\times}(r_{\rm p},\Pi) =  \frac{\rm S_{+,\times} D - S_{+,\times} R_{\rm D}}{R_{\rm S} R_{\rm D}} \,,
\label{eq:xi_g+_LS}
\end{equation}
with
\begin{equation}
 {\rm S}_{+,\times} X = \sum_{i,j} \epsilon_{+,\times} (i|j) \,,
\end{equation}
where $\epsilon_+ (i|j) = \Re \left[ \epsilon \exp{\left( -2 {\rm i} \phi_{ij} \right)} \right]$
is the $(+)$-component of the ellipticity of a galaxy $i$ in sample $\rm S$, defined with respect to
the separation vector pointing to position $j$ in sample $X$. $\phi_{ij}$
is the position angle of this separation vector at the position $i$ and $X$ refers to either $\rm D$ or $\rm R_{\rm D}$. The $(\times)$-component is defined analogously using the imaginary part of the expression above. Since $\xi_{\rm g\times}$ changes sign under parity transformation, it does not contain a physical signal, and we therefore do not consider it further in this study.

For the measurements in the mock BOSS LOWZ samples we follow SM16 by using $25$
logarithmic bins in the interval $0.1 < |r_{\rm p}| / (h^{-1} \textrm{Mpc}) < 200 $
and $20$ linear bins in the range $ |\Pi| < \Pi_{\rm max} $.
Similarly, for the measurements in the mock SDSS Main samples we follow J19 by using $11$
logarithmic bins in the interval $0.1 < |r_{\rm p}| / (h^{-1} \textrm{Mpc}) < 60 $
and $30$ linear bins within $ |\Pi| < \Pi_{\rm max} $. In both cases  $\Pi_{\rm max} = 60\, h^{-1} \textrm{Mpc} $ was chosen.
The projected correlation is then given by
\begin{equation}
w_{\rm gx}(r_{\rm p}) = \int_{-\Pi_{\rm max}}^{\Pi_{\rm max}} \xi_{\rm gx}(r_{\rm p}, \Pi) \ {\rm d}\Pi \,,
\label{eq:w_theo}
\end{equation}
where $w_{\rm gx}$ stands for $w_{\rm gg}$ or $w_{\rm g+}$. In the measurement the integral above is approximated by a Riemann sum over the line-of-sight bins.

Errors on the measurements of \wgg and \wgp are estimated using jackknife resampling.
The \flagship octant is therefore split into $N_{\rm JK} = 88$ HEALPix regions. The covariance is then estimated as  
\begin{equation}
C^{\rm JK}_{ij} = (N_{\rm JK}-1) \langle \Delta_i \Delta_j \rangle,
\label{eq:jk_cov}
\end{equation}
with $\Delta_i = w^{\rm JK}_i - w_i$, where $w_i$ is the projected correlation measured in the $r_{\rm p}$ bin $i$ on the full sky area,
$w^{\rm JK}_i$ is the same measurement, but dropping one 
jackknife sub-region at a time and $\langle \ldots \rangle$
is the average over the $N_{\rm JK}$ measurements of $\Delta_i$.
Further details on the jackknife sampling, the construction of random catalogues and the code for performing the measurements are given in H22.

\subsection{Galaxy clustering: \flagship versus observations}
\label{sec:wgg_fs_vs_obs}

In Fig.~\ref{fig:wgg_lowz_sdss} we compare the measurements of \wgg in the
SDSS Main and BOSS LOWZ samples, derived from observed data by J19 and SM16,
respectively, to the corresponding measurements from \flagship as an additional validation of the mock samples
described in Sect.~\ref{sec:gal_cats}. In the case of LOWZ \wgg refers to the
cross-correlations between the full LOWZ sample and the four luminosity sub-samples,
while for SDSS \wgg is the auto-correlation of the red and blue sub-samples, respectively.

We find the deviations between \flagship and observations at scales above $5$ \mpch
to be consistent with the $1\sigma$ jackknife errors. However, at smaller scales
we find significant deviations, which are highest at roughly $1$\mpch and
increase with the brightness of the sample, reaching almost $200\%$ for the LOWZ-L1
sample. It is interesting to note here that H22 did a similar comparison between
\wgg in the LOWZ sub-sample and measurements from the MICE simulation \citep{lensing-fosalba2016}, which is based
on a similar combined HOD and abundance matching model as \flagship. This comparison revealed
smaller deviations below $60\%$, indicating that the stronger deviations which
we find in \flagship result from the choice of model parameters for the galaxy population,
rather than from the simulation method itself. It is important to note here that the HOD–abundance matching model in \flagship was calibrated against the SDSS Main Sample, but not the LOWZ sample.

\subsection{Three-dimensional correlation function of orientation vectors $\eta$}

When comparing the IA signal in \flagship with the HAGN simulation,
we take advantage of the fact that we have access to the three-dimensional galaxy
positions and orientations in both simulations. This allows us to
measure the three-dimensional alignment without losing information due to projections,
as in the analysis of observational samples. In addition, we can
define the statistics to depend solely on galaxy orientations but not on the shapes.
We expect the shapes of galaxies in HAGN to be, on average, rounder than those found in
observations, due to an absence of thin disks, which results from
resolution effects in the simulation \citep[see][and references therein]{Hoffmann22_cosmos_discs}. This deficiency of disks may bias correlations of
intrinsic ellipticity components, such as $w_{\rm g+}$,
when considering samples that include galaxies of all morphological types, as is the case
in the present work. We measure the IA signal in HAGN independently of galaxy shapes
as the two-point alignment statistics of galaxy orientation
with their surrounding large-scale structure in three dimensions, as
\begin{equation}
\eta_X(r) = \langle |{\bf \hat{X}}_{1} \cdot {\bf \hat r}_{12}|\rangle (r) - 1/2 \ ,
\label{eq:eta}
\end{equation}
where ${\bf \hat{X}}_{1}$ refers to the normalised three-dimensional major or minor axis vectors
(${\hat{\bf A}_{\rm 3D}}$ or ${\hat{\bf C}_{\rm 3D}}$, respectively, as defined
in Sect.~\ref{sec:flagship_sim}) of a galaxy at position $1$ and
${\bf \hat r}_{12}$ is the normalised distance vector that points to a galaxy at
position $2$, and $\langle \dots \rangle$
is the average over all galaxy pairs that are separated by the distance $r = | {\bf r}_{12} |$.
The subtraction of $1/2$ leads to $\eta_X(r)=0$ if the alignment between ${\bf \hat{X}}_{1}$
and ${\bf r}_{12}$ is random.\footnote{In the literature, the 3D alignment statistic is commonly quantified as
$\langle {\bf \hat{X}}_{1} \cdot {\bf \hat r}_{12} \rangle (r) - 1/3$.
Here, we instead adopt Eq.~\ref{eq:eta}, as we find it more intuitive: for a random distribution, $|{\bf \hat{X}}_{1} \cdot {\bf \hat r}_{12}|$ is uniformly distributed between $0$ and $1$ with an average value of $1/2$. Both definitions are valid measures of 3D alignment and convey essentially the same information.}
Errors are estimated by jackknife resampling, using $88$ HEALPix
regions for measurements in the \flagship light-cone and $64$ cubical
regions in HAGN. In the process the jackknife regions are used
to organise the galaxy catalogue in a simple tree structure,
which allowed for a significant acceleration of the search
for galaxy pairs. As in the computation of $w_{\rm g+}$ we increase the signal-to-noise by
computing the cross-correlation between the positions of galaxies in a
density sample and the orientations of galaxies in magnitude-limited sub-samples.
The construction of these samples is described in Sect.~\ref{sec:hagn}.
The code used for the measurements in the \flagship
light-cone and the HAGN simulation box (including error estimation) is publicly
available\footnote{\url{https://github.com/kaidhoffmann/covo}}.
A potential concern for IA constraints from AMR simulations such as HAGN is a systematic effect arising from the alignment between galaxy shapes and the axes of the Cartesian simulation grid, commonly referred to as grid locking. We do not expect the $\eta_X$ statistic to be affected by this effect, based on the results of \citet{Chisari15}, who showed that grid locking has no significant impact on position–orientation correlations
in HAGN.

%% file: sections/modelling_galaxy_shapes.tex
\section{Modelling galaxy shapes}\label{sec:modelling_shapes}

Modelling the intrinsic 3D galaxy shapes is the first out of the two main steps in our IA model. Together with the intrinsic 3D galaxy orientations, which are
discussed in Sect. \ref{sec:modelling_orientations}, the 3D shapes allow us to derive the 2D intrinsic galaxy shear components by projection along the observer's
line of sight.

We model the 3D shapes following the methodology from H22 which takes into account observed dependencies of galaxy shapes on
redshift, magnitude, and colour. In that model, galaxies are approximated as 3D ellipsoids, whose shape is fully characterised by the axis ratios $q_{\rm 3D}$ and $r_{\rm 3D}$, which we have defined in
Eq.~(\ref{eq:3D_axes_ratios}).
At a fixed point in redshift-magnitude-colour space, the 3D axis ratio distribution is described by a Gaussian model,
\eq{
\label{eq:P3d}
\tilde P(q_{\rm 3D},r_{\rm 3D}) =
\exp \Biggl\{
-\frac{1}{2}
\left[
\left( \frac{q_{\rm 3D} - q_0}{\sigma_{qr}} \right)^2 +
\left( \frac{r_{\rm 3D} - r_0}{\sigma_{qr}} \right)^2
\right]
\Biggr\},
}
where $q_0$, $r_0$, and $\sigma_{qr}$ are free model parameters.
The distribution is truncated as
\eqa{
\label{eq:P3dnorm}
    P =
    \begin{cases}
     \tilde P_{\rm 3D} / \mathcal{N} & \text{if} \ q_{\rm 3D} \ , r_{\rm 3D} \in (0,1]\,; \\
    0 & \text{else}\,,
    \end{cases}
\label{eq:qr_pdf}
}
with
\begin{equation}
    \mathcal{N} = \int_{0}^1 \int_{0}^1 \tilde P_{\rm 3D}(q_{\rm 3D},r_{\rm 3D}) \ {\rm d}r_{\rm 3D} \ {\rm d}q_{\rm 3D}
\end{equation}
to ensure $A_{\rm 3D} \geq B_{\rm 3D} \geq C_{\rm 3D}$.
This model is motivated by results from \citet{Hoffmann22_cosmos_discs},
who showed that the distribution of $q_{\rm 3D}$ and $r_{\rm 3D}$ 
of disk galaxies in HAGN and Illustris TNG \citep{IllustrisTNG} can be well approximated
by a truncated diagonal Gaussian distribution\footnote{
Note that a diagonal Gaussian distribution may not be a suitable approximation for other combinations of shape parameters. For instance, the joint distribution of
$q_{\rm 3D}$ and $s_{\rm 3D}$ must satisfy the constraint
$s_{\rm 3D} \lesssim q_{\rm 3D}$, which is not captured by a diagonal Gaussian distribution.}.
H22 simplified this Gaussian model by assuming identical dispersion parameters $\sigma_{qr}$
for $q_{\rm 3D}$ and $r_{\rm 3D}$. This simplification was motivated by their finding that such an approach allows for more robust parameter estimation from small, noisy samples, thereby facilitating the model calibration described below. Furthermore, H22 demonstrated that this model is sufficiently accurate for reproducing the observed two-dimensional axis ratio distributions from COSMOS which included galaxies of all types.

The 3D shape of a given galaxy in the simulation can now be obtained
from Eq. \eqref{eq:P3dnorm} by randomly drawing the axis ratios $q_{\rm 3D}$
and $r_{\rm 3D}$ from $P$.
In order to introduce a dependence of galaxy shapes on galaxy properties, an additional modelling step is necessary. For that purpose, we modify the axis ratio distribution parameters based on each galaxy's redshift, magnitude, and colour.
This parameter dependence has been calibrated at distinct points in redshift-colour-magnitude space, such that the distribution of projected 2D axis ratios for an
ensemble of randomly oriented ellipsoids matches the observed distribution from
the COSMOS catalogue, described in Sect. \ref{sec:gal_cats:cosmos}. The parameters
for a specific galaxy in the simulation are then obtained by interpolation between
the points where the model was calibrated. Further details
on the method, the motivation and the calibration of the galaxy shape model
are given in H22.

The simulated galaxy shapes are publicly available as part of the \Euclid
Flagship Catalogue on CosmoHub\footnote{\label{fn:cosmohub}\url{https://cosmohub.pic.es}}
\citep{Carretero17,Tallada20}. The code used to generate these simulations is
available on GitHub.\footnote{\label{fn:genIAL}\url{https://github.com/kaidhoffmann/genIAL}}

\begin{figure}
        \includegraphics[width=0.48\textwidth]{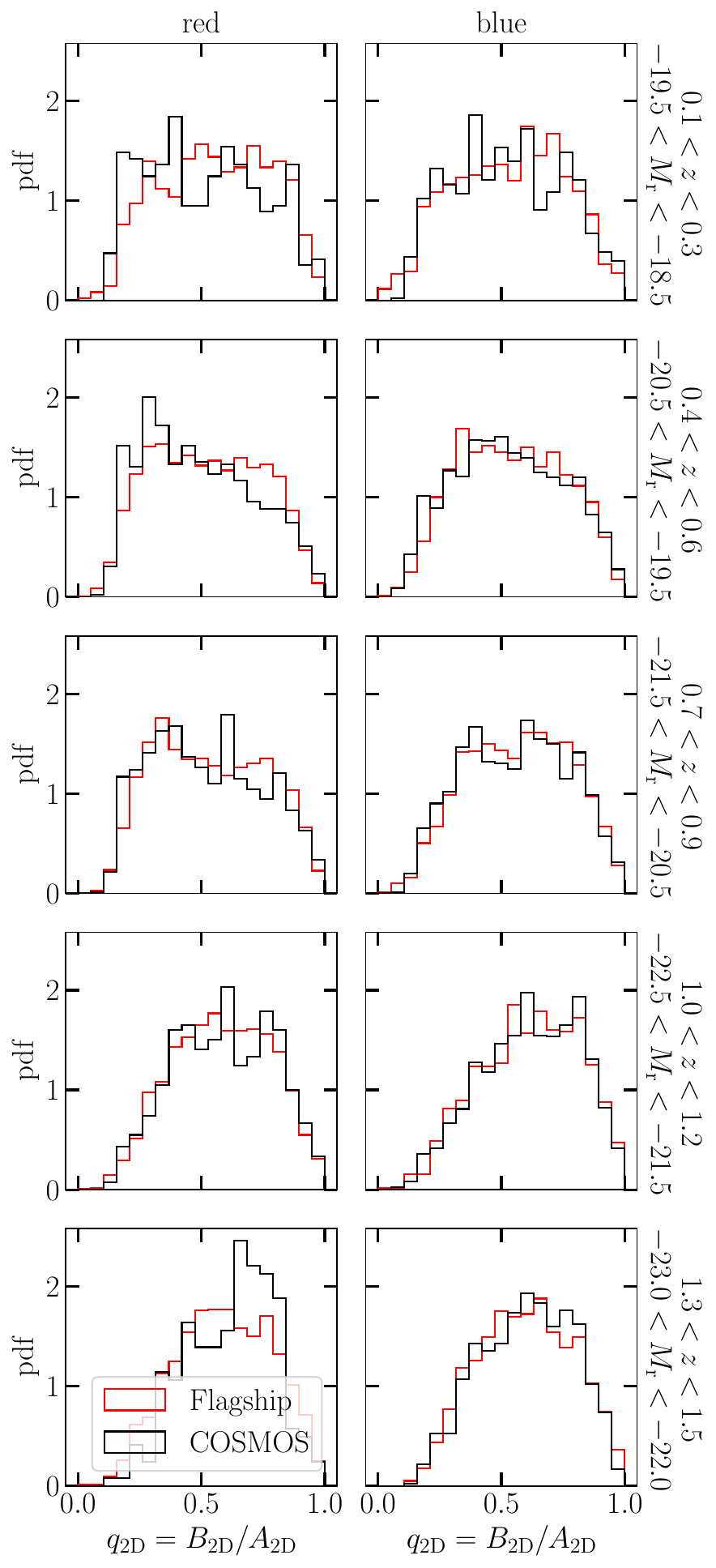}
        \caption{2D galaxy axis ratios measured for red and blue galaxies in different
    volume-limited samples. The redshift and absolute Subaru $r$-band limits of each sample are indicated on the right. 
    Red and black histograms show results from the \flagship simulation and COSMOS observations, respectively.}
        \label{fig:2d_axis_ratio_pdf}
    \end{figure}

In Fig. \ref{fig:2d_axis_ratio_pdf} we compare shape distributions
of the projected 2D axis ratios in \flagship with the COSMOS reference data
for a set of ten volume-limited samples of the red and blue galaxies,
covering absolute magnitudes in the range of $-19.5 < M_r < -23.0$ and redshifts
in the range of $0.2<z<1.5$ as detailed in Sect. \ref{sec:gal_cats:cosmos}.
We find an overall good agreement between \flagship and COSMOS in all samples, demonstrating
that the simulated 2D galaxy shape distribution follows the observed colour-magnitude-redshift dependence.

%% file: sections/modelling_galaxy_orientations.tex
\section{Modelling galaxy orientations}\label{sec:modelling_orientations}

We model the galaxy 3D orientations in two steps. In the first step we set initial orientations, which are then randomised in a second step, such that the IA statistics in the simulations match the reference measurements.

\subsection{Initial orientations}
\label{sec:initial_orientations}
The initial orientations are set separately for central and satellite galaxies, using two distinct methods for each type.

\paragraph{Centrals:}
All central galaxies are initially oriented by aligning their principle axes
with those of their host dark matter halos, i.e. $(\hat{\bf{A}},\hat{\bf{B}},\hat{\bf{C}})_{\rm gal} = (\hat{\bf{A}},\hat{\bf{B}},\hat{\bf{C}})_{\rm halo}$.
This part of the modelling deviates from previous semi-analytic alignment models (e.g. J13, H22). In these models only red central
galaxies are aligned with the principle axes of their host halos while blue centrals are assumed to be disks, whose minor axis is initially aligned with
their host halo's angular momentum vector. We simplify this mixed modelling for red and blue galaxies for three reasons:

1) a colour-dependent modelling of the orientations relies on a colour-based classification of galaxies into disks and ellipticals.
Such a classification is known to be inaccurate, in particular when using a simple colour cut, e.g. disks can be red when being oriented edge-on to the observer due to dust extinction or when having a large central bulge (see e.g. the discussion in \citealt{Xiao12,Lackner12,Hoffmann22_cosmos_discs,Uzeirbegovic22}).
The mixed modelling further neglects the significant fraction of galaxies with irregular morphologies \citep[e.g.][]{Conselice05, Huertas-Company16}. For these reasons it is not obvious that the mixed model is more accurate than the more simple modelling used in this work and described below.

2) The minor axis direction of a dark matter halo correlates with the direction of its angular momentum \citep{Bailin05, Bett07}. In our model this alignment will be weaker compared to the mixed alignment models, where it is set explicitly.
However, since we randomise galaxy orientations in the second step of our modelling, taking galaxy colour into account, we expect the final alignment signal for blue galaxies to be similar to that obtained with the mixed modelling approach.

3) A consistent description of all galaxies reduces the total number of parameters needed for introducing a redshift, magnitude, and colour dependence of the galaxy-halo misalignment.

\paragraph{Satellites:}
All satellite galaxies are initially oriented by pointing their major axis towards their host halo's centre, which is motivated by findings in hydrodynamic simulations \citep{Pereira2008,Pereira2010,Welker2017} and observations \citep{Johnston19,Georgiou2019b}, as well as the success of this modelling approach in reproducing
the observed alignment statistics at small scales \citep[][H22]{fortuna21a}.

Note that this model component could be advanced by aligning satellites with their sub-halos or by introducing a radial dependence to the alignment model \citep{Georgiou2019b}. However, in a comparison of these advanced models with the simpler radial alignment model used in this work, \cite{VanAlfven2023} found that the latter is sufficient for reproducing second-order IA statistics as measured in IllustrisTNG.

\subsection{Random misalignment}
\label{sec:misalignment}
It has been shown in the literature that semi-analytic IA models, which assume perfect alignment between galaxies and their host halos, predict IA statistics that are significantly higher than those observed in the data \citep{Heymans04}. This discrepancy can be removed by introducing a random misalignment between galaxies and their host dark matter halos \citep[e.g.,][]{Heymans06,Okumura09b}, which is also predicted by hydrodynamic simulations \citep[e.g.,][]{vdBosch02}.

To randomise the 3D major and minor axes (${\hat{\bf A}}_{\rm 3D}$ and ${\hat{\bf C}}_{\rm 3D}$, respectively) we follow the approach of H22 by first drawing two misalignment angles $\theta_A$ and $\theta_C$ from the von Mises--Fisher distribution \citep{Fisher93},
\begin{equation}
    P(\cos\theta) = \frac{1}{2 \sigma_{\rm MF}^2 \mathrm{sinh}(\sigma_{\rm MF}^{-2})}\exp\left( \frac{\cos\theta}{\sigma_{\rm MF}^2} \right),
    \label{eq:ms_dist}
\end{equation}
for each galaxy. The randomised principle axis vectors ${\hat{\bf A}}_{\rm 3D}^r$
and ${\hat{\bf C}}_{\rm 3D}^{r,\prime}$ are then constructed
such that ${\hat{\bf A}}_{\rm 3D} \cdot {\hat{\bf A}}_{\rm 3D}^r = \cos(\theta_A)$
and ${\hat{\bf C}}_{\rm 3D} \cdot {\hat{\bf C}}_{\rm 3D}^{r,\prime} = \cos(\theta_C)$.
Here ${\hat{\bf C}}_{\rm 3D}^{r,\prime}$ is a temporary vector which is in general
not perpendicular to ${\hat{\bf A}}_{\rm 3D}^r$ since $\theta_A$ and $\theta_C$ are
drawn independently from each other. The final randomised minor axis orientation
is therefore obtained as ${\bf C}_{\rm 3D}^r = ({\hat{\bf A}}_{\rm 3D}^{r} \times {\hat{\bf C}}_{\rm 3D}^{r,\prime}) \times {\hat{\bf A}}^r_{\rm 3D}$ and is then normalised to ${\hat{\bf C}}_{\rm 3D}^r = {\bf C}_{\rm 3D}^r / |{\bf C}_{\rm 3D}^r|$.

We adopt the von Mises–Fisher distribution, since it successfully employed by H22 to reproduce the observed IA signal from BOSS LOWZ data using a semi-analytical IA model similar to the one adopted in this work. \citet{Bett12} showed that this distribution provides a good fit to the misalignment angles between halo and galaxy angular momentum vectors in simulated haloes including baryons and galaxy formation physics. However, a direct validation of the von Mises–Fisher distribution against misalignment angles between the principal axes of central galaxies and haloes, as well as between the major axes of satellites and the halo-centre direction, remains to be done. \citet{Velliscig15} analysed such distributions in the EAGLE and cosmo-OWLS simulations and proposed an alternative fitting model. While its functional form is similar to the von Mises–Fisher distribution, it requires four free parameters, which would make the parameter calibration in our work substantially more challenging. A further single-parameter alternative to the von Mises–Fisher distribution is the Dimroth–Watson distribution. \citet{VanAlfven2023} demonstrated that a semi-analytical IA model based on this distribution successfully reproduced IA statistics of galaxies in the IllustrisTNG hydrodynamical simulation

The width of the distribution \sigMf is a free parameter that provides control over the alignment amplitude. Increasing (decreasing) \sigMf leads to a higher (lower) randomisation of the initial orientations and hence to a lower (higher) IA signal in the simulation.
We can hence control the dependence of the predicted IA signal on galaxy properties by introducing a dependence of \sigMf on galaxy redshift $z$, absolute rest-frame magnitude $M_r$ and rest-frame colour index $u-r$ (defined in Sect. \ref{sec:gal_cats}),
\begin{equation} \label{eq:sigma_model} 
\sigma_{\mathrm{MF}}=p_0 \underbrace{\left(\frac{z}{z_0}+1\right)^{p_1}}_{\begin{matrix}{\sigma_z\left(p_1\right)}\end{matrix}} \underbrace{\left(\frac{M_r}{M_0}+p_2\right)^{p_3}}_{\begin{matrix}{\sigma_{\text {mag }}\left(p_2, p_3\right)}\end{matrix}} \underbrace{\left(\frac{u-r}{(u-r)_0}+p_4\right)^{p_5}}_{\begin{matrix}{\sigma_{\rm{c o l}}\left(p_4, p_5\right)}\end{matrix}},
\end{equation}
The constants $M_0=-22$, $(u-r)_0=1$ and $z_0=1$ are chosen to avoid large values of the bases of the $\sigma_{\rm z}$, $\sigma_{\rm mag}$ and, $\sigma_{\rm col}$ terms\footnote{The $M_0$ value in particular was motivated by the literature \citep{Joachimi11,Singh15}.}. We chose the functional form of these terms to allow for a large range of dependencies of the alignment signal on galaxy properties while avoiding negative values for \sigMf, as can arise for instance when extrapolating linear relations. In addition to the dependence on galaxy properties, we add a dependence on galaxy type, which can be either central or satellite by using a separate set of the parameters $p_i$ for each type.

As a result, we have two von Mises--Fisher distributions in our model, \sigMfC and \sigMfS, with a total number of $12$ free parameters. By introducing a type dependence to the model we obtain control over the scale dependence of the final IA signal in the simulation (see Sect. \ref{sec:IAstats_sigmaMF}). Another motivation for using two separate sets of misalignment parameters for centrals and satellites is that the initial orientations were set with different methods. Using a type-independent misalignment would therefore be an inconsistency in the model. The final form of Eq. (\ref{eq:sigma_model}) is a result of a development process, in which we gradually added complexity to the model until we were able to match the different constraints simultaneously with the simulation. This matching is achieved by calibrating the parameters $p_i$ as described in Sect. \ref{sec:param_calibration}.

\subsection{Response of IA statistics to galaxy misalignment}
\label{sec:IAstats_sigmaMF}
In preparation for the calibration of the \sigMf parameters from Eq. (\ref{eq:sigma_model}) we study our IA model in a simplified setting, fixing \sigMfC and \sigMfS to constant values that are independent of galaxy redshift, magnitude, and colour.
\begin{figure*}
	\includegraphics[width=\textwidth]{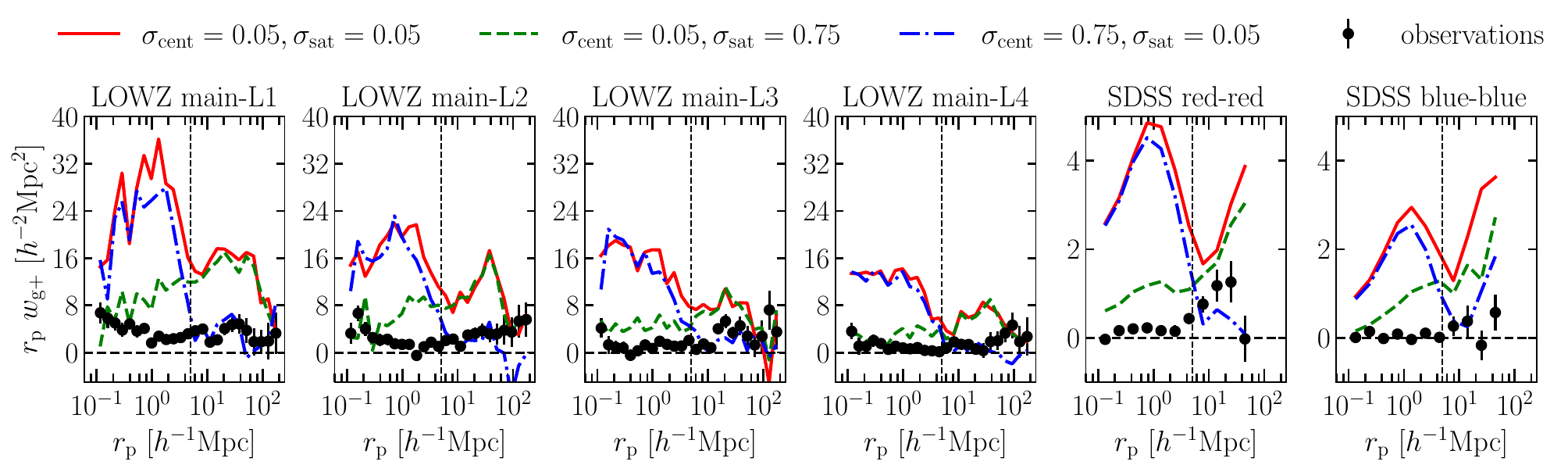}
    \caption{
        Projected galaxy-ellipticity cross-correlation between a density sample and various shape samples. The four left panels show results based on shapes from the luminosity sub-samples, with the full LOWZ sample as the density sample. The two right panels display results for SDSS, using the Main sample as the density sample and colour sub-samples as shape samples. Black dots represent observational data, while coloured lines represent \flagship simulations for three simplified intrinsic alignment (IA) models, each using different fixed values for $\sigma_{\rm MF}^{\rm cent}$ and $\sigma_{\rm MF}^{\rm sat}$.
    }
    \label{fig:wgp_lowz_fixed_sigma}
\end{figure*}
In Fig. \ref{fig:wgp_lowz_fixed_sigma} we show $w_{g+}$ versus the scale $r_{\rm p}$ for the LOWZ and SDSS sub-samples,
comparing the measurements in the observed reference samples with those from the \flagship mock samples for three cases of fixed $\sigma_{\rm MF}^{\rm cent}$ and $\sigma_{\rm MF}^{\rm sat}$ values.

In the first case we apply a minimal misalignment to both, centrals and satellites, by setting
$\sigma_{\rm MF}^{\rm cent}=\sigma_{\rm MF}^{\rm sat}=0.05$ (red lines in Fig. \ref{fig:wgp_lowz_fixed_sigma}). The resulting amplitudes are roughly between $2$ and $20$ times higher than those of the reference samples, depending on the sample and the scale. Besides these strong differences, we also find a similarity between the simulation and the reference samples in that the amplitudes increase the brighter and the redder the samples are. This finding might result from the similar magnitude and colour dependence of the galaxy clustering in the observations and the simulation (see Fig. \ref{fig:wgg_lowz_sdss}). 

In the second case of our simplified model we keep the low misalignment for centrals in \flagship while increasing misalignment for satellites by setting $\sigma_{\rm MF}^{\rm cent}=0.05$, $\sigma_{\rm MF}^{\rm sat}=0.75$ (green lines in Fig. \ref{fig:wgp_lowz_fixed_sigma}). We find that the amplitudes strongly decrease at small scales ($r_{\rm p}\lesssim 5 h^{-1}$Mpc), while we see no significant change at large scales in the LOWZ sub-samples and a relatively weak decrease for the SDSS-sub-samples.
In the third case we keep the low misalignment for satellites, while increasing the misalignment for centrals, setting $\sigma_{\rm MF}^{\rm cent}=0.75$, $\sigma_{\rm MF}^{\rm sat}=0.05$ (blue lines in Fig. \ref{fig:wgp_lowz_fixed_sigma}). We now find a strong decrease of the amplitudes at large scales
($r_{\rm p} \gtrsim 5 \, h^{-1}$ Mpc) compared to the first case and only mild or no significant changes at small scales compared to the first case, depending on the sample. These trends persist across a range of satellite fractions in the galaxy samples; see Table~\ref{tab:mocks_summary}.

\begin{figure*}
	\includegraphics[width=\textwidth]{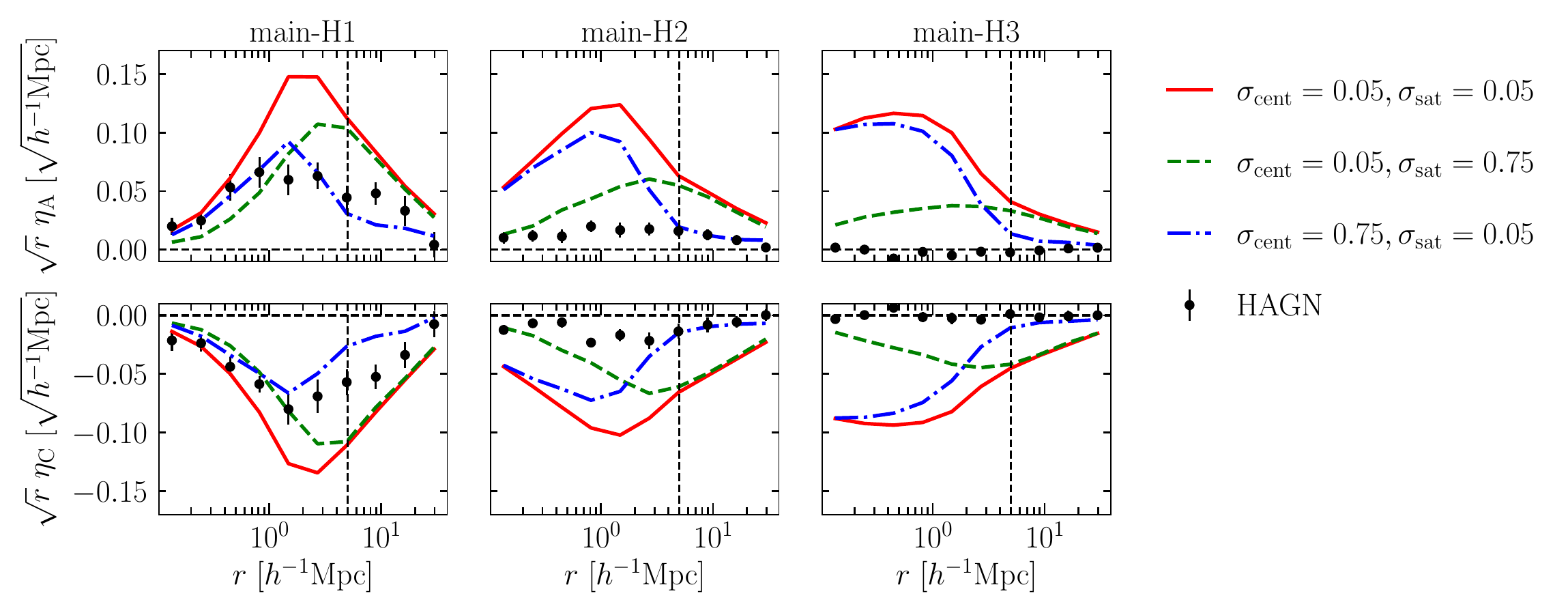}
   \caption{3D IA cross-correlations ($\eta$, as defined in Eq. \ref{eq:eta})
   between a main density sample and the orientations of galaxies in three
   luminosity-limited sub-samples. Orientations are given by the 3D major
   and minor axes $A$ and $C$. Black dots show measurements
   from the HAGN simulation at $z=1.0$ with errors indicating the standard deviation.
   coloured lines show the results from mock HAGN samples from \flagship for three
   simplified IA models with different combinations of fixed values for
   $\sigma_{\rm MF}^{\rm cent}$ and $\sigma_{\rm MF}^{\rm sat}$, analogously to
   Fig. \ref{fig:wgp_lowz_fixed_sigma}. Errors for \flagship measurements are
   similar to those displayed in Fig.~\ref{fig:eta_hagn_flagship} and are not shown for clarity.}
    \label{fig:eta_hagn_flagship_fixed_sigma}
\end{figure*}

We extend this investigation in Fig. \ref{fig:eta_hagn_flagship_fixed_sigma}, where we compare the 3D alignment statistics $\eta_{\rm A}$ and $\eta_{\rm C}$, measured in the three magnitude-limited samples H1 to H3 from HAGN (from bright to faint, as defined in Sect.~\ref{sec:gal_cats})
with the corresponding measurements derived from the mock HAGN samples from \flagship. The \flagship results are shown for the same three \sigMfC, \sigMfS combinations discussed above.

For the measurements in HAGN we find the behaviour expected from results in the literature \citep{Chisari15}, i.e. \etaA (\etaC) have positive (negative) amplitudes, indicating a preference for the galaxies' 3D major (minor) axis to be oriented parallel (perpendicular) to the vector pointing to its neighbours. 

The dependence of the total amplitudes in HAGN on scale and magnitude further follows our expectations, as it decreases with scale and increases with the samples brightness.
The same dependencies on scale and magnitude can be seen in \flagship at large scales ($r_{12} > 5 h^{-1}{\rm Mpc}$) for the three cases of our simplified IA model. However, at small scales the magnitude dependence in \flagship is inverted, as the amplitude decreases with the brightness of the sample.
This finding is interesting, as we have not seen such an inverted small-scale magnitude dependence in the \flagship mocks of the observed samples, where we found \wgp  amplitudes to be lower for the SDSS samples than for the brighter LOWZ samples.

This behaviour poses a challenge
for the calibration of the magnitude dependence of the IA model, as we need to randomise the orientations of faint satellites more than those of bright satellites to obtain an increase of the
signal with brightness in the mock HAGN samples. 
Besides this magnitude dependence, it is interesting to note in Fig.
\ref{fig:eta_hagn_flagship_fixed_sigma}, that increasing the randomisation of orientations only for satellites decreases the signal only at scales below $r_{12} = 5 h^{-1}{\rm Mpc}$
(compare the red and green lines). Increasing the randomisation for centrals on the other hand decreases the signal at large and at small scales
simultaneously (compare the red and blue lines).
Overall we conclude from this preliminary investigation that our IA model provides a high flexibility, as it allows for a significant variation of the IA amplitude in the simulation. The goal is now to
calibrate the dependence of \sigMfC and \sigMfS on galaxy properties, such that the signal matches all reference measurements at all scales simultaneously.

\subsection{Variations across random realisations}\label{sec:noise_from_random_realizations}

The IA model parameters in Eqs. (\ref{eq:P3d}) and (\ref{eq:sigma_model})
determine the probability distributions from which we draw random shapes and
orientations for each galaxy, respectively. For a fixed set of IA model
parameters the shapes and orientations
of individual galaxies will therefore vary across different random realisations. This variation translates into a variation of the resulting
IA signal that is measured in the simulation.

\begin{figure}
	\includegraphics[width=\columnwidth]{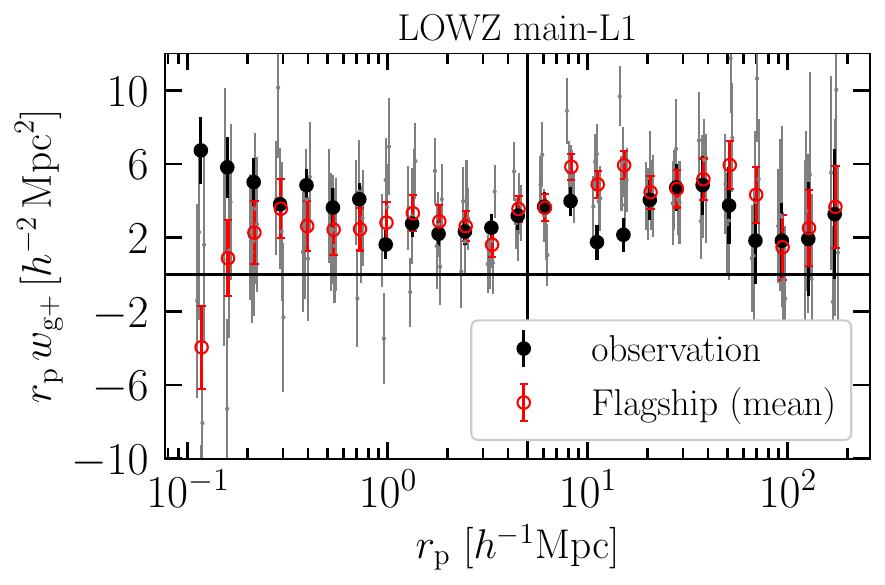}
    \caption{Projected galaxy-ellipticity cross-correlation for the BOSS LOWZ sample L1. Grey dots represent results from five random model realisations using parameters from the final \flagship model. Red dots indicate the average across these random realisations with re-scaled errors, while black dots show observational results.}
    \label{fig:wgp_random_reps}
\end{figure}

We illustrate this effect in Fig. \ref{fig:wgp_random_reps}, where
we show \wgp, measured in five random realisations of the IA model 
for the mock BOSS LOWZ L1 sample. The IA parameters
used for generating these realisations are those used for the
production of the final \flagship IA catalogue. We find a substantial
variation across random realisations, which can reach the same
order of magnitude as the signal itself.
This variation poses a problem when calibrating the model,
as it introduces noise in the cost function that we aim to minimise.
In order to mitigate the impact of that noise
on our calibration, we
run the IA model up to five times
at each point in the parameter space during the calibration.
We obtain the final IA statistics that enter the cost function
by averaging over measurements from the $N_{i}$ random realisations, i.e.
$X_{\rm model} = \sum_i X_{\rm model}^{i} / N_i$,
where $i$ indicates a single random realisation.
These averaged statistics are shown as red symbols in Fig. \ref{fig:wgp_random_reps}.

\section{Misalignment parameter space exploration}\label{sec:param_calibration}

We explore the parameter space of the galaxy misalignment model component following two objectives.
Firstly, we aim to calibrate the model parameters by minimising the deviations between the
IA statistics from the simulation and those from the reference samples.
Our second objective is to trace the probability distribution of the IA model parameters,
which enables us to estimate the uncertainties on the IA contamination in \Euclid-like sample
that is predicted by the simulation.

\subsection{Cost functions}\label{sec:cost_function}

We calibrate the parameters $p_i$ in the \sigMf model from Eq. (\ref{eq:sigma_model}) by minimizing the total
deviation between the IA statistics measured in the different mock IA catalogues from \flagship and those obtained
from the corresponding reference measurements.
For a given sample $\mathrm s$ we define this deviation as
\begin{equation}
\chi_{s}^2 = \sum_{i}^{N_\mathrm{bin}} \frac{(X_\mathrm{model} - X_\mathrm{ref})_i^2}{(\sigma_\mathrm{model}^2 + \sigma_\mathrm{ref}^2)_i} \,,
\label{eq:chisq_red}
\end{equation}
where $X$ is either the projected statistics \wgp in the case of the observational samples from SDSS and LOWZ or the 3D statistic \etaA in the case of the HAGN samples
(defined in Sect. \ref{sec:ia_statistics}). The predictions of the model measured in \flagship and the measurements of the reference samples are denoted by $X_\mathrm{model}$ and $X_\mathrm{ref}$, respectively.
The corresponding standard deviations are denoted as $\sigma_{\rm model}$ and $\sigma_{\rm ref}$, while $N_\mathrm{bin}$ is the number of scale bins in which the statistic $X$ is measured.

When calibrating the misalignment parameters, our aim is to minimise the overall deviation
between \flagship and the reference samples, which we define as the cost function
\begin{equation}
C = \sum_s^{N_{\rm s}} w_s \ \chi_{s, \rm{r}}^2 \,,
\label{eq:cost_function}
\end{equation}
where $N_{\rm s}=9$ is the number of galaxy sub-samples summarised in Table \ref{tab:mocks_summary}, $\chi_{s, \rm{r}}^2 = \chi_{s}^2/N_\mathrm{bin}$ is the reduced $\chi^2$ of sample $s$, 
and $w_s$ is a weight that we assign to each sub-sample.
When calibrating the model parameters for central galaxies, we computed \cost from measurements
of $X$ at small and large scales, respectively, as explained in
Sect.~\ref{sec:param_sampling}, using slightly different weights in both cases.

For large scales ($>5$\mpch), we assign weights of $0.1$ to each of the four BOSS LOWZ sub-samples, resulting in a combined weight of $0.4$. Similarly, we assign weights of $0.2$ to each of the two SDSS samples, which leads to a combined weight of $0.4$ for SDSS and a total observational weight of $0.8$. For the three HAGN sub-samples, we set weights of $0.2/3$ each, which leads to a total weight of $1.0$ for all samples combined. This weighting strategy reduces the overall contribution of the HAGN samples, accounting for the higher confidence we place in the observational constraints and aiming to mitigate potential biases introduced by inaccuracies in the IA signal predicted by HAGN. When computing \cost at small
scales ($<5$\mpch), we neglect the LOWZ samples L1 and L2 by setting their weights to zero.
This is motivated by the strong deviations of the clustering in \flagship from the reference observations (see Fig. \ref{fig:wgg_lowz_sdss}). Since these deviations in clustering also affect the \wgp predictions
from the \flagship IA model, including L1 and L2 may bias the calibration results for satellite parameters.
To maintain the same overall LOWZ contribution to \cost of $0.4$ as on large scales, we increase the weights for the fainter sub-samples, L3 and L4, to $0.2$ each. 

While the cost \cost is a valid measure for the deviations between the \flagship and the reference samples,
it is not suitable for the error propagation technique, which we present in Sect. \ref{sec:error_prop},
since it is not directly related to the likelihood $\mathcal{L}$ that $\sigma_{\rm model}$
and $\sigma_{\rm ref}$ are randomly sampled from the same underlying probability distribution.
In order to establish this relation, we compute, in addition to \cost, the total deviation as
\begin{equation}
\chi^2_\mathrm{tot} = \sum_s^{N_{\rm s}} w_s \ {\chi}_{s}^2.
\label{eq:chisq_tot}
\end{equation}
Assuming a normal distribution of the likelihood for measuring $X$ at a given point $\bf p$ in the parameter space of the model, we can obtain the likelihood as $\mathcal{L} \propto \exp\left\{-\frac{1}{2} \chi_{\text{tot}}^2\right\}$. The weights $w_s$ are the same as those we apply when computing \cost in Eq.~\eqref{eq:cost_function}.

In addition to weighting, our ${\chi}_\mathrm{tot}^2$ definition differs from the common statistics ${\chi}^2$ in two aspects.
1) $X_\mathrm{model}$ is not an analytic prediction but a measurement with non-negligible errors that need to be taken into account when quantifying the significance of the difference between model and reference measurements. 2) We do not take into account the covariance between different scale bins. This simplification is valid if the signals are small and the errors are dominated by shot noise, but future calibration exercises should take into account the full covariance.

\subsection{Parameter sampling method}
\label{sec:param_sampling}

A challenging aspect in the calibration is to efficiently sample the high-dimensional space spanned by
the $2 \times 6$ parameters in Eq. (\ref{eq:sigma_model}) that describe the redshift, luminosity, and colour dependence of the misalignment parameters \sigMfC and \sigMfS in our IA model. This challenge is due to
the numerical cost, which arises in part from generating the simulation, but mainly from the measurements
of the various IA statistics at each sampling point.

In order to reduce the sampling volume in the parameter space and thus lower the computational costs of the calibration, we leverage a finding from our preliminary study in Sect. \ref{sec:IAstats_sigmaMF}. There, we found that on large scales ($> 5$ \mpch), the alignment statistics \wgp and \etaA are most sensitive to the parameters controlling the alignment of centrals ($p_i^{\rm cent}$), while the impact of the parameters controlling satellite alignment ($p_i^{\rm sat}$) is relatively weak. This result allows us to first explore only the six-
dimensional space spanned by $p_i^{\rm cent}$, by comparing the IA statistics from \flagship to those of the reference
samples only at large scales between $5$ and $40$ \mpch
\footnote{Note that the HAGN results may be affected by the limited box size of $100$ \mpch. However, we do not expect this effect to significantly impact our calibration results, as the contributions of measurements to the cost function $\chi_{s}^2$ are down-weighted by their errors, which increase substantially with scale. In addition, we have reduced the weight of the HAGN constraints to the total cost \cost, as detailed in Sect. \ref{sec:cost_function}}. Once we have found a set of best-fit parameters for centrals,
we fix these parameters and calibrate the parameters $p_i^{\rm sat}$ by comparing to the IA reference statistics
at small scales, between $0.1$ and $5$ \mpch.

In both cases, we generate an initial set of sampling points by uniformly sampling the parameter space within
the limits $0 <p_0 < 10$, $-10 < p_1 < 10$, $-0.5 < p_2 < 10$, $-10 < p_3 < 10$, $0.5 < p_4 < 10$, $-10 < p_5 < 0$.
The lower limits for $p_2$ and $p_4$ where chosen to ensure a positive basis of the $\sigma_{\rm mag}$ and $\sigma_{\rm col}$ terms in Eq.~(\ref{eq:sigma_model}) for $M_r<-11$ and $u-r > -0.5$, respectively, hence avoiding
imaginary values for \sigMf within the magnitude and colour range covered by the mock \Euclid sample (see Fig. \ref{fig:zmc_distribution}).
The constraint $p_5 < 0$ is imposed to ensure that \sigMf decreases with the $u-r$ colour index, leading to higher misalignment for bluer galaxies compared to red galaxies, and consequently a lower IA signal.

To further reduce the computational cost of the calibration, we discard sampling points from
the initial distribution for which \sigMf$>2$ within the volume of the redshift-magnitude-colour
space covered by \flagship, i.e.  $0.0<z<2.0$, $-23.0 < M_r < -16.0$; $0.5 < u-r < 3.0$
(see Fig. \ref{fig:zmc_distribution}). Our reasoning behind this choice is that values of
\sigMf$>2$ lead to a distribution of random galaxy orientations that is close to a uniform distribution.
It is therefore not necessary for our \sigMf model to predict higher values in order to include a
zero alignment signal for any object in our \Euclid-like sample as a possible calibration outcome.
This pre-selection allows us to reduce the uniform initial random distribution by over $98\%$ percent.
Note here that we apply this selection only on the initial sampling points, but not on those
that we probe during the subsequent calibration.

Starting from an initial set of $1000$ points, we continue sampling the parameter space
with an iterative procedure, that consists of the following steps:
\begin{enumerate}
    \item Run the IA model on the mock galaxy samples at each sampling point in the parameter space, measure the IA statistics and
    compute the corresponding deviation from the reference samples, quantified by the cost function \cost
    from Eq.~(\ref{eq:cost_function}).
    \item Select the fraction $f_{\rm low}$ of points with the lowest cost and compute their covariance matrix.
    \item Draw $N_{r}$ random points from the Gaussian distribution that is described by the
    covariance matrix from step 2.
\end{enumerate}

For centrals we repeat these steps $12$ times, while gradually decreasing $f_{\rm low}$ from $5\%$ to $0.5\%$ from the first to the final iteration. For satellites we performed only $4$ iterations, decreasing $f_{\rm low}$ from $5\%$ to $1\%$, as we find a faster convergence of the results. In each step of the iteration we generate
$250$ new sampling points, which leads to a total of $4000$ and $2000$ points for centrals and satellites,
respectively, including the initial guess distribution.

As detailed in Sect.~\ref{sec:noise_from_random_realizations}, we average the IA statistics over
$N$ random realisations to mitigate the impact of noise on the model predictions.
We thereby set $N=3$ for an initial `burn-in' phase, which lasts over the first seven iterations
for centrals and is restricted to the initial sampling step for satellites.
Once we find the region with the lowest cost \cost to be visually well covered by new sampling points,
we end the `burn-in' phase and increase the number of iterations to $5$, gaining precision in the model
predictions at the price of higher computational costs.

\subsection{Calibration results}
\begin{table}[t]
\centering
\caption{Parameters of the galaxy-halo misalignment model
used in \flagship; see Eq.~(\ref{eq:sigma_model}). Errors are $1$ sigma uncertainties, derived from the distributions of sampling points.}
\begin{tabular}{cccc}
\toprule
parameter & centrals & satellites & model component \\
\hline \hline
\noalign{\vskip 1pt}
$p_{0}$ & \phantom{0}$1.02^{+2.52}_{-1.02}$ & \phantom{0}$4.74 \pm 1.72 $ & overall amplitude \\
\hline
$p_{1}$ & $-0.41 \pm 0.24$ & \phantom{0}$0.43 \pm 0.34 $ & $\sigma_z$ \\
\hline
$p_{2}$ & \phantom{0}$3.46 \pm 1.85$ &\phantom{0} $5.72 \pm 2.33$ & $\sigma_{\rm mag}$ \\
$p_{3}$ & \phantom{0}$2.71 \pm 0.92 $ &\phantom{0} $2.26 \pm 1.12$ & \\
\hline
$p_{4}$ & \phantom{0}$4.28 \pm 1.27$ &\phantom{0} $4.27 \pm 2.34 $ & $\sigma_{\rm col}$ \\
$p_{5}$ & $-2.43 \pm 0.59$ & $-3.07 \pm 0.88$ & \\
\hline
$\chi_{\rm tot}^2/$d.o.f. & $0.95$ & $1.32$
\end{tabular}
\label{tab:params_misalignment_FS2}
\end{table}

The set of galaxy-halo misalignment parameters for which we find the cost
\cost to be minimal (hereafter referred to
as best-fit parameters) is given in Table~\ref{tab:params_misalignment_FS2}.
These are the parameters which we use to compute the alignment on the full light-cone output
of \flagship.

\begin{figure*}
    \begin{center}
	\includegraphics[width=\textwidth]{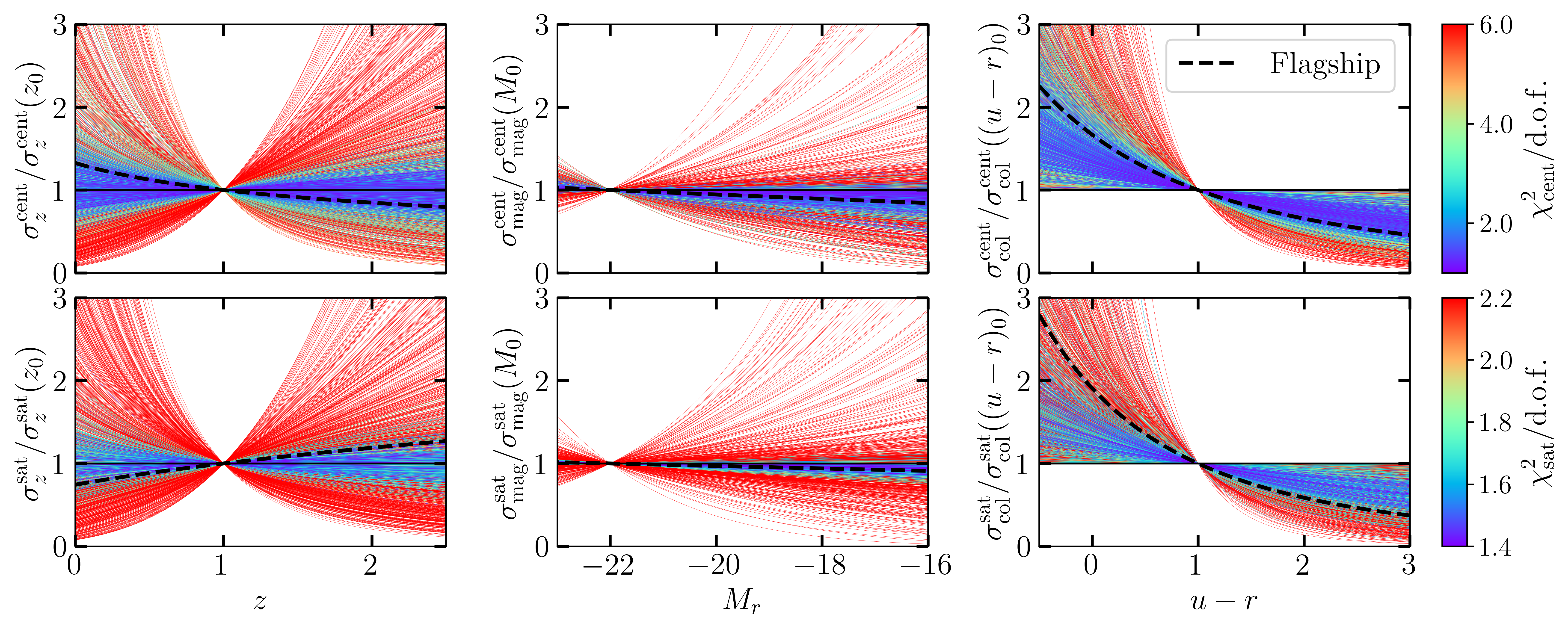}
    \caption{Components of the model for the galaxy misalignment parameter \sigMf from Eq. (\ref{eq:sigma_model})
    for the parameters sampled during the model calibration. Line colours indicate the \chisqtot deviation between simulations
    and reference data (as defined in Eq. \ref{eq:chisq_tot}) per degree of freedom (d.o.f.). The black dashed line corresponds to the model used in the production
    of the \flagship simulation. Top and bottom panels show results for the calibration of satellite and central misalignment, respectively.
    All results are shown
    with respect to the values at $z_0=1.0$, $M_0=-22$, $(u-r)_0=1.0$ to facilitate the comparison of the dependence of the misalignment
    amplitude on the different galaxy properties.}
    \label{fig:sigma_zmc}
    \end{center}
\end{figure*}

In order to obtain a physical understanding of this result,
we show in Fig. \ref{fig:sigma_zmc} the redshift, magnitude, and colour components of our \sigMf model
for the parameter combinations probed during the calibration. Results are shown for central and satellite
galaxies on the top and bottom panels, respectively. They are normalised to the model values
at $z_0=1.0$, $M_0=-22$, $(u-r)_0=1.0$ for a better visual comparison of the scaling with galaxy properties.
The thick black line in this figure shows the model components for our best-fit parameters.
The thin lines show the results for all parameter combinations that were sampled during the 
calibration process. Their colours indicate the corresponding \chisqtot deviations between
the IA statistics in \flagship and the reference samples, defined by Eq. \eqref{eq:chisq_tot}.

Focusing on sampling points with the lowest \chisqtot deviations from the reference samples
($\hat{\chi}_{\rm tot}^2 < 1.5$), we find for both centrals and satellites that the dependence
of the misalignment on redshift and magnitude, quantified by $\sigma_{z}$ and $\sigma_{\rm mag}$,
is weak compared to the colour dependence, which is quantified by $\sigma_{\rm col}$.
Results are distributed around unity, which is consistent with a galaxy misalignment that
is independent on redshift and magnitude. This weak dependence may allow for a reduction
of free parameters in a future version of the IA model, by assuming a constant relationship
of \sigMf with redshift and magnitude. Note that such a constant relationship does not imply
that the alignment signal is independent of these properties. This can be seen in Fig.
~\ref{fig:wgp_lowz_fixed_sigma} and~\ref{fig:eta_hagn_flagship_fixed_sigma}, where
we found a strong variation of the IA signal across different galaxy samples for
fixed misalignment parameters. For centrals, a dependence on magnitude and redshift
can be introduced solely through the mass and redshift dependence of the host halo alignment,
where the former may be affected mass resolution effects, introducing mass dependent
noise in the measured halo orientations
(see discussion in Sect. \ref{sec:orientation_measurements} and H22).
For satellites, such dependencies may instead be introduced via the mass- and redshift-dependence
of host-halo concentrations \citep{Dutton14}.
Therefore, a more detailed physical interpretation of the results may be misleading.

In contrast to $\sigma_{z}$ and $\sigma_{\rm mag}$, $\sigma_{\rm col}$ deviates significantly from a constant
relation, as the values decrease with larger $u-r$ colour.
From our preliminary studies (Fig.~\ref{fig:wgp_lowz_fixed_sigma} and~\ref{fig:eta_hagn_flagship_fixed_sigma}), we expect this decrease of misalignment
to translate into an increase of the alignment signal with galaxy redness.
The trends that we find for the parameter sampling points with the lowest \chisqtot values
are also present for our best-fit model that is defined by the parameters for which the cost \cost is minimal. 
However, the latter shows a stronger dependence on redshift and colour.
We attribute this offset to differences in the definitions of \cost and \chisqtot, as well
as the noise affecting both quantities (see Sect.~\ref{sec:noise_from_random_realizations}),
which may cause a displacement between the sampling point with the lowest \cost value
and the minimum of the underlying distribution.

\begin{figure*}
	\includegraphics[width=\columnwidth]{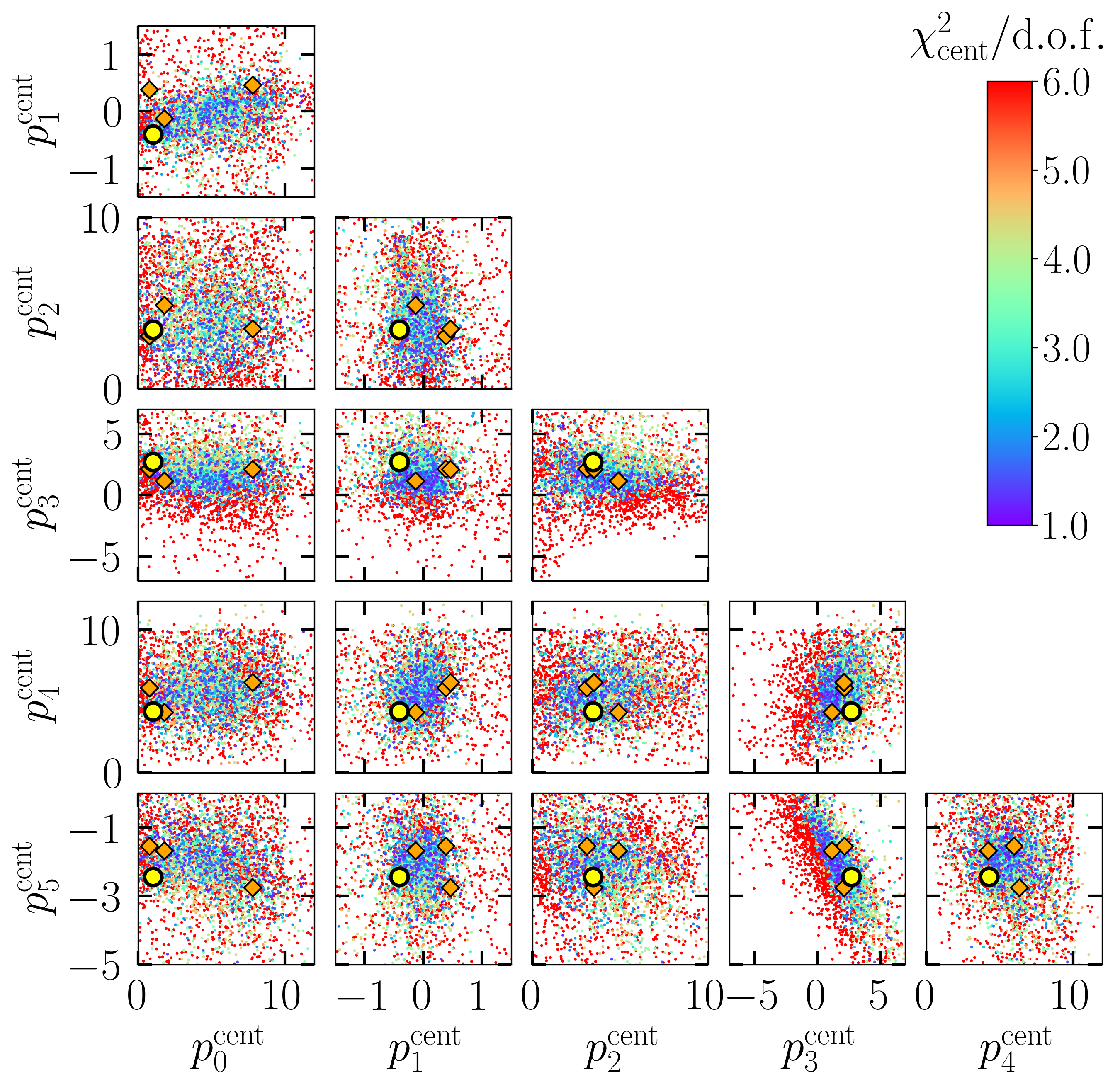}
    \includegraphics[width=\columnwidth]{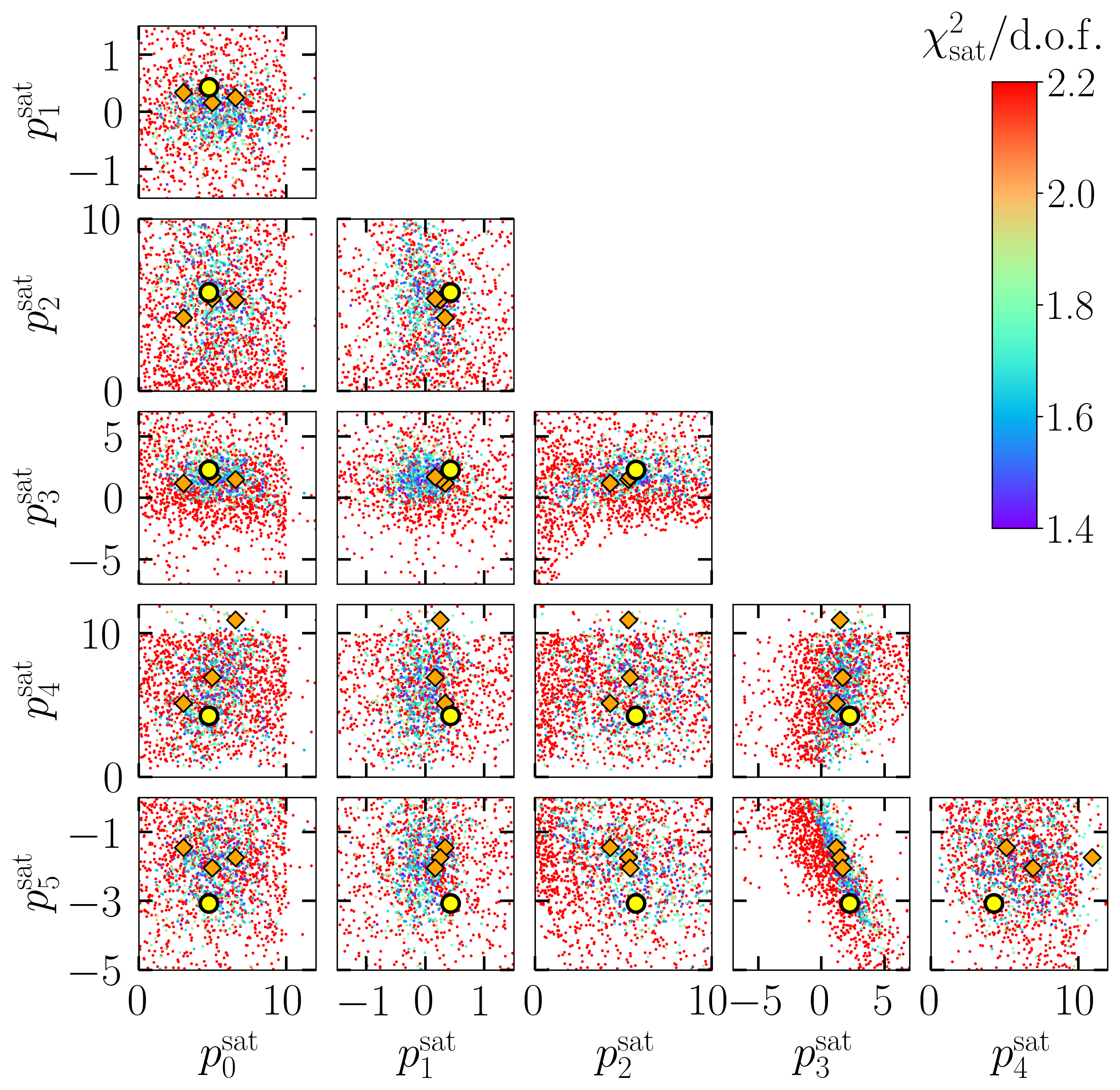}
   \caption{Parameter combinations of the galaxy misalignment model (Eq. \ref{eq:sigma_model})
   at which the simulation was evaluated in the calibration process, described in Sect.~\ref{sec:modelling_orientations}. Colours indicate the \chisqtot (averaged over five realisations, see Sect. \ref{sec:param_sampling}) deviations 
   per degree of freedom (d.o.f.) of different
   IA statistics measured in the simulation from the references measurements (see Eq.~\ref{eq:chisq_tot}).
   Left panels show the parameter combinations for centrals for which the \chisqtot values are derived from
   from IA statistics at scales larger than $5$\mpch. Right panels show the corresponding results for
   parameters of satellites, where \chisqtot is derived at scales below $5$\mpch. The parameters used
   for the IA model in \flagship (see Table \ref{tab:params_misalignment_FS2}) are marked as yellow dots.
   Orange diamonds show the three sampling points with the lowest \chisqtot values.}
    \label{fig:parameter_space}
\end{figure*}

The impact of this noise on our constraints becomes more apparent in Fig.~\ref{fig:parameter_space},
which shows the distribution for the sampling points in the parameter space for centrals and satellites
in the left and right panels, respectively. The colours indicate
the \chisqtot deviation between the IA statistics of the simulation and the reference samples.
Red dots are those with the highest \chisqtot values and indicate the
volume covered by the initial distribution, whereas the bluest dots mark the region in the parameter
space for which we found the best agreement between simulation and reference samples.

There are several interesting aspects to note in Fig.~\ref{fig:parameter_space}.
Firstly, the \chisqtot values are significantly higher for satellites than for centrals,
which could result from the strong deviations between the 3D IA statistics
of \flagship and HAGN at small scales that we see in Fig.~\ref{fig:eta_hagn_flagship_fixed_sigma}.
Second, the shape of the parameter regions with good fits (with \chisqtot$<2$) for centrals and satellites is similar, despite the fact that the misalignment described by these parameters is defined
for very different initial orientations in both cases, as we have discussed in Sect.~\ref{sec:initial_orientations}. One possible explanation for this finding could be that
the constraints are too weak to reveal significant deviations. 
Another potential reason could be a bias introduced by using the same initial guess distribution for centrals and satellites. However, the fact that the best-fit volumes are significantly smaller than the volume spanned by the initial distributions, and are well enclosed by them, argues against a bias imposed by the initial guess.

A third aspect to note is the large uncertainty in the parameter $p_0$, which controls the overall
amplitude of the misalignment and therefore directly the amplitude of the alignment signals.
This is because \chisqtot is affected by the errors
of the IA statistics in the reference samples as well as those from the measurements in the \flagship
simulation (Eq.~\ref{eq:chisq_tot}). Correlations with the parameters $p_1$ and $p_5$ for
centrals and $p_4$ for satellites further contribute to the uncertainty on $p_0$.
Finally, we observe a strong correlation between the parameters $p_3$ and $p_5$, which are the exponents
of the magnitude and colour terms of our misalignment model, and hence describe how
strongly the galaxy misalignment scales with these galaxy properties. A possible explanation
for the correlation of these parameters is the correlation between galaxy colours and
magnitudes in \flagship, which can be seen in projection over redshift in 
Fig.~\ref{fig:zmc_distribution}.

However, we refrain from a detailed physical interpretation of the calibration results, as these are likely influenced by two systematic effects. First, the mass dependence of noise in the measured halo orientations (Sect. \ref{sec:orientation_measurements}) can impact the inferred dependence of galaxy–halo misalignment on galaxy properties. In addition, differences in clustering between \flagship and the observations (Sect.~\ref{sec:wgg_fs_vs_obs}) are expected to further affect the calibration parameters.
Since the clustering also affects the \wgp measurements, an overly high clustering will be 
compensated by an overly low alignment amplitude, and vice versa.
We mitigate this effect by excluding data for which we find the most significant
deviations in \wgg from the calibration, i.e. the measurements from the BOSS
LOWZ samples L1 and L2 at scales below $5$\mpch (Sect.~\ref{sec:param_sampling}).
However, smaller deviations
are still present in the remaining data and their strength depends not only on scale
but also on the luminosity, colour and redshift of the samples. The deviations
in the clustering may therefore affect the dependence of the IA signal
on redshift and galaxy properties, as described by our calibration
parameters. 
Since the statistical uncertainty of \wgp is generally much larger than that of \wgg, significant deviations in the clustering amplitude do not generally translate into significant bias in IA statistics.
Nevertheless, the IA parameters controlling the galaxy-halo misalignment should be regarded
as effective parameters that are specific to \flagship.

\subsection{Validation}
\label{sec:validation}

\begin{figure*}
	\includegraphics[width=\textwidth]{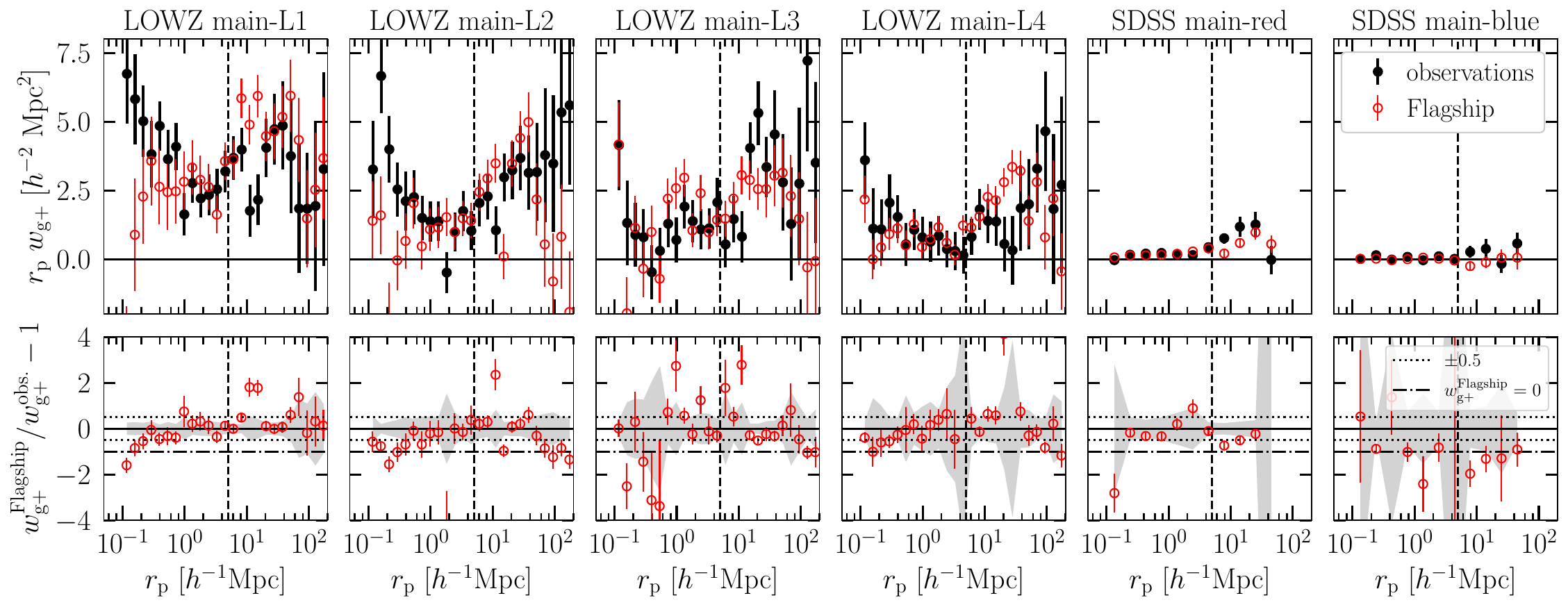}
   \caption{Projected galaxy-ellipticity cross-correlation function \wgp as a function of transverse separation for the observational reference samples (black points) and the corresponding calibrated \flagship mock samples (red points),
   analogous to Fig.~\ref{fig:wgg_lowz_sdss}. The bottom panels show the relative deviation of the \flagship signals from the observations.
   Dashed-dotted horizontal lines at $-1$ in the bottom panels
   indicate the relative deviation expected for a null detection
   of the IA signal in \flagship; dotted horizontal lines indicate a relative deviation of $\pm 0.5$. The dashed vertical lines separate the regimes where central and satellite misalignment parameters were fitted in the simulation. The grey areas indicate the standard deviations of the observations.}
    \label{fig:wgp_lowz_sdss}
\end{figure*}

For validation we compare the predictions from our best-fit \flagship IA model, averaged over five realisations, against the reference samples used for calibration.
In Fig. \ref{fig:wgp_lowz_sdss}, we show the projected cross-correlation \wgp for
the four luminosity sub-samples of the BOSS LOWZ sample and the colour sub-samples
of the SDSS main sample (analogous to Fig. \ref{fig:wgp_lowz_fixed_sigma}).
For the LOWZ sub-samples, we find an overall good agreement between
results from \flagship and the observational reference, as the discrepancies
are within the standard deviations of the measurements.
However, for the brightest samples L1 and L2 and at scales of less than $1\,h^{-1}\,{\rm Mpc}$,
we find the \flagship results to be consistently below the observational results. Note that
for these samples, we also found the strongest deviations from the reference samples
in clustering and have therefore set their weight in the cost function
to zero when calibrating the misalignment parameters for satellites
(Sect.~\ref{sec:param_calibration}). The simulation is hence not calibrated to
match the LOWZ observations in this regime.

For the red SDSS sub-sample we find the deviations between the \flagship results and
the observations to be consistent overall within the errors, although
we note that at scales above $5$ \mpch the \flagship results are consistently
below the observations.
The \flagship results for the blue SDSS sample are consistent with zero
at all scales and therefore below the observational reference.
However, also here the deviations between \flagship and observations are consistent within
the errors.

\begin{figure*}
	\includegraphics[width=\textwidth]{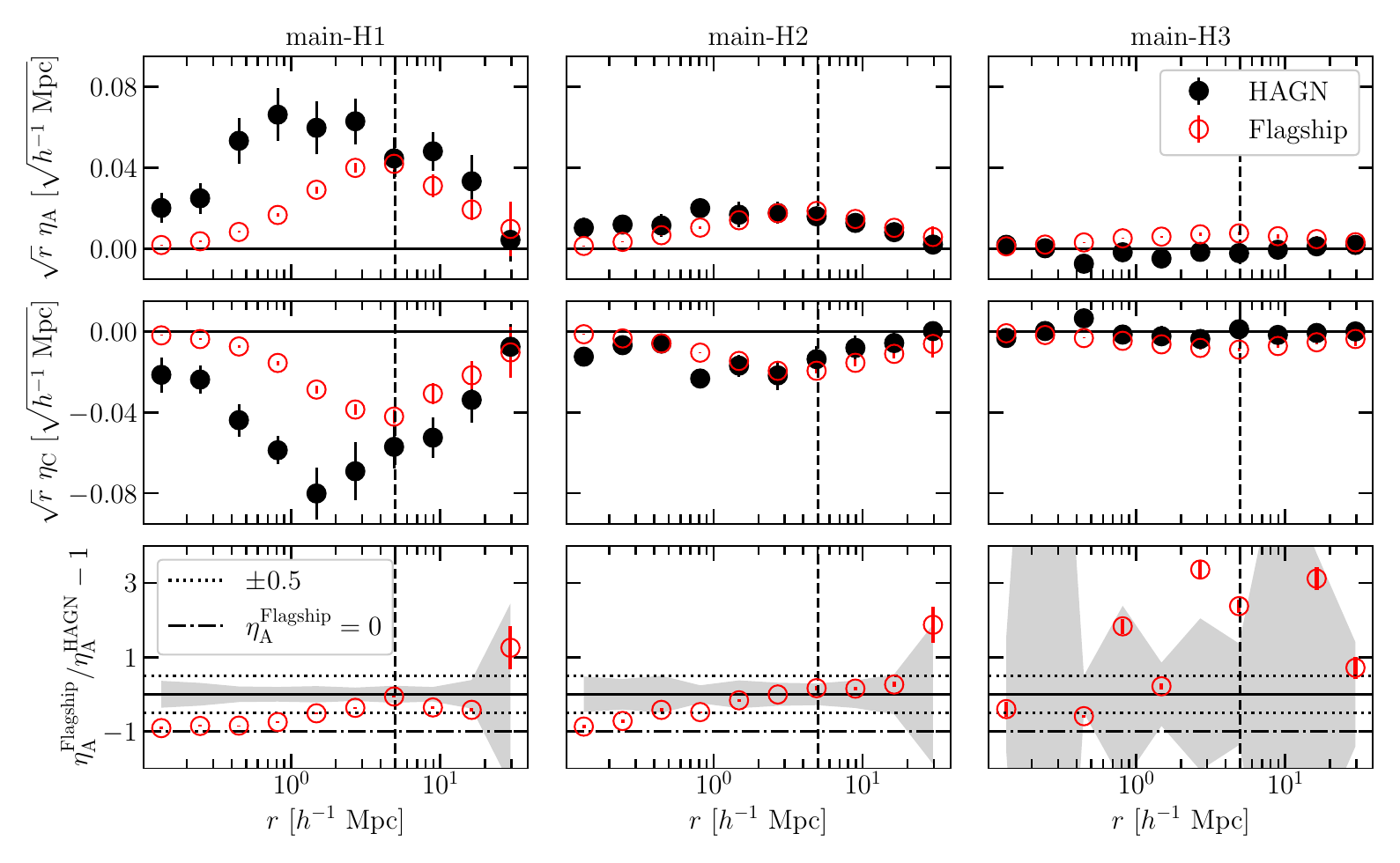}
   \caption{{\it Top:} 3D correlations between galaxy major axis $\bf A$ and the vector {$\bf r$} pointing
   to the position of a neighbouring galaxy (as defined in Eq. \ref{eq:eta}). Black and red symbols
   show measurements from the comoving output of HAGN at $z=1.0$ and \flagship light-cone at the same
   redshift respectively. Results are shown for the magnitude limited samples H1 to H3 (from bright to dim),
   as defined in Table \ref{tab:mocks_summary}. Error bars indicate the Jack-knife estimates
   of the standard deviation. {\it Center:} Analogous to the top panel for the
   correlation of the galaxy minor axis $\bf C$. {\it Bottom:} Relative deviations of the \flagship measurements
   from HAGN. Grey areas indicate the standard deviations of the HAGN measurements. Dashed lines
   mark $100\%$ deviations.}
    \label{fig:eta_hagn_flagship}
\end{figure*}

In Fig. \ref{fig:eta_hagn_flagship} we compare the
3D IA statistics $\eta$, defined in Eq. (\ref{eq:eta}), from \flagship with the reference measurements from HAGN
for the three magnitude-limited samples H1 to H3.
The top panels display \etaA, which measures the average alignment between the major axis of a galaxy 
in a magnitude sub-sample with the direction of a neighbouring galaxy from the main sample. Similarly, the central panels show \etaC, which quantifies the same alignment but for the galaxy's minor axis (analogously to Fig. \ref{fig:eta_hagn_flagship_fixed_sigma}). The bottom panels show the relative deviations between \flagship and HAGN.

We find that \flagship matches the overall trends predicted by HAGN, showing a decrease in amplitude for fainter samples and at both small and large scales. However, our comparison also reveals several discrepancies:
At scales above $5$~\mpch we find the $\eta_A$ measurements from \flagship to deviate
from HAGN by up to $50\%$ for the bright samples H1 and H2. While deviations in the largest-scale bin are greater, they are consistent with the increased measurement errors at that scale. For the faint sample, H3, we observe similar absolute deviations as for the brighter samples. However, due to the low amplitude of the signal, these absolute deviations result in large relative deviations of up to a factor of five.
At scales below $5$~\mpch, we find the \flagship results to be up to a factor 2 below those
from HAGN, as the signal in \flagship decreases towards zero. 
This discrepancy is most notable for the brighter samples, H1 and H2, where the HAGN measurements exhibit a high signal-to-noise ratio.

Interestingly, we achieved better agreement between \flagship and HAGN for the brightest sample, H1, at small scales when fixing the misalignment parameters to \sigMfC$=0.75$ and \sigMfC$=0.05$ in our preliminary analysis (see Fig. \ref{fig:eta_hagn_flagship_fixed_sigma}). This demonstrates that the IA model can replicate the scale dependence of the HAGN signal within a specific luminosity and redshift range.
However, obtaining highly accurate matches between the constraining datasets and \flagship across all luminosity, colour, redshift, and scale ranges simultaneously, while using misalignment parameters that vary based on galaxy properties, remains a challenge for our model.

This difficulty may stem from three primary factors. First, it is possible that the IA signal in HAGN is inconsistent with the observational constraints, making it difficult to match both HAGN and observational data, even with a realistic IA model. Second, there may be limitations in our modelling approach, including: a) insufficient flexibility in the model to capture the full redshift, magnitude, and colour dependence of galaxy misalignment, b) shortcomings in the model calibration process, or c) the relatively low weighting of the HAGN sample in the cost function used for calibration.  Third, inaccuracies of the \flagship properties, such as the clustering, colour, or luminosity distributions may reduce the accuracy of the IA model predictions, even if the latter is sufficiently realistic. These issues, either individually or in combination, may contribute to the discrepancies we observe for the HAGN H1 sample
at small scales.
However, given the overall good agreement between \flagship and HAGN when considering all samples and all
scales, we conclude that our IA model is sufficiently realistic to serve as a test-bed for analytical IA models.

The simulated galaxy orientations and the simulation code are publicly available%
.\footref{fn:cosmohub} \footref{fn:genIAL}

%% file: sections/euclid_predictions.tex
\section{Uncertainty propagation}
\label{sec:error_prop}

\begin{table}[h]
\caption{Best-fit parameters of Eq.~(\ref{eq:aia_sigmamf}), describing the empirical relation between IA amplitude and the scatter of the von Mises--Fisher distribution, \sigMf, for the observational galaxy samples (both central and satellites) used for calibration in this work.}
\begin{center}
\begin{tabular}{llrrr}
\hline
\noalign{\vskip 1pt}
sample & type & $a$ & $b$ & $c$ \\
\hline\hline
\noalign{\vskip 1pt}
LOWZ L1   & cen &   0.01 & 1.10 & $-2.40$ \\     
& sat & 	    8.75 & 68.26 & $-2.33$     \\
LOWZ L2   & cen &   0.01 & 0.77 & $-$2.42    \\  
& sat & 	    $-$0.33 & 52.86 & $-$2.38    \\
LOWZ L3   & cen &   0.01 & 0.56 & $-$2.16    \\  
& sat & 	    2.40 & 54.57 & $-$2.49     \\
LOWZ L4   & cen &   0.01 & 0.47 & $-$2.19    \\  
& sat & 	    $-$0.62 & 44.53 & $-$2.26    \\
SDSS red  & cen &   0.00 & 0.15 & $-$2.18    \\  
& sat & 	    0.29 & 12.96 & $-$2.29     \\
SDSS blue & cen &   0.00 & 0.15 & $-$2.04   \\  
& sat & 	    0.33 & 5.65 & $-$2.32      
\end{tabular}
\label{tab:fit_error_prop}
\end{center}
\end{table}

To eventually derive informative priors on IA model parameters, we need to propagate the uncertainty in the \flagship IA model inherited from the statistical uncertainty in the calibration observations (and simulations), as well as from the finite volume of \flagship. Both of these sources of uncertainty are included in the $\chi^2$ of the calibration fits; see Eq.~(\ref{eq:chisq_red}). Propagated into the IA signals measured from the \flagship mock, these uncertainties can then be translated into the permitted range of the parameters of analytic IA models fitted to the mock datasets, which can serve as informative priors for cosmological inference.
Full error propagation, e.g. via Monte Carlo sampling, is computationally expensive because at each step the semi-analytic model needs to be optimised, and IA summary statistics re-measured. In this work we limit ourselves to order-of-magnitude estimates via an approximate scheme.

As shown in Fig.~\ref{fig:wgp_lowz_fixed_sigma}, to good approximation we can control the small-(large-)scale IA amplitude via the satellite (central) model parameters of the semi-analytic model, with the transition at around $5$ \mpch.
We derive the covariance ${\rm Cov}(\vec{p})$ of the fit parameters $p_{0..5}$ for both centrals and satellites from the distributions of \chisqtot shown in Fig.~\ref{fig:parameter_space}.
We linearly propagate the uncertainty on $p_{0..5}$ into the parameter controlling the misalignment distribution,
\begin{equation}
\label{eq:sigmamf_perr}
    {\rm Var}(\sigma_{\rm MF}) = \sum_{i,j=0}^5 \frac{\partial \sigma_{\rm MF}}{\partial p_i} {\rm Cov}(\vec{p})_{ij} \frac{\partial \sigma_{\rm MF}}{\partial p_j} \;,
\end{equation}
where the derivatives are determined analytically from Eq.~(\ref{eq:sigma_model}).

We generate mock realisations of the SDSS samples used for calibration, systematically varying 
\sigMf separately for central and satellites (while holding the other fixed) and measuring the amplitude of the resulting IA signal on scales above and below $5$~\mpch, which roughly separates the regimes where centrals and satellites drive the IA amplitude. The resulting relations are well fit by the empirical model
\begin{equation}
\label{eq:aia_sigmamf}
    A_{\rm IA} = a + b \exp \left( c \sigma_{\rm MF} \right)\;,
\end{equation}
with the best-fit coefficients for the free parameters $a,b,c$ per galaxy sample listed in Table~\ref{tab:fit_error_prop}. The offset $a$ is generally small for all samples, while $c$ shows only little variation with an average value around $c=-2.3$ because this scaling is a feature of the von Mises--Fisher distribution rather than a sample-dependent property. The parameter $b$ scales with the luminosity of the sample, weakly for centrals, and approximately exponentially for satellites. 
Some residual luminosity trends are to be expected because the IA amplitude is not solely determined by either satellites or centrals, e.g. due to cross-correlations between centrals and satellites (cf. Fig.~\ref{fig:wgp_lowz_fixed_sigma}). Equations ~(\ref{eq:sigmamf_perr})~and~(\ref{eq:aia_sigmamf}) can now be combined to propagate errors from the \flagship IA model to the amplitude of an intrinsic alignment correlation function, separately on small and large transverse scales. Through Eq.~(\ref{eq:sigma_model}) this propagation depends on the galaxy sample's typical redshift, luminosity, and colour.

Applying this scheme using typical values for the galaxy samples used for calibration, we find standard deviations on $A_{\rm IA}$ that exceed the parameter value, even for bright samples with strong signals and on both large and small transverse scales. This is caused by the additional uncertainty due to the finite volume of the \flagship in Eq.~(\ref{eq:chisq_red}). The two sources of statistical error in the denominator have roughly the same order of magnitude: while the real SDSS observations typically cover a larger volume, the \flagship measurements have been averaged over five shape noise realisations. 

This means that, in order to obtain informative priors on IA parameters in the future, we need not only compile as many informative direct IA observations (and hydrodynamic simulation measurements) as possible over a wide range of galaxy sample properties, but also implement the IA models into simulations that exceed the constraining power of the observations by at least an order of magnitude to suppress additional statistical uncertainty beyond those of the observations to less than $10\,\%$, e.g. by using multiple light-cone realisations. While the HOD parameters in \flagship are held fixed here, their scatter against observational calibrators is likely to constitute another significant source of uncertainty in the overall IA model, especially on small scales. In conclusion, we are currently unable to derive meaningful, informative prior constraints on the amplitude of IA signals in \Euclid's weak lensing samples directly from \flagship.

\section{IA contamination in \Euclid-like samples}
 \label{sec:ia_contamination_euclid}

As a first application of our semi-analytic IA model, we predict the expected contribution of IA signals to the observed cosmic shear signals in a \Euclid-like survey. We select all galaxies brighter than $I_{\rm E}=24.5$ and with valid photometric redshift estimates in the full octant of \flagship. The latter were obtained with a template-based method (see Paltani et al., in prep.) applied to the mock photometry of the near-infrared \Euclid bands and LSST $ugriz$ fluxes to a depth expected by the time of \Euclid's final data release. The resulting galaxy sample has a number density of $26\,{\rm arcmin}^{-2}$ on the sky, close to, but slightly higher than, \Euclid's nominal wide survey expectation.

We divide the sample into 13 equipopulated tomographic bins via the first mode of the photometric redshift probability density function. This choice ensures good redshift resolution and a fairly even distribution of statistical errors across the signals we extract. For every combination of tomographic bins we measure angular pseudo-power spectra of galaxy ellipticity correlations up to a maximum multipole of $\ell=3000$. Details of the methodology are described in \cite{EP-Tessore}; briefly, the power spectra are calculated from \texttt{Healpix} \citep{gorski05} maps at a resolution of $N_{\rm side}=2048$, shape noise-subtracted, and not corrected for the effect of the limited footprint of the mock. Since data is available over a full octant, the mixing between multipoles, and between $E$- and $B$-modes, is limited to large scales and has minor impact, particularly in the power spectrum ratios that we consider.

\begin{figure*}%
    \centering%
	\includegraphics[width=\textwidth]{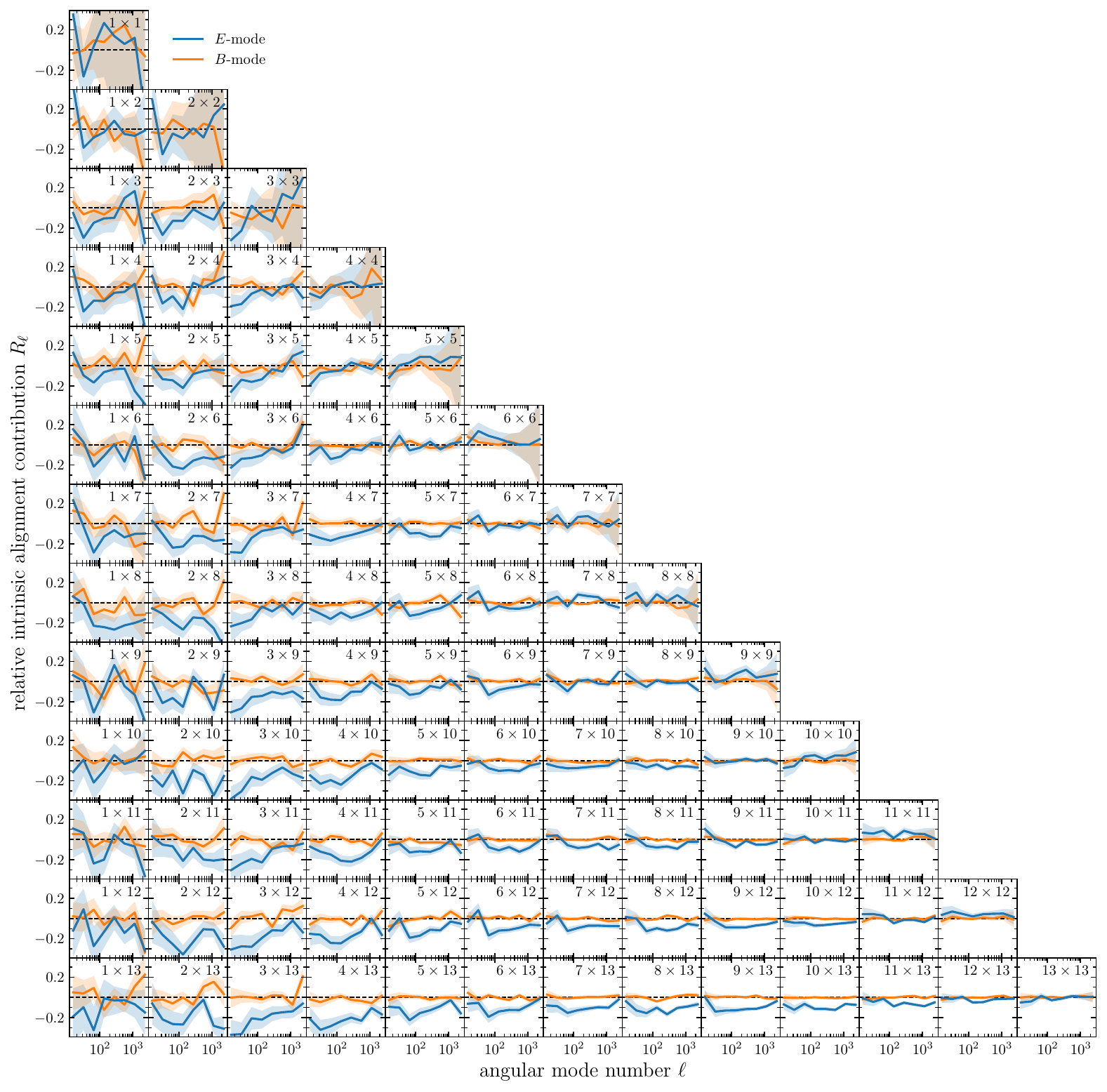}%
    \caption{Relative intrinsic alignment (IA) contribution $R^{ij}_{\ell}=C^{ij}_{\rm IA}/(C^{ij}_{\rm IA}+C^{ij}_{\rm GG})$ to a tomographic cosmic shear analysis in a \Euclid-like galaxy sample divided into 13 photometric redshift bins. Panels correspond to the different tomographic bin combinations, with auto-correlations along the diagonal and higher-redshift bins towards the bottom and right. Blue (orange) lines show the relative change to due intrinsic alignments to the $E$-($B$-)mode angular shear power spectra. Bands around the signals indicate the standard deviations as determined from a Jackknife covariance.}
    \label{fig:IA_euclid_sample}
\end{figure*}

We separate the contributions to the ellipticity correlations by the cosmological signal (i.e. the gravitational shear correlations, denoted by GG) and the combined signal from intrinsic ellipticity correlations (II) and shear-intrinsic cross-correlations (GI). Figure$\,$\ref{fig:IA_euclid_sample} shows the relative contribution of the combined II and GI signal to the total ellipticity correlation in each tomographic bin combination.

As expected, the $B$-mode contribution is consistent with zero throughout as our model is based on tidal, as well as radial satellite, alignments, which only generate $B$-modes at higher order \citep{Hirata04,schneider10}.
In the $E$-mode, the IA contamination is fairly strong overall, reaching the $10\,\%$ level and beyond over a broad range of redshifts and angular scales. This result should be interpreted with caution, not only because of the expected uncertainties in the IA model parameters discussed in Sect. \ref{sec:error_prop}, but also owing to potential inaccuracies in the HOD model, for which we found indications when comparing the colour–magnitude distribution in Flagship with COSMOS observations (Sect. \ref{sec:gal_cats:cosmos}). Nevertheless, it is worth noting that our model predictions are of the same magnitude as the analytic forecasts made by \citet{fortuna21a} who also considered a calibrated IA model for both central and satellite galaxies. Depending on modelling choices, they found a systematic shift in the parameter $S_8$ by up to $5\,\%$, which roughly corresponds to an average bias of $10\,\%$ in power spectrum amplitude.

The figure shows that IA makes a positive contribution in tomographic auto-correlations due to the II term, and a negative contribution to tomographic cross-correlations where the GI term dominates. The GI term tends to be most significant when the tomographic bins are far apart in redshift so that the lensing efficiency of the foreground structures that align nearby galaxies is large. It is noteworthy that the opposing trends of the II and GI contributions are capable of breaking any degeneracy with shifts in the mean of the redshift distributions in the tomographic bins, which would move all signals coherently to larger or smaller amplitudes. This degeneracy breaking is not observed when binning more coarsely into fewer tomographic bins, as is done in current surveys mostly due to the lower overall galaxy number densities \citep[see e.g. the discussion in][]{DESpKiDS23}. Nevertheless, cross-talk between residual uncertainty in the modelling of IA and of tomographic redshift distributions will remain a major issue for \Euclid and other Stage-IV surveys \citep{fischbacher23,leonard24}.

%% file: sections/conclusions.tex
\section{Summary and conclusions}\label{sec:Conclusions}

This work describes the addition of realistic intrinsic alignments (IA) of galaxies to \Euclid's \flagship simulation and mock galaxy catalogue \citep{EuclidSkyFlagship}, providing a test bed for the analysis choices and methods applied to forthcoming weak gravitational lensing analyses of \Euclid.

Our semi-analytic IA model consists of two main components: firstly, we randomly sample galaxy ellipticities conditional on the redshift, luminosity, and colour in the simulation such that they match the corresponding distributions observed in the COSMOS Survey \citep{scoville07}. Second, to determine alignment, central galaxies follow the orientation of their host halo, while satellites align radially towards the halo centre.
Unlike previous work \citep[e.g.][]{Heymans06,joachimi13b,MICEIA}, we apply the same
model to both early- and late-type galaxies instead of aligning the minor axis of blue central galaxies with the halos' angular momentum vector, assuming they are
rotationally supported discs. 
The strength of alignment is controlled separately for centrals and satellites via random misalignments, which depend via power laws on galaxy redshift, luminosity, and colour.
To calibrate these dependencies, we employ constraints from multiple IA observations from the Sloan Digital Sky Survey (SDSS; \citealp{Singh16,Johnston19}), providing constraints up to redshift $z=0.36$. Since \Euclid weak lensing samples will reach $z\simeq2$ and beyond, we compensate for the current dearth of observational IA constraints at high redshifts by using measurements in three luminosity sub-samples from the Horizon AGN hydrodynamic simulation \citep{Dubois14} at $z=1$. However, the latter samples are downweighted in the model optimisation as there is still considerable uncertainty in IA predictions from simulations.

Optimising a total of $12$ alignment parameters, we found an overall good agreement between \flagship and the different reference data sets, with close matches in the distributions of axis ratios to all COSMOS samples and agreement in alignment statistics mostly within the $1\sigma$ errors.
However, significant discrepancies occurred when comparing \flagship against the brightest Horizon AGN sub-sample on small scales. These deviations may result from the low weight that we assigned to Horizon AGN in the optimisation, or might indicate that the IA observations and Horizon AGN simulations are difficult to reconcile with a simple model, despite the long redshift baseline. A future, more detailed comparison of semi-analytic IA models against a range of hydrodynamic simulations will be required to shed light on these model limitations. 
Significant discrepancies also occur in the small-scale clustering of the brightest SDSS-LOWZ sub-sample. Since we exclude these scales from the IA parameter optimisation, low-level differences are also apparent in the corresponding alignment signals. Improvements here require a revision of the hybrid halo occupation and abundance matching approach for the \flagship galaxy mock.

For both centrals and satellites the model matches the constraints best for parameters that set a strong dependence of the galaxy misalignment on colour and a relatively weak dependence on magnitude and redshift, in agreement with the results from \citet{MICEIA} who employed a similar model.
This result suggests that our IA model could potentially be simplified by eliminating the direct dependence on magnitude and redshift in the galaxy misalignment. Such a simplification would reduce the dimension of the model parameter space and remove parameter degeneracies. A simpler model is desirable to increase the predictiveness of the IA model given the limited number and constraining power of the calibration data sets.

Propagating the statistical uncertainty due to the calibration data and the limited precision of the \flagship mock measurements, we found the combined uncertainties on the IA amplitude currently still too large to provide informative bounds on IA parameter ranges.
Nevertheless, we conclude that the best-fit calibrated IA model presented in this work is a useful tool for creating realistic mock IA data sets for a wide range of observed one- and two-point statistics of galaxy ellipticities. As a first application, we predicted the IA contamination to a tomographic weak lensing measurement akin to what can be expected for the final \Euclid data release and found it to be significant, with $\sim 10\,\%$ IA contribution in some tomographic bin combinations, in line with earlier predictions.

A companion paper (Paviot et al., in prep.) will investigate how well common analytic IA models can fit the combined clustering and IA measurements from \flagship, informing some of the modelling choices for weak lensing inference from \Euclid's first data release as well as comparing simulation predictions with observational samples not used in the calibration procedure.
In line with forthcoming updates of the \flagship galaxy models, we will also improve the IA generation. This will include an expanded observational calibration data set, hydrodynamic simulation measurements from several projects \citep[e.g.][]{nelson2019,pakmor2023,schaye2023} and over a wider redshift range, multiple lightcone realisations, and tests of current modelling assumptions such as the independence of galaxy shapes and orientations, and the spatial isotropy of misalignments.